\documentclass[11pt]{article}

	\usepackage{fullpage}
	\usepackage{graphicx}
	\usepackage{amsfonts}
	\usepackage{amsmath}  
	\usepackage{amsthm}
	\usepackage{amssymb}
	\usepackage[applemac]{inputenc} 
	\usepackage{fancyvrb}
	\usepackage{multicol}
	\usepackage{courier}
	\usepackage{hyperref}
	\usepackage[font={sl}]{caption}
	\newcounter{theorem_c}
	\numberwithin{theorem_c}{section}
	\newtheorem{theorem}[theorem_c]{Theorem}

	\newtheorem{hyp}[theorem_c]{Hypothesis}

	\newcommand{\naturals}{\mathbb{N}}
	\newcommand{\integers}{\mathbb{Z}}
	
	\newcommand{\reals}{\mathbb{R}}

	\newcommand{\eqdef}{\stackrel{def}{=}}
	\newcommand{\eqdeftemp}{\stackrel{\Delta}{=}}
	\newcommand{\imply}{\Rightarrow}
	\newcommand{\suchthat}[2]{\{#1 \text{ s.t. } #2\}}

	\newcommand{\expect}[1]{\mathbb{E} \, #1}
	\newcommand{\avg}[1]{\bigl\langle #1 \bigr\rangle}
	\newcommand{\sgn}{\operatorname{sgn}}
	\newcommand{\Av}[1]{\operatorname{Av} \, #1}
\begin{document}

	\title{\textbf{Aspects of Statistical Physics\\ in Computational Complexity}}
	\author{Stefano Gogioso\\
		\footnotesize{Quantum Group},
		\footnotesize{Department of Computer Science}\\
		\footnotesize{University of Oxford, UK}\\
		\href{mailto: stefano.gogioso@cs.ox.ac.uk}{\small{stefano.gogioso@cs.ox.ac.uk}}
	}
	\date{24 June 2013}
	\maketitle

	\setlength{\parindent}{0pt}
	\numberwithin{equation}{section}

	\begin{abstract}

		The aim of this review paper is to give a panoramic of the impact of spin glass theory and statistical physics in the study of the K-sat problem, as summarised by the words of Amin Coja-Oghlan (\cite{survey_cojaOghlan}, Warwick 2010)
		\begin{center}
		\textit{"Random K-sat is a spin glass problem (with a combinatorial flavor)"}.\\
		\end{center}

		The introduction of spin glass theory in the study of the random K-sat problem has had profound effects on the field, leading to some groundbreaking descriptions of the geometry of its solution space and helping to shed light on why it seems to be so hard to solve. Most of the geometrical intuitions have their roots in the Sherrington-Kirkpatrick model of spin glass: its simple formulation and complex free-energy landscape make it the ideal place to start our exploration of the statistical physics of random K-sat.\\

		We'll start Chapter \ref{chapter_SpinGlassFundamentals} by introducing the SK model from a mathematical point of view, presenting some rigorous results on free-entropy density and factorisation of the Gibbs measure and giving a first intuition about the cavity method. We'll then switch to a physical perspective and start exploring concepts like pure states, hierarchical clustering and replica symmetry breaking in the sandbox provided by the SK model.\\

		Chapter \ref{chapter_FromSKtoKsat} will be devoted to the spin glass formulation of K-sat: we'll introduce factor graphs, draw the connection between pure states and clusters of solutions, and define the complexity. The most important phase transitions of K-sat (clustering, condensation, freezing and SAT/UNSAT) will be extensively discussed in Chapter \ref{section_RSB}, with respect their complexity, free-entropy density and the so-called \textit{Parisi 1RSB parameter}: rigorous results and physically-inspired analysis will blend together to give as much intuition as possible about the geometry of the phase space in the various regimes, with a special focus on clustering and condensation.\\

		The so-called algorithmic barrier will be presented in Chapter \ref{chapter_KsatAlgorithms} and exemplified in detail on the Belief Propagation (BP) algorithm. The BP algorithm will be introduced and motivated, and numerical analysis of a BP-guided decimation algorithm will be used to show the role of the clustering, condensation and freezing phase transitions in creating an algorithmic barrier for BP.\\

		Taking from the failure of BP in the clustered and condensed phases, Chapter \ref{cavityMethod} will finally introduce the Cavity Method to deal with the shattering of the solution space, and present its application to the development of the Survey Propagation algorithm.
	\end{abstract}

	\newpage

	\tableofcontents
	\newpage
%% END TITLES

%% BEGIN Introduction
\section{Introduction}
	\label{chapter_Introduction}

	\subsection{The K-sat problem}

		The K-satisfiability problem, better known as \textbf{K-sat}, asks whether one can satisfy $M$ given constraints over $N$ boolean variables, i.e. whether there is an assignment $b$ of boolean values to the $N$ variables that satisfies a given K-CNF $\mathcal{I}$ (the assignment is then called a \textbf{satisfying assignment}, or a \textbf{solution} for $\mathcal{I}$). 
		A \textbf{CNF} is a boolean formula involving only connectives $\vee$, $\wedge$ and $\neg$ and written in conjunctive normal form, i.e. as the conjunction of many \textbf{clauses}, each clause being the disjunction of many literals (a literal is either a variable $x_i$ or the negation $\neg x_i$ of a variable). 
		A \textbf{K-CNF} is a CNF where each clause has exactly K literals.\\

		Despite the simplicity of its formulation, there is no known (deterministic) algorithm which can solve the K-sat problem in polynomial time.
		There are, on the other hand, efficient solvers that can find solutions with high probability for reasonably low \textbf{constraint density} $\alpha \eqdef M/N$: an understanding of the properties of these algorithms goes through an understanding of statistical properties of the K-sat solution space.\\

		We call \textbf{random K-sat} the (conceptual) variant of the K-sat problem where the instance $\mathcal{I}$ is allowed to be a random variable
			\footnote{From now on we'll write \textbf{r.v.} for \textbf{random variable}.} 
		(and thus so is its solution space), governed by some probability measure over the space of instances of K-sat. 
		Two commonly used measures are
		\begin{enumerate}
			\item[(a)] the \textbf{uniform model}: the instance $\mathcal{I}$ is chosen with the uniform probability in the space of instances of K-sat (and the uniform probability measure is introduced on its solution space).
			\item[(b)] the \textbf{planted model}: the issue with the uniform model is that the instance could be unsatisfiable, so one first choses a random assignment $b$ of boolean values to the variables, and then chooses M clauses satisfied by it, uniformly over all such clauses.
		\end{enumerate}

		The two models induce different measures on the solution space $\suchthat{(\mathcal{I},b)}{b \text{ satisfies }\mathcal{I}}$: the former yields immediate results on algorithms but is very hard to work with, while the second is easy to work with but hard to connect to algorithms.\\

		The idea behind the key paper \cite{core_algorithmicBarriers} (which will prove the existence of a clustering phase transition in the solution space of $\mathcal{I}$) is to work in the planted model (where the slightly higher abundance of solution-rich instances allows a successful use of the so-called \textit{2nd moment method}), and then transfer the results to the uniform model via the following 
		\begin{theorem} (Transfer theorem)\\
			There is a sequence $\xi_K \rightarrow 0$ s.t. if a property holds with probability $1-\exp[-\xi_K \, N]$ in the planted model then it holds with high probability (see below) in the uniform model.
		\end{theorem}
		When talking about an event $\mathcal{E}(\mathcal{I})$ depending on a random instance $\mathcal{I} \equiv \mathcal{I}(N,M)$ of K-sat, we'll say that $\mathcal{E}(\mathcal{I})$ happens \textbf{with high probability (w.h.p.)} iff $\lim\limits_{N \rightarrow \infty, M/N \rightarrow \alpha} \mathbb{P}(\mathcal{E}(\mathcal{I})) = 1$ .\\

	\subsection{Statistical mechanics}
		\label{section_statisticalMechanics}

		Unless otherwise stated, this section is based on \cite{gibbs_tong}.\\

		In statistical physics one considers systems of many
			\footnote{We're usually talking about $N \approx 10^{23}$ particles, plus or minus a handful of orders of magnitude.} 
		identical particles: the focus is on the \textit{ensemble} properties of the system, i.e. in the statistical properties of the system as a whole, ignoring the details of the single microstates.
		As the number $N$ of particles is big, we'll usually be working in the so-called \textbf{thermodynamic limit} $N \rightarrow \infty$, and we'll often write $f(N) \approx g(N)$ for $\lim\limits_{N \rightarrow \infty} f(N) = \lim\limits_{N\rightarrow \infty} g(N)$.\\

		Our starting point is the definition of a \textbf{temperature} $T > 0$ for our system, and a much more useful \textbf{inverse temperature} 
			\footnote{Where $K_B$ is the Boltzmann constant, making $\frac{1}{\beta}$ a measure of energy.}
		$\beta \eqdef \frac{1}{K_B \, T}$: high temperature corresponds to $\beta \ll 1$ while low temperature corresponds to $\beta \gg 1$, the \textbf{zero-temperature limit} being $\beta \rightarrow \infty$.\\

		The system will have a \textbf	{Hamiltonian} $H$, an operator on the space of states describing the energy $E_\sigma$ of each state $\sigma$ of the system: $H(\sigma) = E_\sigma$. We define the following probability measure on the space of states, called the \textbf{Gibbs measure}:
		\begin{equation}
		\begin{aligned}
			\mu(\sigma) &\eqdef \frac{1}{Z}\exp[-\beta \, H(\sigma)]\\
			Z & \eqdef \sum_{\sigma}  \exp[-\beta \, H(\sigma)]
		\end{aligned}
		\end{equation} 
		where $Z$ is called the \textbf{partition function for the system}, and $\sum_{\sigma}$ stands for the sum with $\sigma$ ranging over the full space of states.\\

		The quantity $\log Z$, called the \textbf{log-partition function} is fundamental, as it allows to reconstruct all the moments of the energy of the system by taking derivatives w.r.t. $\beta$. For example mean and variance of $E$ are reconstructed as
		\begin{equation}
		\begin{aligned}
			\avg{E} & = \sum\limits_{\sigma} \mu(\sigma) H(\sigma) = - \frac{\partial}{\partial \beta} \log Z \\
			\operatorname{Var}[E] & = - \frac{\partial}{\partial \beta} \avg{E} =  \frac{\partial^2}{\partial \beta ^2} \log Z
		\end{aligned}
		\end{equation}
		The log-partition function is also connected to the \textbf{free-energy}
			\footnote{Representing the amount of energy in a system that can be used to do physical work.} 
		by $F = -\frac{1}{\beta} \log Z$.\\

		An ubiquitous example is provided by two state systems (also called spin-$\frac{1}{2}$) systems, and we'll refer to such system as \textbf{spins}. The state space of a spin is given by its two possible configurations $\{\pm\}$; its Hamiltonian is $H(\pm) = \mp \epsilon$, and the Gibbs measure becomes
		\begin{equation}
			\mu(\pm) = \frac{1}{\exp[+\beta \, \epsilon] + \exp[-\beta \, \epsilon]} \exp[\pm\beta \, \epsilon] = \frac{\exp[\pm\beta \, \epsilon]}{2\cosh[\beta \, \epsilon]}
		\end{equation}
		Indeed the first, humble step in the application of statistical physics to the K-sat problem will be realising that, at the end of the day, a boolean variable is nothing but a spin.\\

		In what follows we'll be interested in systems of many interacting
			\footnote{As opposed to many free spins, where the Gibbs measure is given by the normalised product of the individual spin measures.}
		spins $\sigma_1,...,\sigma_N$. One iconic system of interacting spins is the Ising model:
		\begin{equation}
			H(\sigma) = - \frac{1}{\sqrt{N}}\sum_{i<j} g_{ij} \sigma_i \sigma_j - \sum_{i} h_i \sigma_i 
		\end{equation}
		where the spins lie on a $\integers^d$ lattice and the \textit{spin couplings} $g_{ij}$ are non-zero only for $i,j$ nearest neighbours on the lattice, and have the same value over all the nearest neighbour pairs. 
		The Ising model is the simplest model for magnetic systems which exhibits non-trivial behaviour: 
		\begin{enumerate}
			\item the spins correspond to magnetic orientations along some fixed axis;
			\item the couplings correspond to the ferromagnetic ($g_{ij}>0$, i.e. tendency to align) or antiferromagnetic ($g_{ij}<0$, i.e. tendency to antialign) interaction of nearby spins;
			\item the \textit{external fields} $h_i$ correspond to external magnetic fields that influence the single spins (in the classical model $h_i = h$ has the same value for all spins, i.e. there is a uniform external field).
		\end{enumerate}
		We'll be interested in the much more complex case of \textbf{spin glasses}, where we'll allow couplings and fields to become random variables. 
				
	\subsection{The Hamiltonian for random K-sat}
		\label{section_KSATHamiltonian}
		
		Given an instance of K-sat $\mathcal{I} = \bigwedge\limits_{a = 1}^{M} C_a$ and an assignment $x_1=b_1,...,x_n=b_N$ of boolean values to the variables, a natural way to define the energy of an assignment $b$ is 
		\begin{equation}
			H_N(b) = \text{ \# of clauses of }\mathcal{I}\text{ violated by } b
		\end{equation}
		Thus minimising the energy is equivalent to minimising the number of violated clauses: if the instance is satisfiable, the satisfying assignments will be exactly those with zero energy.\\

		The first step towards spin glasses is to go from boolean assignments $b = (b_1,...,b_N) \in \{0,1\}^N$ to spin configurations $\sigma = (\sigma_1,...,\sigma_N) \in \{\pm 1\}^N$: the correspondence is a matter of convention, and we'll take it to be $b_i=0,1 \leftrightarrow \sigma_i=-1,+1$. 
		From now on we'll treat the K-sat problem as if it was formulated in terms of spins: when talking of \textbf{variable} $i$ or $x_i$ we'll usually refer to the index of the spin, when talking of spin $\sigma_i$ we'll be talking about the r.v. encoding the boolean assignment of a variable.\\

		Given clause $C_a = z_{i(a,1)},...,z_{i(a,K)}$, where $z_i \in \{x_i, \neg x_i \}$, we're interested in having an indicator function $W_a(\sigma)$ for its violation (dependence on $N$ is kept implicit): 
		\begin{equation}
			W_a(\sigma) \: \eqdef \: 
			\begin{cases} 
				1, & \text{if } \sigma \text{ violates clause }C_a \\ 
				0, & \text{if } \sigma \text{ satisfies clause }C_a
			\end{cases}
		\end{equation}

		Define the spins $J_a = (J_a^1,...,J_a^K)$ to be 
		\begin{equation}
			J^r_a \: \eqdef \: 
			\begin{cases} 
				+1, & \text{if } z_{i(a,r)} = \neg x_{i(a,r)} \\ 
				-1, & \text{if } z_{i(a,r)} = x_{i(a,r)} 
			\end{cases}
		\end{equation}
		i.e. $J^r_a$ is the spin that makes $z_{i(a,r)}$ false. 
		Then $W_a(\sigma) = 1$ 
		$\iff$ each spin $\sigma_{i(a,r)}$ of $\sigma$ involved in clause $C_a$ falsifies its the corresponding literal $z_{i(a,r)}$ 
		$\iff$ each spin $\sigma_{i(a,r)}$ is aligned with $J_a^r$:
		\begin{equation}
			W_a(\sigma) \: = \: \prod_{r=1}^K \dfrac{(1+J_a^r \sigma_{i(a,r)})}{2}
		\end{equation}

		The Hamiltonian for instance $\mathcal{I}$ is then 
		\begin{equation}
			H_N(\sigma) \: = \: \sum_{a=1}^{M} W_a(\sigma)
		\end{equation}
		which is a random function of $\sigma$ since $\mathcal{I}$ is a random instance of K-sat.\\

		The partition function is 
		\begin{equation}
			Z_N = \sum_{\sigma} \exp\left[-\beta \,(\text{\# of clauses of }\mathcal{I}\text{ violated by } \sigma)\right]
		\end{equation}
		Thus in the zero temperature limit $\beta \rightarrow \infty$ only the satisfying assignments contribute to the sum and we have the following neat way of counting them
		\begin{equation}
			\left. Z_N \right|_{\beta = \infty} = \text{ \# of satisfying assignments for }\mathcal{I}
		\end{equation}
		Indeed at zero temperature the Gibbs measure is concentrated on the satisfying assignments only.
%% END Introduction

\newpage

%% BEGIN Spin glass fundamentals
\section{Spin glass fundamentals}
	\label{chapter_SpinGlassFundamentals}

	\subsection{The SK model, for mathematicians}
		\label{section_SKmodelForMathematicians}
		
		Except where otherwise stated, this section is based on \cite{survey_talagrandNew}\cite{survey_talagrandOld}.

	\subsubsection{Formulation}		

		Consider the spin space of an N-spin system 
		\begin{equation}
			\Sigma_N \eqdef \{-1,+1\}^N
		\end{equation}
		equipped with the Hamming distance 
		\begin{equation}
			\label{hammingDistanceDef}
			d(\sigma, \tau) \eqdef \#\suchthat{1 \leq i \leq N}{\sigma_i \neq \tau_i} 
		\end{equation}
		Consider also a family of iid standard gaussian r.v.s (the \textbf{spin couplings}, or the \textbf{disorder})
		\begin{equation}
			\begin{array}{rcl}
				\{g_{ij}\}_{1\leq i < j \leq N} & \text{ with } & \expect{g_{ij}} = 0\\
				& \text{ and } & \operatorname{Cov}[g_{ij},g_{i'j'}] = \expect{g_{ij} \: g_{i'j'}} = \delta_{ii'}\delta_{jj'}
			\end{array}
		\end{equation}
		and a vector $h \in \reals^N$ (the components are the \textbf{external fields} for the spins).\\

		The Hamiltonian $H_N$ of the SK model is the following random function over $\Sigma_N$
		\begin{equation}
		\label{SK_hamiltonian}
			H_N(\sigma) \eqdef -\frac{1}{\sqrt{N}} \sum\limits_{i<j} g_{ij} \sigma_i \sigma_j - \sum\limits_{i} h_i \sigma_i
		\end{equation}

		Eq'n \ref{SK_hamiltonian} shows how the model assigns energy to spin configurations:
		\begin{itemize}

			\item An individual spin contributes to a lower system energy when aligned with its external field 
		
			\item A pair of spins $\sigma_i, \sigma_j$ with positive coupling (i.e. $g_{ij} > 0$) contribute to a lower system energy when aligned (i.e. $\sgn [\sigma_i] = \sgn [\sigma_j]$)
		
			\item A pair of spins $\sigma_i, \sigma_j$ with negative coupling (i.e. $g_{ij} < 0$) contribute to a lower system energy when anti-aligned (i.e. $\sgn [\sigma_i] \neq \sgn [\sigma_j]$)
		
			\item A pair of spins with null coupling (i.e. $g_{ij} = 0$) does not contribute to the system energy
		
		\end{itemize}

		The SK model introduces not one but two sources of randomness on the spin system:
		\begin{enumerate}
			
			\item[(a)] the space $\Sigma_N$ is endowed with a Gibbs measure
			\begin{equation}
				\label{equation_GibbsMeasureSKSpinSpace}
				\mu_N(\sigma) \eqdef \frac{1}{Z_N}exp[-\beta H_N(\sigma)]
			\end{equation}
			where $Z_N \eqdef \sum\limits_{\sigma} \exp[-\beta H_N(\sigma)]$ is the partition function as usual.
			
			\item[(b)] the measure $\mu_N$ is itself random, as it depends on the $\frac{N(N-1)}{2}$ r.v.s $\{g_{ij}\}_{ij}$.
		
		\end{enumerate}
		
		Given a function $f$ on $\Sigma_N$, we'll use the following notation (the \textbf{brackets}) for the expectation taken over the Gibbs measure
		\begin{equation}
			\avg{f} \eqdef \frac{1}{Z_N}\sum\limits_{\sigma} f(\sigma) \exp[-\beta H_N(\sigma)]
		\end{equation}
		
		Then $\avg{f}$ is still a random variable, depending on the random couplings. On the other hand we'll denote the expectation taken over the couplings by the usual $\expect$ symbol, and we'll end up writing $\expect{\avg{f}}$ for the full expectation over all sources of randomness. 
		Please note that $\expect{f}$ is also a random variable, this time depending on the random spin configuration (governed by the Gibbs measure and usually denoted by $\sigma$).\\

		When dealing with several spin configurations at a time we'll assume to have at our disposal a sequence of iid r.v.s $(\sigma^k)_k\geq1$, the \textbf{replicas}, all governed by the Gibbs measure of eq'n \ref{equation_GibbsMeasureSKSpinSpace}.\\

		Replicas are primarily used to linearise products of brackets, as in
		\begin{equation}
			\avg{g}^n = \avg{g(\sigma^1)\cdot...\cdot g(\sigma^n)}
		\end{equation}
		where we've extended brackets in the obvious way
		\begin{equation}
			\avg{f(\sigma^1,...,\sigma^n)} \eqdef \frac{1}{(Z_N)^n}\sum\limits_{\sigma^1,...,\sigma^n} f(\sigma^1,...,\sigma^n) \exp[-\beta \sum\limits_{1\leq k \leq n} H_N(\sigma^k)]
		\end{equation}

		Despite this linearisation of brackets, the energies of different spin configurations are not independent: their covariance is given by 
		\begin{equation}
			\label{SK_covH}
			\operatorname{Cov}[H_N(\sigma^1), H_N(\sigma^2)] = \frac{1}{N} \sum\limits_{i<j}\sigma^1_i \sigma^1_j \sigma^2_i \sigma^2_j = \frac{N}{2}R^2(\sigma^1,\sigma^2)-\frac{1}{2}
		\end{equation}
		where  $R(\sigma^1,\sigma^2)$ is the \textbf{overlap of the two spin configurations}
		\begin{equation}
			R(\sigma^1,\sigma^2) \; \eqdef \; \frac{1}{N} \sum\limits_{i}\sigma^1_i \sigma^2_i \; = \; 1-2d(\sigma^1,\sigma^2)
		\end{equation}

		Because all $R(\sigma_i,\sigma_j)$ for $i \neq j$ have the same distribution, when under expectation (and/or not concerned with which pair of spins we choose) we'll usually just write $R$ for the overlap.

	\subsubsection{The free-entropy density: replica symmetric (RS) case}

		As we've seen in section \ref{section_statisticalMechanics}, the log-partition function $\log Z_N$ gives us a lot of information about the system (e.g. the energy distribution or, in the case of K-sat, the distribution of the number of violated constraints). 
		In the SK model it is a random variable, and we also consider its \textbf{quenched average}
			\footnote{As opposed to the \textbf{annealed average} $\frac{1}{N} \log\expect{Z_N}$, which is much easier to compute.}
			\footnote{To get some physical intuition, notice that the quantity $\frac{1}{\beta} p_N$ is the expected free energy per spin.} 
		(also known, in the thermodynamic limit, as the \textbf{free-entropy density})
		\begin{equation}
			p_N \eqdef \frac{1}{N} \expect{\log Z_N}
		\end{equation}

		Physical considerations (similar to \textit{equipartition of energy}) suggest that the r.v. $\frac{1}{N}\log Z_N$ should be self-averaging in the thermodynamic limit
			\footnote{i.e. its fluctuations around its expected value $p_N$ should become small, with variance falling as $\frac{1}{N}$): this is the case for the SK model, as shown by Theorem \ref{guerraRSSolution}.}
		and we expect the quenched average to capture most of the information the log-partition function gave us in the classical non-random cases.\\

		Since $p_N$ captures so much information about the system, we expect its exact computation to be far from trivial (especially in the low temperature regime where spin glasses are characterised by a complex free-energy landscape): the quest for its value has given birth to some of the most interesting tools of spin glass theory, like the replica method and the cavity method. 
		The rest of this section \ref{section_SKmodelForMathematicians} is dedicated to a concise survey of the main results on the free-entropy density of the SK model.

		\begin{theorem} (Guerra's replica-symmetric bound)\\
			For all $\beta,q > 0$, and any choice of law for the external fields $h$ (assume them iid), we have
			\begin{equation}
				\label{guerraBound}
				p_N(\beta,h) \leq \log 2 + \expect{\log \cosh [\beta (g\sqrt{q}+h)]} + \frac{\beta^2}{4}(1-q)^2
			\end{equation}
			The RHS is usually denoted $SK(\beta,q,h)$,), and $g$ is a r.v. distributed like the spin couplings.
		\end{theorem}

		Optimising over $q$ one obtains the tightest possible bound for
		\begin{equation}
			\label{guerraBound_qRelation}
			q = \expect{\tanh^2[\beta(g\sqrt{q}+h)]}
		\end{equation}
		A theorem by Latala and Guerra then guarantees the existence of a unique such $q$, and from now on we'll consider that specific value (and we'll write $SK(\beta,h)$).\\

		The overlaps can be shown, in the high-temperature region $\beta < \frac{1}{2}$, to converge in distribution to the $q$ of \ref{guerraBound_qRelation}, and this implies the convergence of $p_N(\beta,h)$ to $SK(\beta,h)$ in the same high-temperature region
			\footnote{A different solution is needed for the low temperature region.} 
		\begin{theorem} (Guerra's replica-symmetric solution)
			\label{guerraRSSolution}\\
			Assume $\beta < \frac{1}{2}$ and any choice of law for the external fields $h$, then
			\begin{equation}
				|p_N(\beta,h)-SK(\beta,h)| \leq O(N^{-1})
			\end{equation}
			The constants hidden by the big-O notation are allowed to depend on $\beta$ and the law of $h$. In fact the authors prove convergence in distribution, not just in first moment.
		\end{theorem}
					
		Control of the overlaps (the variance of which bounds the correlation of finite sets of spins) also allows to show that the Gibbs measure fully factorises in the thermodynamic limit: 
		\begin{theorem} (Factorisation of the Gibbs measure)\\
		\label{gibbsMeasureFactorisationTheorem}
		Assume $\beta < \frac{1}{2}$, and let $\mu_{N,p}$ be the marginal distribution of $(\sigma_1,...,\sigma_p)$ under $\mu_N$. Then we have
		\begin{equation}
		\expect{|| \mu_{N,p} - \nu_p ||^2} \leq O(N^{-1})
		\end{equation}
		where $||\,\mu-\nu\,|| \; \eqdef \sum\limits_{s \in \Sigma_p} |\,\mu(s)-\nu(s)\,|$ is the total variational distance, and $\nu_p$ is the product probability measure on $\Sigma_p$:
		\begin{equation}
		\nu_p(s_1,...,s_p) \eqdeftemp \frac{1}{2^p} \prod\limits_{i\leq p} (1+\avg{\sigma_i} s_i)
		\end{equation}
		\end{theorem}

	\subsubsection{A first look at the cavity method}
		\label{cavityMethodAFirstLook}
		The cavity method is arguably the most important tool in spin glass theory: its applications encompass all the results we've seen and a good share of the ones to come. Here we introduce its basic formulation in the SK model, and in section \ref{cavityMethod} we'll develop its full potential within the factor graph formalism.\\ 

		In short, the cavity method is induction over $N$ for Gibbs averages: it reduces the computation of brackets for an $N$ spin system to that for an $N-1$ spin system by 

		\begin{enumerate}
			
			\item[(1a)] fixing the value of a spin $\sigma_N$ to $+1$
			
			\item[(2a)] removing the spin from the system, thus creating a \textit{cavity} and raising the temperature\footnote{Raising is good, as our results tend to hold from a given temperature up, rather than down.}
			
			\item[(3a)] measuring the Gibbs average in the smaller system
			
			\item[(1b)-(3b)] replacing the spin and do (1a)-(3a) again, fixing its value to $-1$ this time
			
			\item [(4)] taking the mean of the two Gibbs averages obtained in (1a)-(3b)
		
		\end{enumerate}

		Mathematically this is the observation that Hamiltonian $H_N$ can be written as that of a smaller $N-1$ spin system plus a term coupling the smaller system with the last spin $\sigma_N$: 
		\begin{equation}
			\label{cavityHamiltonianInduction}
			H_N(\sigma) = \frac{\sqrt{N-1}}{\sqrt{N}} H_{N-1}(\sigma_1,...,\sigma_{N-1}) + \alpha_N\sigma_N
		\end{equation} 
		where the coupling $\alpha_N$ is given by
		\begin{equation}
		\label{cavityHamiltonianInductionCoupling}
		\alpha_N \eqdeftemp \frac{1}{\sqrt{N}} \sum\limits_{i < N} g_{iN}\sigma_i + h_N
		\end{equation}

		Absorbing the $\frac{\sqrt{N-1}}{\sqrt{N}}$ factor into the temperature of the $N-1$ spin system, denoting by $\Av$ the average over $\sigma_N = \pm 1$, and expanding the brackets $\avg{.}$ of the original system in terms of the brackets $\avg{.}_{-}$ of the smaller system, we get the desired result
		\begin{equation}
			\avg{f} = \dfrac{\avg{\Av \bigl( \: f(\sigma) \, \exp[\,\beta \: \alpha_N \sigma_N\,] \: \bigr)}_{-}}{\avg{\Av \bigl( \: \exp[\,\beta \: \alpha_N \sigma_N\,] \: \bigr)}_{-}}
		\end{equation}

		\newpage
	
	\subsection{The SK model, for physicists}
		\label{SKmodelPhysicists}
		
		Except where otherwise stated, this section is based on \cite{RSB_denef}\cite{survey_mezardParisiVirasoro}.\\

		In order to investigate the low temperature region (where the \textit{replica-symmetric} solution breaks), we'll continue our survey of the SK model with a shift in perspective: we'll put mathematical rigour aside, and try to gain some intuition about replica symmetry (and replica symmetry breaking) through the eyes of a physicist.\\

		From now on we'll assume no external field is present in our model, and throughout this section \ref{SKmodelPhysicists} we'll use the following simplified Hamiltonian
		\begin{equation}
			\label{SK_hamiltonian_physicists}
			H_N(\sigma) \eqdef -\frac{1}{\sqrt{N}} \sum\limits_{i\neq j} g_{ij} \sigma_i \sigma_j
		\end{equation} 

	\subsubsection{Pure state}
		\label{physicistsPureStates}
	
		In analogy with quantum mechanics, physicists refer to ergodic components of the Gibbs measure as \textit{pure states}. Suppose the Gibbs measure can be decomposed as 
		\begin{equation}
			\mu_N(\sigma) = \sum\limits_{\psi} \omega_\psi \mu_N^{(\psi)}(\sigma)
		\end{equation}
		where $\mu_N^{(\psi)}$ are probability measures on $\Sigma_N$. Then the $ \mu_N^{(\psi)}$ are \textbf{pure states} iff the following holds:
		
		\begin{enumerate}
		
			\item[(a)] the $\mu_N^{(\psi)}$ satisfy \textbf{cluster decomposition}: for any set of \textbf{local}
				\footnote{i.e. only involving a finite number of spins.} 
			functions $A_1,..., A_p$ we have that $\avg{A_1\cdot...\cdot A_p}_\psi \approx \avg{A_1}_\psi\cdot ... \cdot \avg{A_p}_\psi$ in the large $N$ limit
				\footnote{Technically a remainder $r(\sigma_{i_1},...,\sigma_{i_r})$ (which will be a function of the spins $\sigma_{i_1},...,\sigma_{i_r}$ involved in the expression) is allowed on the RHS of the approx equation, as long as it its magnitude vanishes in the large $N$ limit when averaged over all the spins, i.e. as long as $\lim\limits_{N \rightarrow \infty. } \dfrac{1}{N^r}\sum\limits_{i_1,...,i_r} |r(\sigma_{i_1},...,\sigma_{i_r})| = 0$. We'll not mention this $r$ anywhere else.}
			, where $\avg{.}_\psi$ is the expectation taken w.r.t. measure $\mu_N^{(\psi)}$. This property is also called \textbf{correlation-decay property}, and it can be rephrased as \textit{large distance implies vanishing correlation}.
		
			\item[(b)] the \textbf{Gibbs probabilities} $\omega_\psi$ are all positive.
		
			\item[(c)] the $\mu_N^{(\psi)}$ themselves cannot be further decomposed in measures satisfying condition (a)-(b).
		
		\end{enumerate}
		
		Sets of positive measure for a pure state have measure zero for all the other pure states. Also, due to theorem \ref{gibbsMeasureFactorisationTheorem}, the Gibbs measure for the SK model in the high temperature region $\beta < \frac{1}{2}$ only has one pure state.

	\subsubsection{Pure state overlaps}
	
		The notion of overlap is now transferred to pure states, where the \textbf{overlap} $q_{\phi\psi}$ \textbf{of two pure states} $\phi$ and $\psi$ is defined as:
		\begin{equation}
			q_{\phi\psi} \eqdef \frac{1}{N}\sum\limits_{i} \avg{\sigma_i}_\phi\avg{\sigma_i}_\psi
		\end{equation}

		Then in particular $q_{\psi\psi} = \frac{1}{N}\sum\limits_{i} \avg{\sigma_i}_\psi^2$ is the \textit{average magnetisation} in pure state $\psi$. It turns out \cite{microstructureUltrametricity} that $q_{\psi\psi}$ is independent of both $\psi$ and the spin couplings: it is thus an invariant of the system depending only on $\beta$, and is sometimes called the \textbf{Edwards-Andersen order parameter} $q_{EA}$.\\

		One can study the distribution of the pure state overlaps by considering the following \textbf{pure state overlap probability density function (PDF)}
		\begin{equation}
			P(q') \eqdef \sum\limits_{\phi,\psi} \omega_\phi \omega_\psi \, \delta(q'-q_{\phi\psi}) = \avg{\delta(q'-\frac{1}{N}\sum\limits_{i} \sigma_i^1 \sigma_i^2)}
		\end{equation}

		Analytical computation of $P(q)$ at fixed disorder is in general not possible, but the cavity method allows computation of the \textbf{disorder-averaged overlap PDF} $\expect{P(q)}$.

		\subsubsection{Ultrametricity and hierarchical clustering}
		A key result comes from studying the following \textbf{overlap triangle PDF} 
		\begin{equation}
			P(q_1,q_2,q_3) \eqdef \sum\limits_{\phi,\psi,\varphi} \omega_\phi \omega_\psi \omega_\varphi \, \delta(q_1-q_{\phi\psi})\delta(q_2-q_{\phi\varphi})\delta(q_3-q_{\psi\varphi})
		\end{equation}

		Using replicas and restricting to positive pure state overlaps this can \cite{natureSpinGlassPhase} be written as
		\begin{equation}
			\label{overlapTrianglePDFeqn}
			\begin{array}{rcl}
				P(q_1,q_2,q_3) 	& = 	& \frac{1}{2} \int\limits_{0}^{q_1} P(q') dq' P(q_1)\delta(q_1-q_2)\delta(q_2-q_3) + 						\\
								&		& + \frac{1}{2}\bigl[\: P(q_1)P(q_2)\theta(q_1-q_2)\delta(q_2-q3) + \text{ two more permutations} \:\bigr]
			\end{array}
		\end{equation}
		where $\theta(x)$ is the Heaviside step function. The first line of eq'n \ref{overlapTrianglePDFeqn} makes $(q_1,q_2,q_3)$ into the sides of an equilateral triangle, while the three terms on the second line cover the case of non-equilateral, isosceles triangles. If we now define a metric on pure states by
		\begin{equation}
			d(\phi,\psi) \eqdef \frac{1}{N}\sum\limits_{i} \bigl( \avg{\sigma_i}_\phi - \avg{\sigma_i}_\psi \bigr)^2 = 2(q_{EA}-q_{\phi\psi})
		\end{equation}
		then eq'n \ref{overlapTrianglePDFeqn} implies that under $d$ all triangles are either isosceles or equilateral, i.e. that the space of pure states exhibits \textbf{ultrametricity}.\\

\begin{figure}[hp]
			\begin{center}
				\includegraphics[width=6cm]{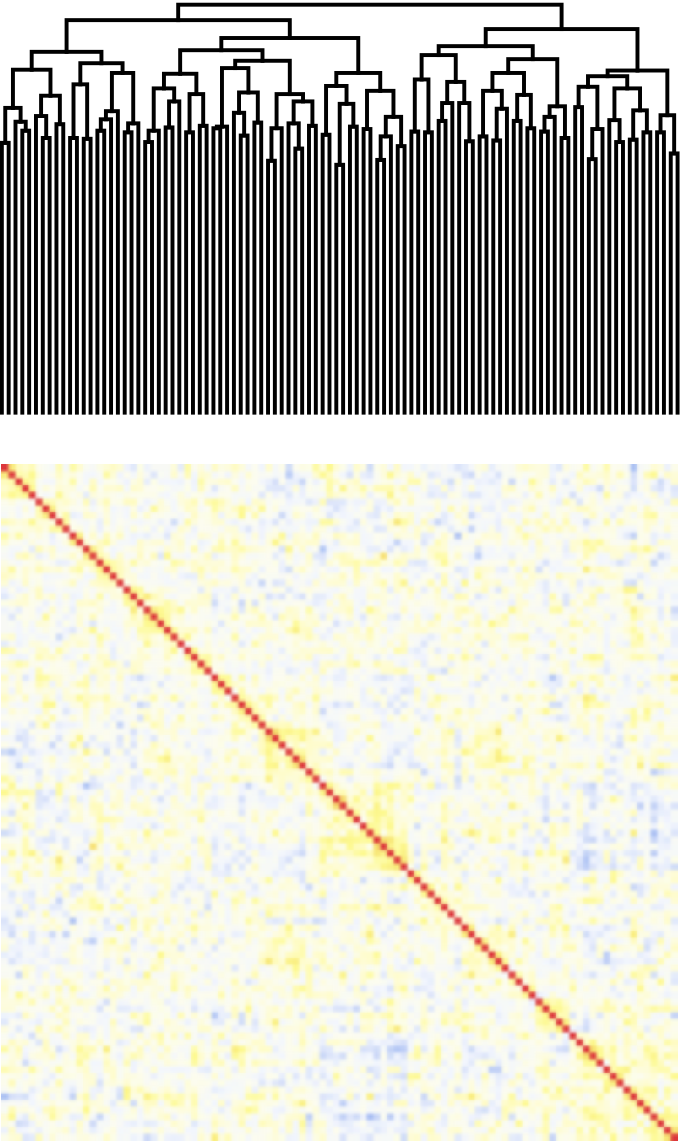}
				\includegraphics[width=6cm]{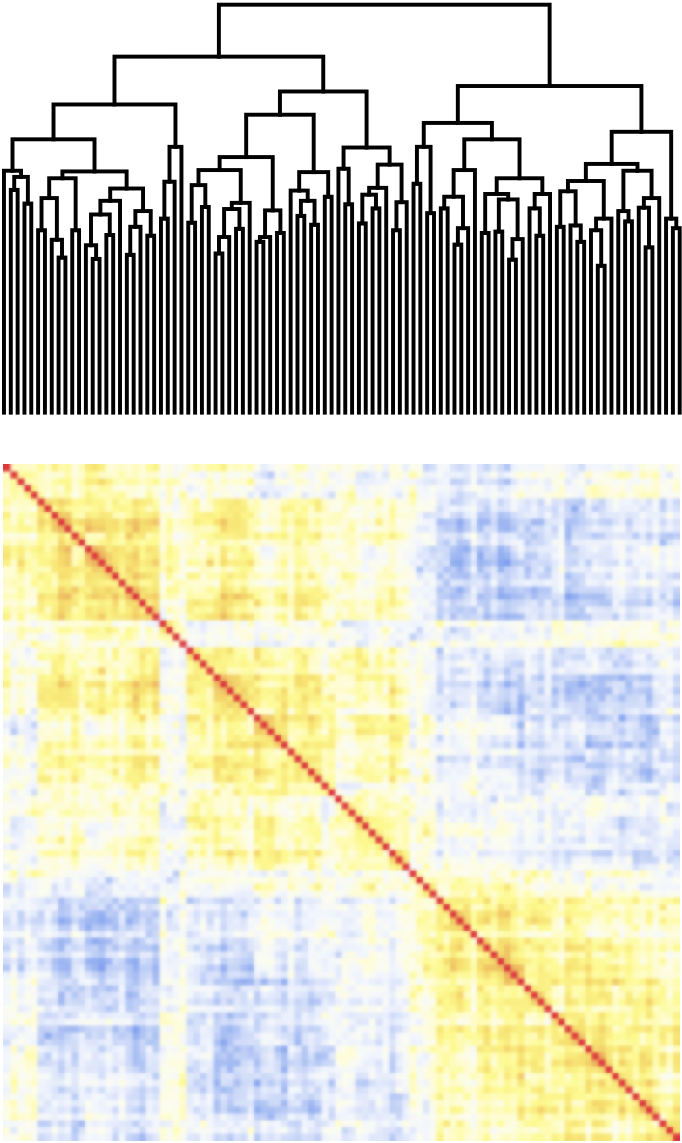}\\
				\includegraphics[width=6cm]{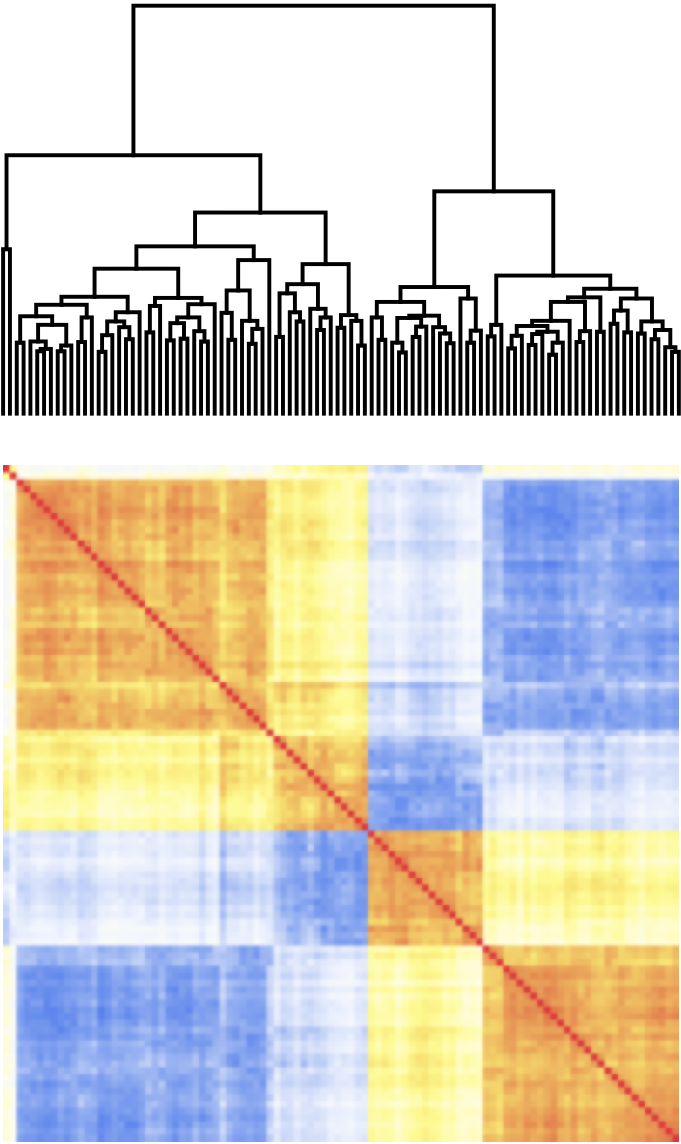}
				\includegraphics[width=6cm]{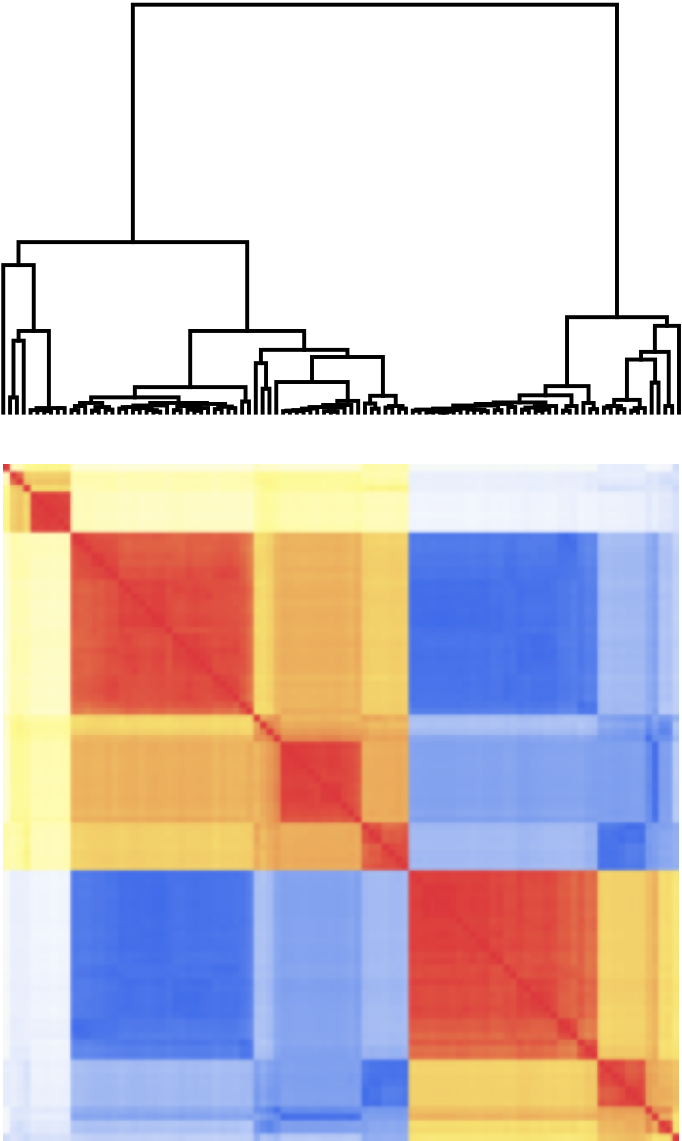}\\
			\end{center}
				\caption{Dendrograms and overlap matrices for 100 randomly sampled pure states, at $T/T_c = 1.2,0.86,0.55,0.12$. Red is maximally positive overlap $q_{\phi\psi}=+1$, blue is maximally negative overlap $q_{\phi\psi}=-1$, white is no overlap $q_{\phi\psi}=0$. Figure from \cite{RSB_denef}.}
			\label{im_overlapClustering}
		\end{figure}

		Ultrametricity is ubiquitous in biology, where it appears in all metrics based on some concept of \textit{most recent common ancestor}
			\footnote{e.g. distance between species, difference of DNA or representative sets of proteins, evolutionary trees, etc.}
		, and in fact ultrametricity is equivalent to the possibility of clustering objects into a hierarchical tree. The ultrametric nature of the space of pure states is most evident in figure \ref{im_overlapClustering} (p. \pageref{im_overlapClustering}), coming from the yet-to-appear \cite{stringGlasses} and obtained as follows.	

		\begin{enumerate}
		
			\item 100 spin configurations (of a SK model with $N=800$ spins) are sampled 100 times each with a Monte Carlo method and then averaged: the Monte Carlo method starts the system at a high temperature
			\footnote{I.e. above the critical temperature $T_c$ corresponding to the so called \textit{spin glass phase transition}, where the replica-symmetric solution breaks.} 
			$T = 1.2 \: T_c$ and cools it down to low temperature $T=0.1 \: T_c$.
		
			\item each configuration of spin averages is taken to be representative of the $(\avg{\sigma_i}_\psi)_{i=1,...,N}$ for some pure state $\psi$ of the system (so to effectively sample 100 pure states from the system). The overlaps of the pure states are then computed and clustered. The 100x100 heatmaps show the overlaps, with red indicating maximum positive overlap $q_{\phi\psi}=+1$, blue indicating maximum negative overlap $q_\phi\psi = -1$ and white indicating no overlap $q_{\phi\psi} = 0$.
		
			\item the dendrogram plots on top of the heatmaps show the clustering of the samples, with vertical height being proportional to temperature. Each pair of pure states $\phi,\psi$ then defines a specific temperature (or \textit{time}) in the cooling (or \textit{evolution}) of the system, the temperature of their most-recent common ancestor in the dendrogram. The difference $\tau(\phi,\psi)$ between that temperature and the final temperature $T_f=0.1 \cdot T_c$ is then a metric on the space of pure states, itself showing ultrametricity
				\footnote{It is closely related to the original metric $d$, but the details of the relation depend on the clustering method used. Unless differently stated we'll assume the metric on the space of pure states to be $d$.}.
		
		\end{enumerate}

		Figure \ref{im_overlapClustering} (p.\pageref{im_overlapClustering}) shows snapshots at $T/T_c = 1.2, 0.86, 0.55, 0.12$ of the evolution of the system:
		
		\begin{enumerate}
		
			\item At high temperature $T = 1.2 \cdot T_c$ the pure states show little or no overlap, and all lie in a single big cluster.
		
			\item Having just crossed the critical temperature, at $T = 0.86 \cdot T_c$ the pure states start showing visible overlaps and shatter in two recognisable clusters (the bigger of which features two barely-recognisable subclusters).

			\item At about half the critical temperature ($T = 0.55 \cdot T_c$) the overlaps continue to increase and the pure states are clustered into two sharply separated clusters, each of which features two clearly recognisable subclusters.
			
			\item At very low temperature $T = 0.12 \cdot T_c$ the pure states are organised into 8 sharp clusters\footnote{With the exception of 9 isolated states.}, hierarchically organised in what is to good approximation a binary tree\footnote{A rough way to obtain a hierarchical tree from the dendrogram tree is by fixing a temperature difference $\Delta T$, discarding all the isolated states and progressively merging all nodes of the dendrogram tree that are nearer in temperature than $\Delta T$, in order of increasing difference of temperature. This procedure applied to the $T=0.12\cdot T_c$ dendrogram tree with $\Delta T \lessapprox 0.06 \cdot T_c$ yields a binary hierarchy of clusters.}. 
		
		\end{enumerate}
		
		For two states $\phi,\psi$ in different clusters, $\tau(\phi,\psi)$ is the temperature below which the two clusters \textit{shatter} and rapidly become separated by impassably high energy barriers.
		\newpage

	\subsubsection{The replica method}
		
		The \textbf{Replica Method} has found, for both its simplicity and effectiveness, widespread use in physical literature as a tool for computing quenched averages in the theory of spin glasses. It is based on the observation that

		\begin{enumerate}
			
			\item[(a)] the problem of computing the quenched average $\expect{\log Z}$ could be reduced (with lots of care in interchanging $\expect{}$ and $\lim\limits_{n \rightarrow 0}$) to the problem of computing $\expect{Z^{n}}$ via
			\begin{equation}
				\label{replicaTrick}
				\begin{array}{rcl}
					\expect{\log Z} & = & \lim\limits_{n \rightarrow 0} \dfrac{\expect{Z^n}-1}{n}
				\end{array}
			\end{equation}
			
			\item[(b)] for $n \in \naturals$, $\expect{Z_N^n}$ can be computed as an annealed average by using $n$ replicas
		
		\end{enumerate}
		
		The procedure implementing the replica method is never fully justified in the literature, but has served the field well; it is certainly worth presenting here, as it encompasses all the core ideas behind replica symmetry breaking (RSB):
		\begin{enumerate}
			
			\item evaluate $\expect{Z_N^n}$ in the thermodynamic limit as in point (b) above, obtaining an expression in $n$ that holds for $n\in\naturals$ and can be extended analytically to the positive reals (or at least a neighbourhood of $n=0$)
			
			\item postulate that the expression for $\lim\limits_{N\rightarrow \infty}Z_N^{n}$ continues to hold when analytically extended, postulate/prove that the limits $N\rightarrow \infty$ and $n \rightarrow 0$ can be exchanged and obtain the quenched average as suggested in point (a) above: $\lim\limits_{N \rightarrow \infty} \dfrac{1}{N} \expect{\log Z_N} = \lim\limits_{n \rightarrow 0} \lim\limits_{N \rightarrow \infty} \dfrac{1}{N}\dfrac{\expect{Z_N^n}-1}{n}$
		
		\end{enumerate}

		Employing $n$ replicas $\{\sigma^{a}\}_a$ (we'll use indices from the first letters of the alphabet $1 \leq a,b,... \leq n$) and integrating\footnote{For reason of clarity, we have absorbed the distribution normalisation factors into the differentials.} away the couplings one gets

		\begin{equation}
		\begin{aligned}
			\label{ZNn_spinIntegral}
			\expect{Z_N^{n}} & = \int dg_{ij} \, e^{-\frac{1}{2}\sum\limits_{i \neq j}g_{ij}^2}\: \sum_{\sigma \in \{\pm 1\}^{nN}} \exp\bigl[ -\beta\frac{1}{\sqrt{N}} \sum_{a} \sum_{i \neq j} g_{ij} \sigma_i^a \sigma_j^a \bigr] \\
			& =  \sum\limits_{\sigma^{1},...,\sigma^{n}} \: \exp\bigl[ \dfrac{\beta^2}{2N} \, \sum_{a,b}\sum_{i \neq j} \sigma_i^a \sigma_j^a \sigma_i^b \sigma_j^b  \bigr]
		\end{aligned}
		\end{equation}

		Something interesting happened here: the $n$ replicas, originally independent copies of the system, got coupled because they shared the same spin couplings.\\

		The second line of q'n \ref{ZNn_spinIntegral} shows formal symmetry between replica indices $a,b$ and spin indices $i,j$: the process of eq'n \ref{ZNn_spinIntegral} is then inverted\footnote{Again we have absorbed the distribution normalisation factors into the differentials.}, this time considering \textbf{replica couplings}
			\footnote{The $N\beta^2$ is there because the $Q_{ab}$ couplings correspond to the $\frac{1}{\sqrt{N}}g_{ij}$ couplings, which had variance $\frac{1}{N}$.}
		$\{Q_{ab}\}_{1 \leq a,b \leq n}$, to decouple the spins: 
		\begin{equation} 
			\label{ZNn_replicaIntegral}
			\expect{Z_N^{n}} = \int dQ_{ab} \,e^{-\frac{1}{2} N\beta^2 \sum\limits_{a \neq b}Q_{ab}^2} \: \sum_{\sigma \in \{\pm 1\}^{nN}} \exp\bigl[ -\beta^2 \sum_{i}\sum_{a \neq b} Q_{ab} \sigma_i^a \sigma_i^b \bigr] 
		\end{equation}

		The spin indices finally disappear by rewriting eq'n \ref{ZNn_replicaIntegral} as
		\begin{equation} 
			\label{ZNn_replicaIntegralRewritten}
			\expect{Z_N^{n}} = \int dQ_{ab} \,e^{-\frac{1}{2} N\beta^2 \sum\limits_{a \neq b}Q_{ab}^2} \: \left(\sum_{S \in \{\pm 1\}^{n}} \exp\bigl[ -\beta^2 \sum_{a \neq b} Q_{ab} S^a S^b \bigr] \right)^N 
		\end{equation}

		In the thermodynamic limit $N \rightarrow \infty$ the integral in eq'n \ref{ZNn_replicaIntegralRewritten} is evaluated by saddle point method, finding the dominant critical point $Q^* \equiv \{Q_{ab}^*\}_{ab}$ of eq'n \ref{saddlePointFreeEnergy} for all $n$. The expression for $\expect{Z_N^n}$ valid at large $N$ is then obtained from eq'n \ref{expectZn}:
		\begin{eqnarray}
			\label{saddlePointFreeEnergy}
				\mathcal{F}(Q) = \frac{1}{2} \beta \sum\limits_{a, b}Q_{ab}^2 -  \frac{1}{\beta} \log \mathcal{Z}(Q)\\
				\mathcal{Z}(Q) = \sum_{S \in \{\pm 1\}^{n}} \exp\bigl[ \beta^2 \sum_{a,b} Q_{ab} S^a S^b \bigr]\\
			\label{expectZn}
				\expect{Z_N^n} = \exp \left[ -\beta N \mathcal{F}(Q^*(n)) \right]
		\end{eqnarray}

		Finally the trick from eq'n \ref{replicaTrick} gives 
			\footnote{Along with the observation that consistency with $Z^0=1$ implies the requirement $\left. \mathcal{F}(Q^*) \right|_{n=0} = 0$.}
		the quenched average in terms of the \textit{saddle-point free energy functional} $\mathcal{F}(Q)$:
		\begin{eqnarray}
			\label{so}
			\lim\limits_{N \rightarrow \infty} \dfrac{1}{N} \expect{\log Z_N} = -\frac{1}{\beta} \left. \dfrac{\partial}{\partial n} \mathcal{F}(Q^*(n)) \right|_{n=0}
		\end{eqnarray}

	\subsubsection{Replica symmetric (RS) solution}
		\label{RSsolution}
		
		The first step on the way to getting an ansatz for the matrix $Q^*_{ab}$ is to understand its physical meaning, which is hidden in the saddle point equations $\mathcal{F}'(Q^*) = 0$
		\begin{equation}
			\label{saddlePointEquations}
			Q^*_{ab} = \dfrac{1}{\mathcal{Z}(Q^*)} \sum_{S \in \{\pm 1\}^{n}} S^a S^b \,\exp\bigl[ \beta^2 \sum_{c,d} Q_{cd} S^c S^d \bigr] 
		\end{equation}

		The RHS of eq'n \ref{saddlePointEquations} is the Gibbs average $\avg{S^aS^b}_{Q^*}$ in an n-spin system, and the $Q^{*}_{ab}$ will thus be called the \textbf{overlaps of the replicas}.\\

		Sherrington and Kirkpatrick give in \cite{SKSpinGlass} their ansatz for $Q^*_{ab}$, the only ansatz that leaves \textbf{replica symmetry}\footnote{I.e. permutational symmetry of the replicas.} unbroken:
		\begin{equation}
			\label{QabRSansatz}
			Q^*_{ab} = u \, \delta_{ab} + v \, (1-\delta_{ab})
		\end{equation}
		A visualisation of the resulting matrix is given on the left in figure \ref{im_ReplicaCouplings} (p.\pageref{im_ReplicaCouplings}).\\

		Constrained extremisation over $u,v$ (yielding $u=1$ and $v=q$) then reproduces Guerra's replica-symmetric solution from theorem \ref{guerraRSSolution} (p.\pageref{guerraRSSolution}):
		\begin{equation}
			\lim\limits_{N \rightarrow \infty} p_N =  -\frac{1}{\beta} \left. \dfrac{\partial}{\partial n} \mathcal{F}(Q^*(n)) \right|_{n=0} = \log 2 + \expect{\log[\cosh(\beta g \sqrt{v})]} + \frac{\beta^2}{4}(u-v)^2
		\end{equation}

	\subsubsection{Replica symmetry breaking (RSB)}
		\label{SK_RSB}

		Unless otherwise stated, this section is based on \cite{parisiOrderParamSummary}\cite{parisiOrderParam}\cite{parisiOrderParamFunction}.\\

		The ansatz of eq'n \ref{QabRSansatz} corresponds to full $S_n$ permutational symmetry of the $n$ replicas: each fixed replica $a$ has overlap $Q^*_{aa} = 1$ with itself and $Q^*_{ab} = q$ with any other replica $b$. But in the low temperature region this ansatz stops holding, and we have to break the $S_n$ symmetry.\\

		The minimal symmetry breaking scheme considered is 
		\begin{equation}
			S_n \rightarrow S_{m_1} \times S_{n/m_1}
		\end{equation}
		corresponding to the $n$ replicas splitting into $n/m_1$ identical clusters of size $m_1$, with full permutation symmetry within each cluster. Each fixed replica $a$ has the following overlaps: 
		\begin{itemize}
			
			\item $Q^*_{aa}=1$ with itself
			
			\item $Q^*_{ab_0} = q_0$ with any of the $m_1-1$ replicas $b_0$ in its cluster
			
			\item $Q^*_{ab_1}=q_1$ with any of the other $n-m_1$ replicas $b_1$ 
		
		\end{itemize}
		This scheme is called \textbf{1-step replica symmetry breaking} or \textbf{1RSB}, and is the one we'll work with when talking of K-sat later on.\\

		The replica breaking scheme can be generalised, introducing a ultrametric hierarchy of clusters: the general case is called \textbf{K-step replica symmetry breaking} scheme, or \textbf{K-RSB}
			\footnote{This K has nothing to do with the K of K-sat, it just follows the convention in the literature.}
		, and consists of a K-level hierarchy of clusters of sizes $1 < m_1 < m_2 < ... < m_K < n$, where each cluster of size $m_{i+1}$ contains $m_{i+1} / m_i$ clusters of size $m_i$. It corresponds to the symmetry breaking scheme
		 \begin{equation}
			S_n \rightarrow S_{m_1/m_0} \times S_{m_2/m_1} \times S_{m_3/m_2} \times ... \times S_{m_{K+1}/m_K}
		\end{equation}
		where by convention we define $m_0 = 1$ (the trivial cluster containing a single replica) and $m_{K+1} = n$ (the trivial cluster containing all the $n$ replicas). The overlaps $1,q_0,q_1,...,q_K$ generalise in the obvious way.\\
	
		The K-RSB scheme is encoded by the non-increasing \textbf{cluster size function} (we set $0 = q_{K+1}$ and $1 = q_{-1}$ by convention):
		\begin{equation}
			\begin{array}{rcl}
				m(q) &\eqdef& m_i \text{ when } q_i \leq q < q_{i-1} \text{, for all } 0\leq i \leq K+1\\
				&=& \text{ max size of clusters with overlaps all of absolute value}>q\\
			\end{array}
		\end{equation}

\begin{figure}[h!]
			\begin{center}
				\includegraphics[width=8cm]{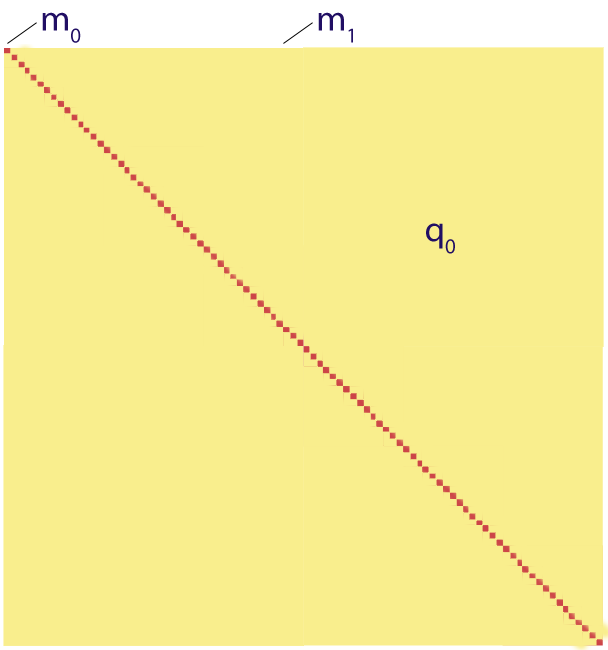}
				\includegraphics[width=8cm]{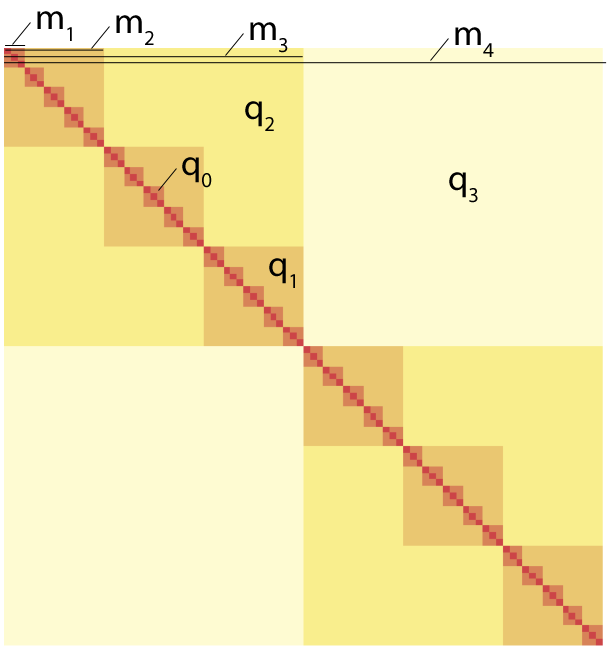}\\
			\end{center}
				\caption{An example of RS Parisi matrix (on the left) and 3RSB Parisi matrix (on the right), both with $n=90$. The RS case has only the trivial cluster sizes $m_0=1$, $m_1=n$; its only overlap is $q_0 = q = 0.4$. The 3RSB case, on the other hand, has the trivial cluster sizes $m_0=1$ (not visualised) and $m_4 = n$, and the non-trivial cluster sizes $m_1,m_2,m_3 = 3,15,45$; its overlaps are $q_0 = q = 0.9$ and $q_1,q_2,q_3 = 0.9,0.6,0.4,0.1$.. Original figure from \cite{RSB_denef}, modified the by author.}
			\label{im_ReplicaCouplings}
		\end{figure}

		In the context of RSB the $n \times n$ matrix $Q^*_{ab}$ is called the \textbf{Parisi matrix}, and the terminology is extended to the RS case. Figure \ref{im_ReplicaCouplings} (p.\pageref{im_ReplicaCouplings}) shows an example of RS Parisi matrix (on the left) and 3RSB Parisi matrix (on the right).\\

		Similarly to the RS solution, also the 1RSB solution for $\lim\limits_{N \rightarrow \infty} p_N $ stops holding as $\beta$ grows bigger, but the full family of K-RSB ansatzes for $K \geq 1$ is enough to cover the entire low-temperature region.
		We'll talk about \textbf{K-RSB phase}
			\footnote{For a specific value of $K$, and we'll talk about the \textbf{RS phase} similarly.}
		 when referring to regions of the phase space where the K-RSB ansatz holds.

		\newpage

	\subsubsection{The $n \rightarrow 0$ limit in RSB (a.k.a. the messy part)}

		RSB concerns clustering of replicas, but what does it say about clustering of pure states? A hint to the answer is given by the pure state overlap PDF, which in the K-RSB phase is
		\begin{equation}
			\expect{P(q')} = \lim\limits_{n \rightarrow 0} \sum_{i=0}^{K}\left(m_i-m_{i+1} \right)\delta(q'+q_i)
		\end{equation}

		The Parisi matrix construction has to be modified if we want some sort of $n \rightarrow 0$ limit to exist: we need $1 = m_0 > m_1 > ... > m_K > 0 = m_{K+1}$, and the $m_i$ will assume a probabilistic interpretation along with the function $m(q)$ which encodes them. Formally
		\begin{equation}
			\expect{P(q')} = \lim\limits_{n\rightarrow 0} \frac{\partial}{\partial q} m(q)
		\end{equation}
		and the function $x(q) \eqdef \lim\limits_{n\rightarrow0} m(q)$ is then akin to a \textbf{pure state overlap cumulative distribution function}.
		This way the hierarchical K-level clustering of replicas will transfer to pure state space, as we wanted. But to do this properly one has \cite{survey_talagrandNew} to understand what a \textit{space of} $0 \times 0$ \textit{matrices} would look like, and that's a story for another time.
%% END Spin glass fundamentals 			
			
\newpage

%% BEGIN From SK to K-sat
\section{From SK to K-sat}
	\label{chapter_FromSKtoKsat}
	Unless otherwise stated, this section is based on \cite{survey_talagrandNew}\cite{core_algorithmicBarriers}\cite{core_clustersAndRSB}\cite{core_mezardZecchina}\cite{gibbs_monassonZecchina}\cite{gibbs_factor}.

	\subsection{Diluted SK and the K-sat Hamiltonian}

		The connection between SK and K-sat goes through the so-called \textbf{diluted SK model}: the Hamiltonian is obtained by multiplying the spin couplings in the SK model by iid Bernoulli r.v.s $\gamma_{ij}$ taking values in $\{0,1\}$:
		\begin{equation}
			\label{dilutedSK_hamiltonian}
			H_N(\sigma) \eqdef -\sum\limits_{i<j} g_{ij} \gamma_{ij} \sigma_i \sigma_j
		\end{equation}
		Taking $\mathbb{E}{\gamma_{ij}} = \mathbb{P}(\gamma_{ij}=1)=\gamma / \frac{N}{2}$ for some fixed constant $\gamma>0$, we get that each spin interacts on average with $\gamma$ other spins (independently of $N$)
			\footnote{The $\frac{1}{\sqrt{N}}$ factor in front of the SK Hamiltonian, which gave the spin couplings a variance of $\frac{1}{N}$, has been absorbed into the expectation of the Bernoulli r.v.s}.\\

		Luckily it turns out \cite{survey_talagrandNew} that a number of results from the SK model can be transferred to the diluted SK model, and more in general to a broader class of Hamiltonians in the form
		\begin{equation}
			\label{generalSK_hamiltonian}
			H_N(\sigma) \eqdef \sum\limits_{a \leq M} W_a(\sigma_{i(a,1)},...,\sigma_{i(a,p)})
		\end{equation}
		where 
		\begin{enumerate}
			
			\item the $W_a: \{\pm 1\}^p\rightarrow \reals$ are iid random functions
			
			\item for some fixed constant $\alpha$ we have $M = \alpha N$ or, alternatively, $M$ is Poisson with mean $\alpha N$
			
			\item the sets $\{1 \leq i(a,1)<...<i(a,p) \leq N\}$ are iid and uniformly distributed
			
			\item the three sources of randomness above are independent of each other
		
		\end{enumerate}

		The diluted SK model is then given by $p = 2$ and 
		\begin{equation}
			W_a(\sigma_{i(a,1)},\sigma_{i(a,2)}) = - g_{i(a,1)i(a,2)} \gamma_{i(a,1)i(a,2)} \sigma_{i(a,1)}\sigma_{i(a,2)}
		\end{equation}

		K-sat belongs to said class, since its Hamiltonian from section \ref{section_KSATHamiltonian} (p.\pageref{section_KSATHamiltonian}) is given by $p=K$ and 
		\begin{equation}
			W_a(\sigma_{i(a,1)},...,\sigma_{i(a,K)}) = \prod\limits_{r=1}^{K} \dfrac{(1+J_a^r \sigma_{i(a,r)})}{2}
		\end{equation}
		In the case of K-sat, the $W_a$ functions are called the \textbf{indicator functions} of the clauses.\\

		Therefore the hope is that good part of the behaviour of the SK model will transfer, at least qualitatively, to random K-sat: this observation has been used throughout the years to gain intuition on what to expect, and what to look for, in the statistical treatment of the problem.
			
	\subsection{Factor graphs}
		
		A natural way to study spin systems with sparse interactions (like the Ising model, diluted SK or K-sat) is to focus on the geometry of the interactions themselves: this is done by studying the so-called \textit{factor graph} of the Gibbs measure.\\

		Consider a collection of $N$ \textbf{variables} $\sigma = (\sigma_1,...,\sigma_N)$ and a function $\mu(\sigma)$ that factors as
		\begin{equation}
			\label{factorisationEquation}
			\mu(\sigma) = \dfrac{1}{Z} \prod\limits_{a \in F}\left.\mu\right|_{a}(\left. \sigma \right|_{\partial_a})
		\end{equation}
		where 
		\begin{enumerate}
			
			\item $Z$ is the \textbf{normalisation constant}
			
			\item the $\left.\mu\right|_{a}$ functions are called the \textbf{factors}
			
			\item $\partial_a \eqdef \suchthat{1 \leq i \leq N}{\left.\mu\right|_{a} \text{ depends on }\sigma_i}$ is the \textbf{neighbourhood} of factor $\left.\mu\right|_{a}$ 
			
			\item $\left. \sigma \right|_{\partial_a} = (\sigma_{i(a,1)},...,\sigma_{i(a,p_a)})$ where we wrote $\partial_a = \{i(a,1),...,i(a,p_a)\}$
		
		\end{enumerate}

		We'll use indices $a,b,c,d,...$ from the beginning of the alphabet for the factors, and indices $i,j,k,l,...$ from the middle of the alphabet for the variables
		\footnote{And we'll confuse indices with the corresponding variables and factors}
		. $F$ is the set of factor indices and we'll let $V$ denote the set of variable indices.\\

		The \textbf{factor graph} for eq'n \ref{factorisationEquation} is then defined to be the bipartite graph with node classes 
		\begin{enumerate}
			\item variable nodes $\sigma_1,...,\sigma_N$
			\item factor nodes $\left.\mu\right|_{1},...,\left.\mu\right|_{M}$
		\end{enumerate}
		and edge $\sigma_i \longleftrightarrow \left.\mu\right|_{a}$ if and only if $i \in \partial_a$. 
		Then $\suchthat{\sigma_i}{i \in \partial_a}$ is the graph neighbourhood of factor node $W_a$. 
		Also we can define $\partial_i \eqdef \suchthat{1 \leq a \leq M}{\left.\mu\right|_{a} \text{ depends on }\sigma_i}$, the \textbf{neighbourhood} of variable $\sigma_i$, so that $\suchthat{\left.\mu\right|_{a}}{a \in \partial_i}$ is the graph neighbourhood of variable node $\sigma_i$.\\

		The \textit{Hammersley-Clifford theorem} \cite{gibbs_factor}\cite{gibbs_grimmett} guarantees that all positive Markov fields and Gibbs ensembles can be represented by factor graphs, and the Gibbs measure for the K-sat problem takes indeed the form of eq'n \ref{factorisationEquation}
		\begin{equation}
			\mu_{N}(\sigma) = \dfrac{1}{Z_N} \exp\left[-\beta \, \sum\limits_{a \in F} \, W_a(\left. \sigma \right|_{\partial_a})\right] = \dfrac{1}{Z_N} \prod\limits_{a \in F} \exp\left[-\beta \, W_a(\left. \sigma \right|_{\partial_a})\right]
		\end{equation}
		The factors nodes correspond to clauses, and the neighbourhood of a factor node is composed of the variables involved in the clauses; the variable node correspond to variables/spins.\\

\begin{figure}[h!]
		\begin{center}
		\includegraphics[width=13cm]{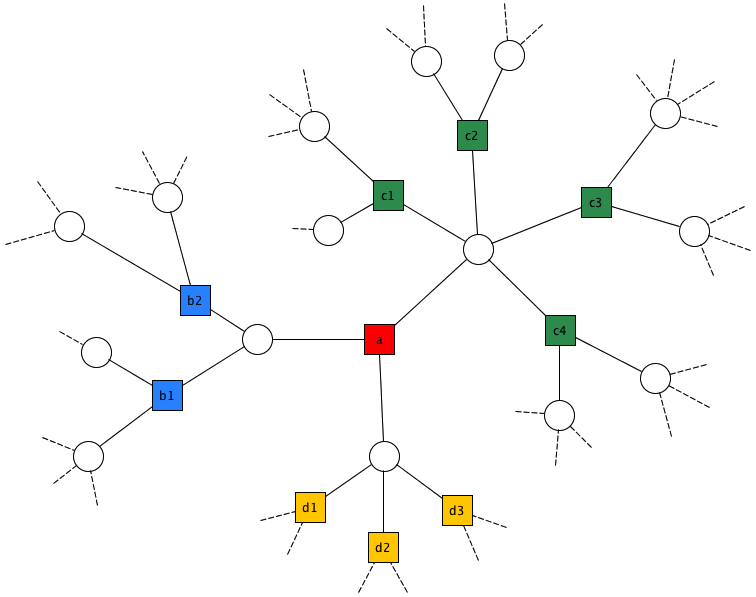}
		\end{center}
		\caption{A small region of a factor graph for 3-sat. Factor nodes are depicted as squares, spin nodes are depicted as circles. Inspired by \cite{core_mezardParisiZecchina}.}
		\label{treelikeFactorGraph}
		\end{figure}

		An example of factor graph for a 3-sat instance can be found in figure \ref{treelikeFactorGraph} (p.\pageref{treelikeFactorGraph}). No distinction is made in the graph topology between positive and negated occurrence of variables in clauses: that information is encoded in the indicator function associated to each factor node.

		The connectivity of factor nodes in K-sat is always K; the connectivities of the variable nodes, on the other hand, depend on the generative model chosen, but become iid poisson r.v.s with mean $K \alpha$ in the thermodynamic limit 
		\begin{equation}\nonumber
			M,N \rightarrow \infty \text{ with } M/N \rightarrow \alpha \text{ w.h.p.}
		\end{equation}

		We expect, from theorem \ref{gibbsMeasureFactorisationTheorem} (p.\pageref{gibbsMeasureFactorisationTheorem}), that the Gibbs measure will factor in the thermodynamic limit.  
		%The question of actual \textit{correlation decay at large distances} has some little complications to do with pure states\footnote{Remember? Cluster decomposition property was one of our requirements when we defined pure states in section \ref{physicistsPureStates}(p.\pageref{physicistsPureStates}).}, but in the thermodynamic limit the full factorisation of the Gibbs measure overcomes this.
		This results in the following behaviour of the factor graph at loop level
		\begin{equation}\nonumber
		\text{ typical size of a loop } = O\left(\log(N)\right)
		\end{equation}
		and has two key consequences:
		\begin{enumerate}
			
			\item the factor graph is almost tree-like, i.e. breadth first exploration of the graph from any node will typically produce trees of diameter $O\left(\log(N)\right)$ before encountering a loop.
			\footnote{This is part of the reason why Belief Propagation and Survey Propagation work so well on random K-sat.}
			
			\item consider a factor node and any two of the K spins connected to it: the spins have initially distance 2, but upon removal of the factor node their distance will typically jump up to $O\left(\log(N)\right)$.
			\footnote{This is part of the reason why the Cavity method is so effective: removal of a single factor node, i.e. creation of a cavity, will make its neighbouring spins to good approximation independent of each other.}
		
		\end{enumerate}

	\subsection{Pure states and clusters of solutions}
		Now that we have factor graphs in our hands we are able to give a better definition of pure states, and it will become evident how pure states, rather than spin configurations, are the correct way to study the random K-sat problem.\\

		A probability measure\footnote{To be precise, a \textit{family} of probability measures $(\mu^{(\psi)}_N)_N$.} $\mu^{(\psi)}_N$ on $\Sigma_N$ is a \textbf{pure state} iff its correlation function
		\begin{equation}
			\label{pureStateCorrelationFunction}
			C_N(r) = \sup\limits_{\substack{I,J\subseteq V \text{ s.t. }\\ \inf\limits_{i \in I, j \in J} d(\sigma_i,\sigma_j)\geq r}} \; \sum_{i \in I, j \in J} \left| \mu^{(\psi)}_N(\{\sigma_i,\sigma_j\})-\mu^{(\psi)}_N(\sigma_i)\mu^{(\psi)}_N(\sigma_j) \right|
		\end{equation} 
		decays at large $r$ in the thermodynamic limit, i.e. iff we have
		\begin{equation}
			\label{pureStateCorrelationFunctionDecay}
			\left[\limsup\limits_{N \rightarrow \infty} C_N(r)\right] \rightarrow 0 \text{ as } r \rightarrow \infty
		\end{equation}

		This definition, although conceptually similar to the one given in section \ref{physicistsPureStates}, is rather cumbersome, and we'll not use it directly: instead  we'll restrict our attention to the zero temperature limit of the so-called \textbf{satisfiable phase}.
			\footnote{I.e. the region where at least a solution exists}\\

		In the zero temperature limit, the measure concentrates on the solutions of $\mathcal{I}$: it can be shown \cite{core_mezardZecchina} that, at zero temperature, a \textbf{pure state} $\psi$ is \textit{composed of}
			\footnote{I.e the measure is supported by.} 
		a set of spin configurations that are
		\begin{enumerate}
		
			\item all of same energy $E^{(\psi)}$
		
			\item \textbf{connected} by \textbf{1-spin flips}, i.e. connected in $\Sigma_N$ with graph structure given by the Hamming distance
				\footnote{I.e. an edge will connect a pair of spin configurations $(\sigma, \tau)$ if and only if $d(\sigma,\tau) = 1$ as per eq'n \ref{hammingDistanceDef}.}
		
			\item \textbf{locally stable}, in the sense that the energy cannot be decreased by any 1-spin flip
		
		\end{enumerate}
		
		In \cite{core_algorithmicBarriers} these are called \textbf{clusters of solutions}, as it is noted that, in the satisfiable phase, they are nothing but the connected components of the space $\mathcal{S}(\mathcal{I}) \subset \Sigma_N$ of \textbf{solutions} (i.e. \textit{satisfying assignments}) to our instance $\mathcal{I}$.\\ 

		From now on when talking of a pure state $\psi$ we'll use notation $\psi$ to denote the cluster of solutions as well as labelling the state, and we'll adopt the zero temperature definition above as our working definition.
		
		\newpage

	\subsection{The complexity}
		Physical experience with spin glasses and other frustrated systems (see e.g. \cite{survey_mezardParisiVirasoro}) suggests that, in the RSB phases, pure states should grow exponentially in $N$, at least to leading order.\\

		If we denote by $\mathcal{N}(\omega; \alpha)$ the expected number of pure states $\psi$ with \textbf{free-entropy density} $\omega$
		\begin{equation}
			\label{freeEntropyDensity}
			\omega \equiv \frac{1}{N} \log Z^{(\psi)}_N
		\end{equation}
		then $\mathcal{N}(\omega; \alpha)$ allows to define the \textbf{complexity} $\Sigma(\omega; \alpha)$ as the unique function satisfying
		\begin{equation}
		\mathcal{N}(\omega) \approx \exp\left[ N \, \Sigma(\omega; \alpha) \right]
		\end{equation}

		What is $\omega$ exactly? Eq'n \ref{freeEntropyDensity} tells us that at zero temperature and in the satisfiable phase 
		\begin{equation}
			\label{omegaIsEntropyDensityAtZeroTempo}
			\omega = \frac{1}{N}\log\left[\;\text{\# of solutions in }\psi \; \right]
		\end{equation}
		and thus $\omega$ coincides with the \textbf{entropy density} $s$ (by definition the RHS of eq'n \ref{omegaIsEntropyDensityAtZeroTempo}). \\

\begin{figure}[h!]
		\begin{center}
		\includegraphics[width=12cm]{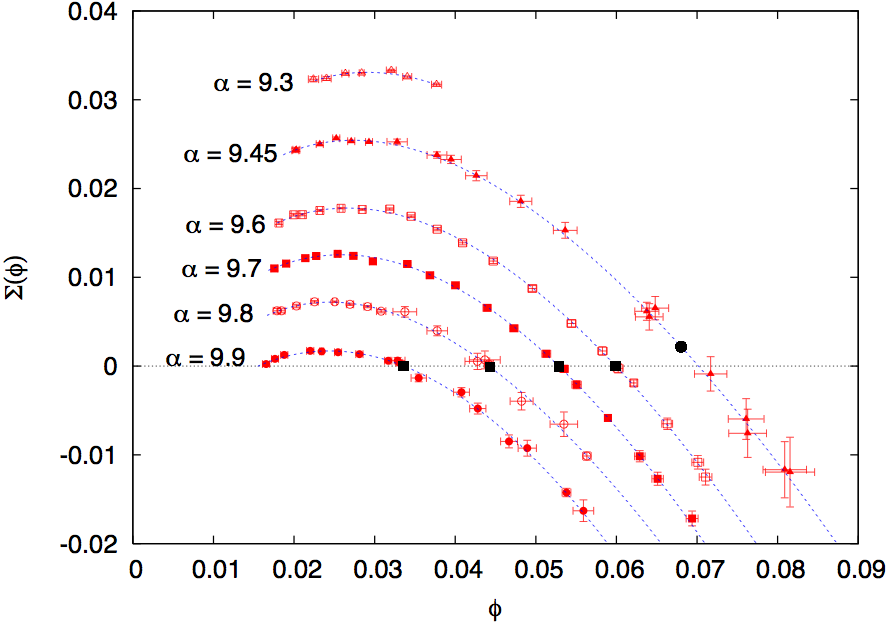}
		\end{center}
		\caption{Plot of the complexity $\Sigma(\omega;\alpha)$ of 4-sat, as a function of the free-entropy density $\omega$ (here denoted $\phi$) and for various values of $\alpha$. Figure from \cite{core_clustersAndRSB}.}
		\label{im_complexityFreeEntropyDensity}
		\end{figure}

		A plot of the complexity $\Sigma(\phi;\alpha)$ for 4-sat is given in figure \ref{im_complexityFreeEntropyDensity}: negative complexity for a value $\omega$ means that w.h.p. there is no cluster with that free-entropy density, while vanishing of the complexity just implies a sub-exponential number of clusters. Notice also that the complexity curve only covers a reduced range of $\omega$ for $\alpha$ small enough.

%% END From SK to K-sat

\newpage

%% BEGIN The phases of K-sat
\section{The phases of K-sat}
	\label{section_RSB}	
	
	\subsection{The Parisi 1RSB parameter}
		\label{Parisi1RSBParameterSection}

		This section is based on \cite{core_clustersAndRSB}.\\
		The definition of pure state given in section \ref{physicistsPureStates} (p.\pageref{physicistsPureStates}) is recovered for K-sat if we write
		\begin{eqnarray}
			\label{ZPureStates2}
				Z^{(\psi)}_N \eqdef \sum\limits_{\sigma \in \psi} \prod\limits_{a \in F} \exp\left[-\beta \, W_a(\left. \sigma \right|_{\partial_a})\right]\\
			\label{muPureStates2}
				\mu^{(\psi)}_N(\sigma) \eqdef \dfrac{1}{Z^{(\psi)}_N}\prod\limits_{a \in F} \exp\left[-\beta \, W_a(\left. \sigma \right|_{\partial_a})\right]\\
			\label{muDecompositionPureStates2}
				\mu_N(\sigma) = \sum\limits_{\psi} w_{\psi} \mu^{(\psi)}_N(\sigma), \; \text{ with } \;w_{\psi} \eqdef \dfrac{Z^{(\psi)}_N}{Z_N} = \dfrac{\text{\# of solutions in }\psi}{\text{total \# of solutions}}
		\end{eqnarray}

		The presentation of the phases of K-sat will cover the RS and 1RSB phases only, as the higher RSB phases are poorly understood in terms of sparse factor graphs. The relevant parameters from section \ref{SK_RSB} will be the \textbf{intra-state overlap} $q_0$, the \textbf{inter-state overlap} $q_1$ and the \textbf{Parisi 1RSB parameter} $m_1 \in [0,1]$, which from now on we'll denote $m$.\\

		The main quantity to study in the context of RSB is the \textbf{replicated free-entropy density}
		\begin{equation}
			\Phi(m) \eqdef \lim\limits_{N \rightarrow \infty} \frac{1}{N} \expect{ \log \sum\limits_{\psi} \left(Z_N^{(\psi)}\right)^m }
		\end{equation}
			
		The replicated free-entropy is directly related to the complexity by the equations
		\begin{equation}
			\label{replicatedFreeEntropyFromComplexity}
			\Phi(m)  = \sup\limits_{\omega \, \in \, [\omega_{-},\omega_{+}]} \left( \Sigma(\omega) + m \cdot \omega \right)
		\end{equation}
		\begin{equation}
			\label{complexityFromReplicatedFreeEntropy}
			\Sigma(\omega(m))  = \Phi(m) - m \Phi'(m) =  \Phi(m) - m\cdot \omega(m)
		\end{equation}
		where $\Sigma(\omega)$ is defined and positive on $[\omega_{-},\omega_{+}]$. Eq'n \ref{complexityFromReplicatedFreeEntropy} is a Legendre inversion and requires $\Sigma(\omega)$ to be concave and $m$ to be in a range $[m_{-},m_{+}]$ s.t. the supremum of eq'n \ref{replicatedFreeEntropyFromComplexity} is found in the interior of $[\omega_{-},\omega_{+}]$. The total free-entropy density is, in the 1RSB approximation, the minimum of $\Phi(m)/m$. \\

		We'll overload the notation and write $\Sigma(m=m_0) \equiv \Sigma(\omega(m_0))$. Note that the slope of curve $\Sigma(m=m_0)$ at any particular value of $m_0$ of $m$ is  $\Sigma(m=m_0)'  =  -m_0$: the point $m=0$ marks the maximum\footnote{The complexity is always concave for K-sat.} of the complexity curve, while the point $m=1$ is, when $\Sigma$ is defined there, the point where the complexity attains slope $\Sigma(m=1)' = -1$.\\

		Figure \ref{im_complexityFreeEntropyDensity} (p.\pageref{im_complexityFreeEntropyDensity}) shows the complexity for 4-sat at different values of $\alpha$: for $\alpha = 9.3$ the complexity is defined only on $[\omega_{-},\omega_{+}] \approx [0.02,0.04]$, and has no point of slope $-1$; for $\alpha = 9.45$ the complexity attains slope $-1$ at a value of $m$ where it is still positive (the point $(\Sigma(m=1),\omega(m=1))$ is marked with the black circle); for higher values of alpha the complexity vanishes before reaching slope $-1$, at some $m=m_s < 1$ (the points $(\Sigma(m=m_s),\omega(m=m_s))$ are marked with black squares). More in general we'll denote by $m_s$ the Parisi 1RSB parameter describing the thermodynamically relevant clusters in RSB phases (this will become clear in the coming section). 

		\newpage

	\subsection{Replica symmetry breaking}
		This section is based on \cite{core_clustersAndRSB}. Replica symmetry breaking manifests in K-sat in the following phases:
		\begin{enumerate}

			\item[RS] In the \textbf{replica symmetric phase} the measure is concentrated
				\footnote{From now on by \textbf{the measure is concentrated} we'll mean that the states we're ignoring are thermodynamically irrelevant, i.e. their collective measure vanishes in the thermodynamic limit.} 
			in a single thermodynamically relevant cluster, that is to say
			\begin{equation}
				\max\limits_{\psi} w_{\psi} \rightarrow 1 \text{ as } N \rightarrow \infty \text{ w.h.p. }
			\end{equation}

			The replicated free-entropy is given by $\Phi_{RS}(m) = m \cdot \omega_\star$, where $\omega_{\star}$ is the contribution of the single dominant cluster. We also have $Z_{N} \approx \exp[N \, \omega_\star]$.\\

			\item[d1RSB] In the \textbf{dynamical 1RSB phase} the measure is concentrated into 
			\begin{equation}
				\mathcal{N}(\omega_\star) \approx \exp[N \, \Sigma_\star] \text{  clusters}
			\end{equation}
			all with the same weight
				\footnote{This is 1RSB, so we expect a single layer of clusters, all statistically identical in the thermodynamic limit.} 
			$w_\psi \approx \exp[-N \, \Sigma_\star]$ and free-entropy density $\omega_\star$.\\

			$\Phi(m)/m$ is minimised at $m=1$, with $\Sigma_\star \equiv \Sigma(\omega_\star) = \Phi(1) - \Phi'(1) > 0$ and $\omega_\star = \Phi'(1)$. The thermodynamically relevant clusters are thus described by a 1RSB solution with Parisi parameter $m=m_s=1$, which means they indeed all have the same size.\\

			\item[1RSB] In the \textbf{1RSB phase} the measure is concentrated into a sub-exponential number of clusters $\psi_1, \psi_2,...$ (w.l.o.g. consider them in order of decreasing weight). The sequence of weights $w_{\psi_n}$ converges to a Poisson-Dirichlet Point process\footnote{Unsurprisingly, the same process that governs allele frequences in infinite coalescent trees \cite{DNAEvolution}.} of parameter $m_s \in (0,1)$
			\begin{equation}
			\begin{split}
				1 \geq w_{\psi_1} \geq w_{\psi_2} \geq ... \geq 0 & \text{, satisfying } \sum_{i=1}^{\infty} w_{\psi_j} = 1\\
				&\text{ and } w_{\psi_i} = z_i \cdot \prod_{1 \leq j < i}(1-z_j)\\
				&\text{ for } (z_i)_i \text{ i.i.d. with density } m_s (1-z)^{m_s-1} 
			\end{split}
			\end{equation}
			For more details about Poisson-Dirichlet Point processes see \cite{survey_talagrandNew}\cite{mezardMontanariInformationPhysicsComputation}\cite{griffithsPDP}.

			The thermodynamically relevant clusters are thus described by a 1RSB solution with a Parisi parameter $m=m_s < 1$ minimising $\Phi(m)/m$. The free-entropy density of these states is $\omega_\star = \Phi'(m_s)$, and the complexity $\Sigma_\star$ vanishes as expected from the sub-exponential number of states. 

		\end{enumerate}

		Note that, in the d1RSB and 1RSB phases, the value of $m$ describing the thermodynamically relevant clusters is always the one minimizing $\Phi(m)/m$.

		\newpage

	\subsection{Phase transitions in K-sat}	
		
		\begin{figure}[hp]
			\begin{center}
				\includegraphics[width=16cm]{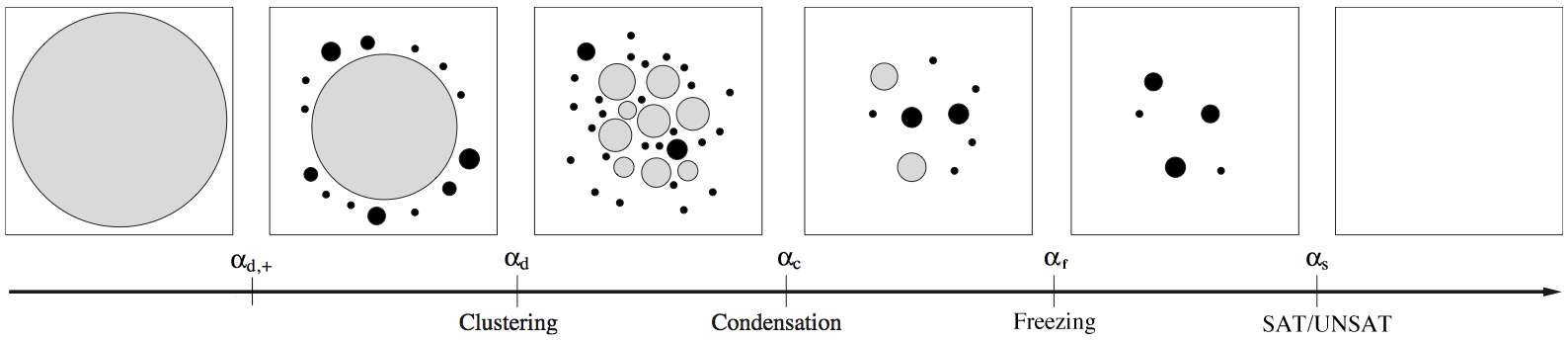}
			\end{center}
			\label{phaseTransitionsFreezing}
		\end{figure}

		This and the following sections are based on \cite{core_clustersAndRSB}\cite{core_gibbsStates}\cite{core_algorithmicBarriers}\cite{survey_cojaOghlan}, with some insights from \cite{survey_randomSatisfiability}\cite{core_mezardZecchina}\cite{mezardMontanariInformationPhysicsComputation} (the figure on top is from \cite{zdebrodovaPhaseTransColouring}). As $\alpha$ grows, K-sat undergoes the following structural phase transitions, some of which will be presented in detail in the next sections:

		\begin{itemize}
		
			\item[$\alpha_{d,+}$] A positive complexity appears for some values of $\omega$, but K-sat is still in the RS phase: there is a single dominant cluster which at this phase transition starts shedding an exponential number of thermodynamically irrelevant clusters described by a 1RSB solution with $m=1$.
			
			\item[$\alpha_{d}$] K-sat undergoes the \textbf{clustering phase transition}\footnote{Also known as \textbf{dynamic phase transition}.} from the RS phase to the d1RSB phase (from now on also called the \textbf{clustered phase}): the main cluster shatters into an exponential number of exponentially small, almost identical, thermodynamically relevant clusters (described by a 1RSB solution with $m=m_s=1$).
			
			\item[$\alpha_{c}$] K-sat undergoes the \textbf{condensation phase transition} from the d1RSB phase to the 1RSB phase (from now on also called the \textbf{condensed phase}): the solutions tend to \textit{condensate} into a sub-exponential number of thermodynamically relevant clusters (described by a 1RSB solution with $m=m_s<1$). In 3-sat this coincides with $\alpha_d$ (i.e. no clustered phase in 3-sat).
			
			\item[$\alpha_{f}$] Frozen variables appear in thermodynamically dominating clusters.
			
			\item[$\alpha_{s}$] K-sat undergoes the SAT/UNSAT phase transition, with a random instance $\mathcal{I}$ going from satisfiable w.h.p. to unsatisfiable w.h.p.
		
		\end{itemize}  

		\newpage

	\subsection{Clustering: the Dynamical phase transition RS $\rightarrow$ d1RSB }
		\label{section_clustering}

\begin{figure}[h!]
		\begin{center}
		\includegraphics[width=16cm]{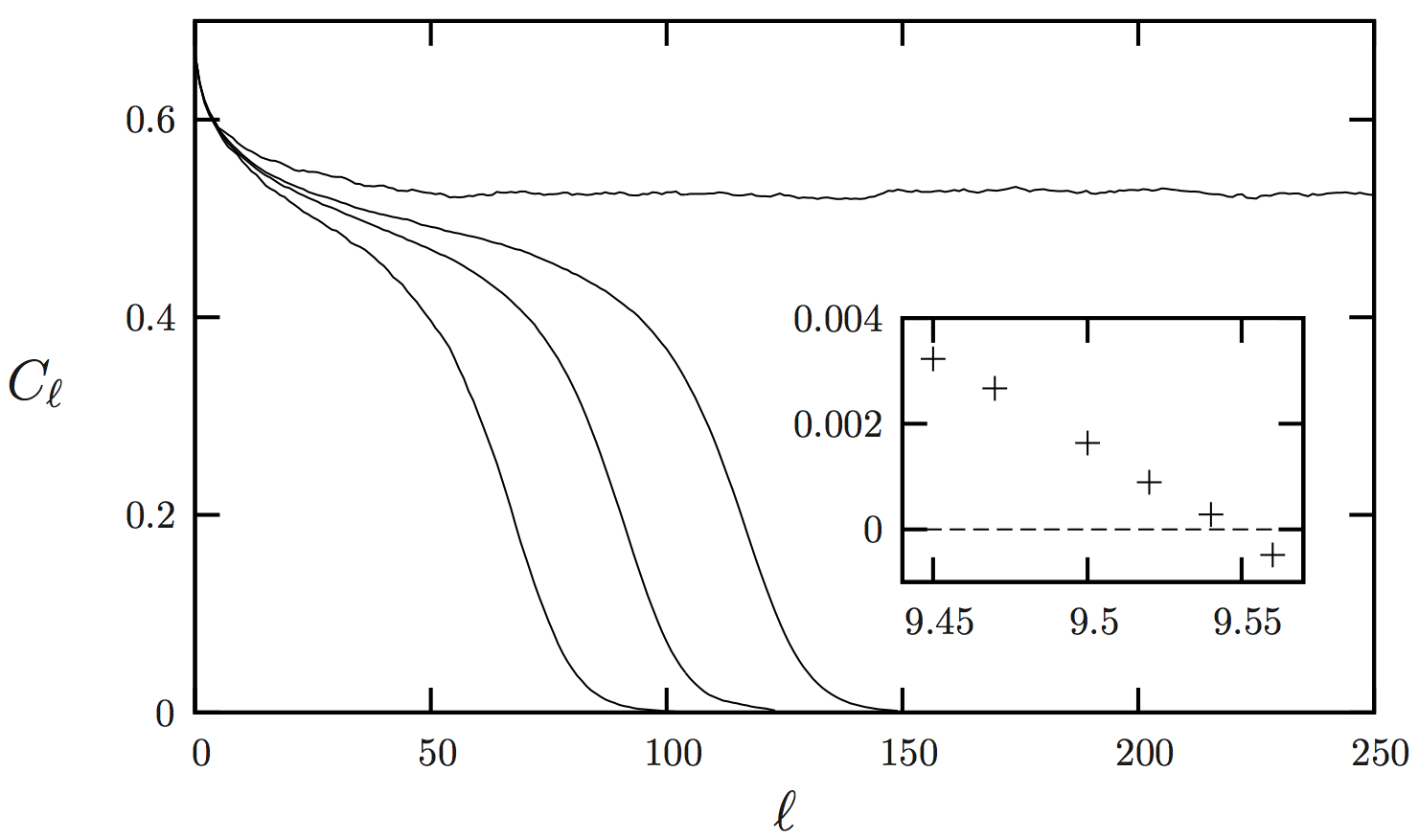}
		\end{center}
		\caption{The point-set correlation $C_l$  for 4-sat: left to right $\alpha = 9.30,9.33,9.35 < \alpha_d \approx 9.38$ (where it vanishes in the long-range limit $l \rightarrow \infty$) and $\alpha = 9.40>\alpha_d$ (where it stops vanishing). Ignore the inset. Figure from \cite{core_gibbsStates}.}
		\label{im_pointSetCorrelation}
		\end{figure}

		Take $D$ to be the ball of radius $l$ around some random variable node in the factor graph, and $\tau \equiv \left. \sigma \right|_{V \backslash D}$ to be its complement. As $\alpha$ crosses $\alpha_d$, the spins become \cite{core_gibbsStates} globally correlated under the Gibbs measure, i.e. the following \textbf{point-set correlation} stops vanishing in the $l \rightarrow \infty$ limit 
		\begin{equation}
			C_l \eqdef \expect{\sum_{\tau}\mu_N( \tau ) } \;  \left|| \mu_N( \cdot  | \tau )- \mu_N( \cdot ) \right||
		\end{equation}
		\cite{survey_cojaOghlan} mentions a 2010+ result from Coja-Oghlan and Gerke by which $\lim\limits_{l \rightarrow \infty}C_l = 1/2$ once crossed $\alpha_d$. Figure \ref{im_pointSetCorrelation} (p.\pageref{im_pointSetCorrelation}) shows this for 4-sat.\\

		This phase transition is best characterised by the clustering of solution space, which the following groundbreaking results from \cite{core_algorithmicBarriers} presents at its finest.

		\begin{theorem} (Shattering)\\
			We define a \textbf{region} to be any non-empty union of clusters of solutions, and we also define the \textbf{height} of a path $\sigma(1),...,\sigma(T)$ to be
			\begin{equation}
				\operatorname{height}(\sigma(1),...,\sigma(T)) \eqdef \max\limits_{1 \leq t \leq t} H_N(\sigma(t))
			\end{equation} 
			Then there is a sequence $\epsilon_K \rightarrow 0$ s.t. for all $\alpha$ in the region\footnote{For K small enough the region might be empty. \cite{core_algorithmicBarriers} mentions quick calculations suggesting the result to hold at least for $K \geq 8$, and the evidence presented in the rest of this chapter would suggest the result to hold for $K \geq 4$.}
			\begin{equation}
				\label{achlioptasShatteringRegion}
				\alpha_d = (1+\epsilon_K)\dfrac{2^K}{K}\log K \leq \alpha \leq (1-\epsilon_K) 2^{K} \log 2 
			\end{equation} 
			the solution space of a random K-sat instance $\mathcal{I}$ \textbf{shatters}, i.e. w.h.p. there exists at least $\exp[O(N)]$ regions with the following properties:
			\begin{enumerate}
				\item each region contains at least an $\exp[-O(N)]$ fraction of all solutions
				\item the distance between any two vertices in distinct regions is at least $O(N)$
				\item every path between any two vertices in distinct regions has height at least $O(N)$
			\end{enumerate}
		\end{theorem}

		The picture is the following: for low $\alpha$ the solution space is a single big cluster, but as it crosses $\alpha_d$ the big cluster shatters into an exponential amount of exponentially small regions, very far from each other\footnote{The farthest they can be is indeed $O(N)$, more precisely $N$.} and separated by very high energy barriers\footnote{The highest they can be is $O(M) = O(N)$, more precisely $M$}.\\

		The proof found in \cite{core_algorithmicBarriers} is probabilistic and very rigorous, based on the establishment of a connection between the planted model and the uniform model that enables the authors to work with typical solutions but still leave space of manoeuvre by allowing a (small but exponential) number of atypical solutions. This level of rigour is relevant because, up until then, similar results had been based on the physics-inspired cavity method, which has not yet received a proper mathematical formalisation.\\

		A rigorous value for the clustering phase transition is given in \cite{core_algorithmicBarriers} to be 
		\begin{equation}
			\alpha_d = (1+o(1))\frac{2^K}{K}\log K
		\end{equation}
		while \cite{core_gibbsStates} and \cite{core_clustersAndRSB} use the cavity method to obtain the more refined (but less rigorous) 
		\begin{equation}
			\alpha_d = \frac{2^K}{K}\left(\log K + \log \log K + 1 + O\left( \frac{\log\log K}{\log K } \right)\right)
		\end{equation}

		Now it's time to give a look at the statistical mechanics of the clustering phase transition.\\

		Figure \ref{core_clustersAndRSBfig3} (p.\pageref{core_clustersAndRSBfig3}) shows the clustering phase transition $\alpha_d$ to be the point at which a positive complexity $\Sigma(m=1)>0$ starts existing: this makes $m=1$ the new minimum for $\Phi(m)/m$, and corresponds to the appearance of an exponential number of thermodynamically relevant clusters with hierarchical structure described by a 1RSB solution $m=m_s=1$.\\

		Figure \ref{core_clustersAndRSBfig3} also shows that the RS estimate $\omega_{RS}$ for the free-entropy density can be expressed as
		\begin{equation}
			\omega_{RS} = \Sigma(m=1) + \omega(m=1)
		\end{equation}
		which confirms the picture of the clustering phase transition as shattering of a single solution cluster of entropy density $\omega_{RS}$ into $\exp[N \, \Sigma(m=1)]$ clusters of entropy density $\omega(m=1)$, without any discontinuous loss of global entropy (or, equivalently, of number of solutions) in the process. The free-entropy density of the typical solution, though, has a discontinuous jump down.\\
		 
		It should be noted that the maximum $\Sigma(m=0)$ of the complexity curve is defined way before the clustering phase transition: this corresponds to the existence of an exponential number of clusters described by a 1RSB solution with parameter $m=0$. But in the RS phase these solutions are thermodynamically irrelevant: this is indeed the $\alpha_{d,+}$ phase transition. Figure \ref{core_clustersAndRSBfig3} shows this for 4-sat, at $\alpha_{d,+} \approx 8.297$.\\

		Figure \ref{im_complexityFreeEntropyDensity} (p.\pageref{im_complexityFreeEntropyDensity}) shows more in detail this evolution of the complexity curve: for $\alpha < \alpha_d$ only a small part of it is defined, around $\omega(m=0)$, while for $\alpha > \alpha_d$ the curve is defined at least up to $\omega(m=1)$ (and much further): it is indeed the appearance of a point $\Sigma'(m=1)$ of slope $-1$ that marks the transition (it's what sets the minimum for $\Phi(m)/m$ at $m=1$).\\

		Figure \ref{core_clustersAndRSBfig5} (p.\pageref{core_clustersAndRSBfig5}) confirms that $m_s=1$ constantly in the region $\alpha_d \leq \alpha \leq  \alpha_c$, i.e. that in the d1RSB phase the thermodynamically relevant clusters are indeed described by a 1RSB solution with parameter $m=1$ (and are thus all approximately equal in size). \\
		
		Finally the overlaps in the d1RSB phase are given by $q_0(m=1)$ and $q_1(m=1)$, and are shown in red in figure \ref{core_clustersAndRSBfig6} (p.\pageref{core_clustersAndRSBfig6}).

		\newpage

	\subsection{Condensation phase transition d1RSB $\rightarrow$ 1RSB}
		\label{section_condensation}
		
		Similarly to the clustering phase transition, the condensation phase transition can be formulated in terms of some notion of correlation decay. 
		Specifically the factorisation result of theorem \ref{gibbsMeasureFactorisationTheorem} (p.\pageref{gibbsMeasureFactorisationTheorem}), which holds for K-sat in the region $\alpha < \alpha_c$, fails in the condensed phase: for $\alpha > \alpha_c$ the following quantity stops vanishing \cite{core_gibbsStates} in the thermodynamic limit
		\begin{equation}
			\expect{ \sum_{\sigma_{i(\cdot)}} \, \left| \; \mu_N(\sigma_{i(1)},...,\sigma_{i(n)})-\mu_N(\sigma_{i(1)})\cdot ... \cdot \mu_N(\sigma_{i(n)}) \; \right|} 
		\end{equation}

		The condensation phase transition is shown in \cite{core_gibbsStates} and \cite{core_clustersAndRSB} to happen at
		\begin{equation}
			\alpha_c = 2^K \log 2 - \frac{3}{2} \log 2 + O(2^{-K})
		\end{equation}

		The best way to understand what's happening this time is to go straight to the statistical mechanics.\\

\begin{figure}[h!]
			\begin{center}
			\includegraphics[width=16cm]{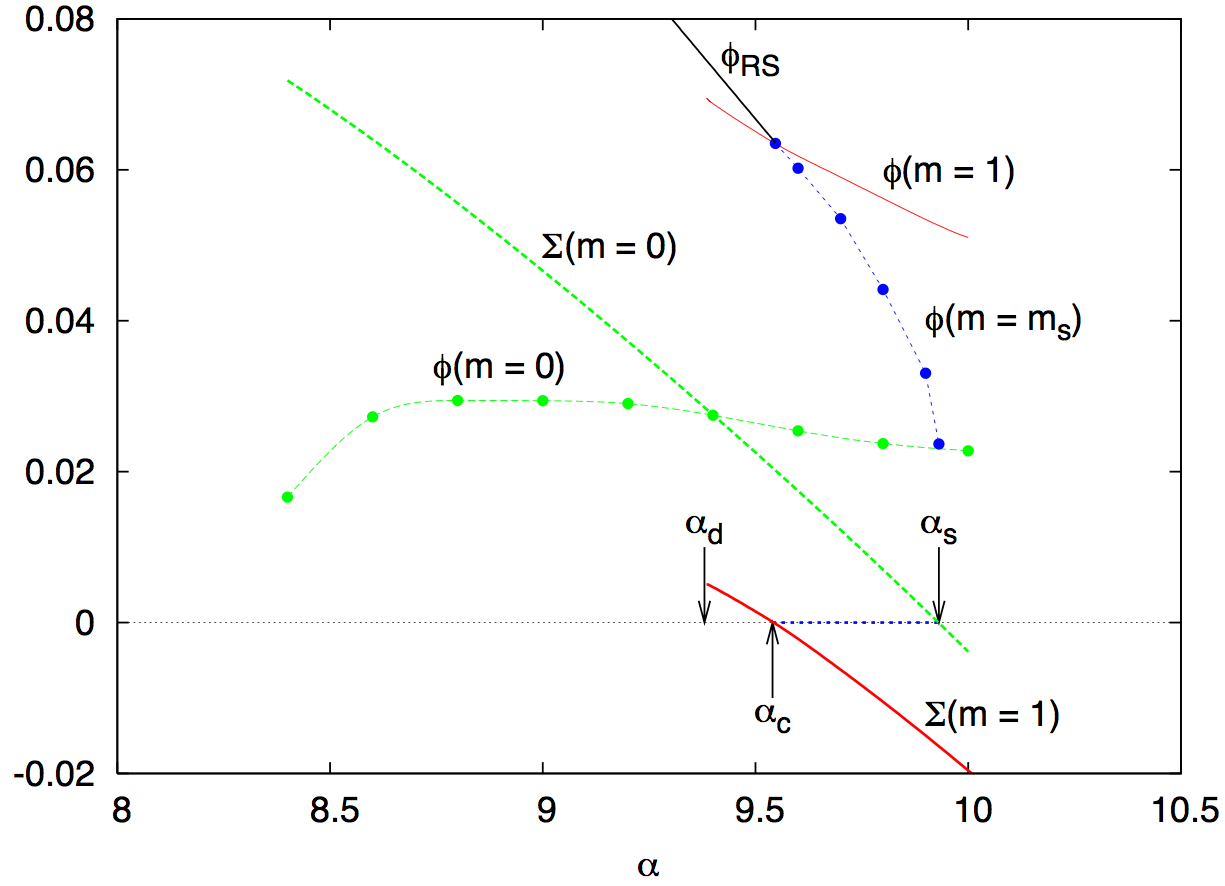}
			\end{center}
			\caption{The complexity $\Sigma$ and free-entropy density $\omega$ (here denoted $\phi$) of 4-sat. The dynamical phase transition is at $\alpha_{d} \approx 9.38$, the condensation phase transition is at $\alpha_c \approx 9.547$ and the SAT/UNSAT phase transition is at $\alpha_s \approx 9.931$. Quantities are colour-coded based on the phase for which they are of interest: green are of interest for the RS phase, red are of interest for the clustering phase, and blue are of interest for the condensation phase. Figure from \cite{core_clustersAndRSB}.}
			\label{core_clustersAndRSBfig3}
		\end{figure}

		Figure \ref{core_clustersAndRSBfig3} (p.\pageref{core_clustersAndRSBfig3}) shows the condensation phase transition to be the point at which $\Sigma(m=1)$ vanishes. Above $\alpha_c$ the quantity $\Phi(m)/m$ is maximised by the value $m = m_s < 1$ s.t. $\Sigma(m=m_s) = 0$, i.e. the highest value at which the complexity is non-negative: this means that the measure is concentrated into a sub-exponential (since $\Sigma = 0$) number of clusters described by a 1RSB solution of parameter $m=m_s<1$. \\

		Figure \ref{core_clustersAndRSBfig3} also confirms that at $\alpha_c$ we get $\omega_{RS} = \omega(m=1)$, and that the decrease in free-entropy density for the typical solution is continuous at $\alpha_c$. It also shows, though, that the free-entropy density has discontinuous derivative at condensation, a phenomenon that we'll encounter again when talking about the residual free-entropy density for BP-guided decimation in section \ref{condensationPhaseTransitionResidulaEntropy}.\\

		Figure \ref{core_clustersAndRSBfig5} (p.\pageref{core_clustersAndRSBfig5}) shows the value of $m_s$ decreasing continuously from $m_s(\alpha_c) = 1$ to $m_s(\alpha_s) = 0$, where it vanishes as $m_s(\alpha) \approx \sqrt{\alpha_s-\alpha}$: the cluster weights oscillate wildly up until the point where all clusters vanish at $\alpha_s$.\\

		Figure \ref{core_clustersAndRSBfig6} (p.\pageref{core_clustersAndRSBfig6}) finally shows, in blue, the overlaps $q_0(m=m_s)$ and $q_1(m=m_s)$ for the clusters in the 1RSB phase.

	\subsection{Freezing phase transition}
		\label{section_frozen}

		A question of relevance for K-sat solvers is: how free am I to set an arbitrary value for a spin? 
		We'll see in section \ref{chapter_KsatAlgorithms} that the general case is related to the so-called \textit{residual free-entropy density} and will require us to compute the marginal distribution of the single spins (using Belief Propagation). 
		The extreme case, though, is that where spins get \textit{frozen}, i.e. they take only one value within a cluster, and can be understood in terms of the \textbf{freezing phase transition}.\\

		\begin{figure}[h!]
			\begin{center}
				\includegraphics[width=16cm]{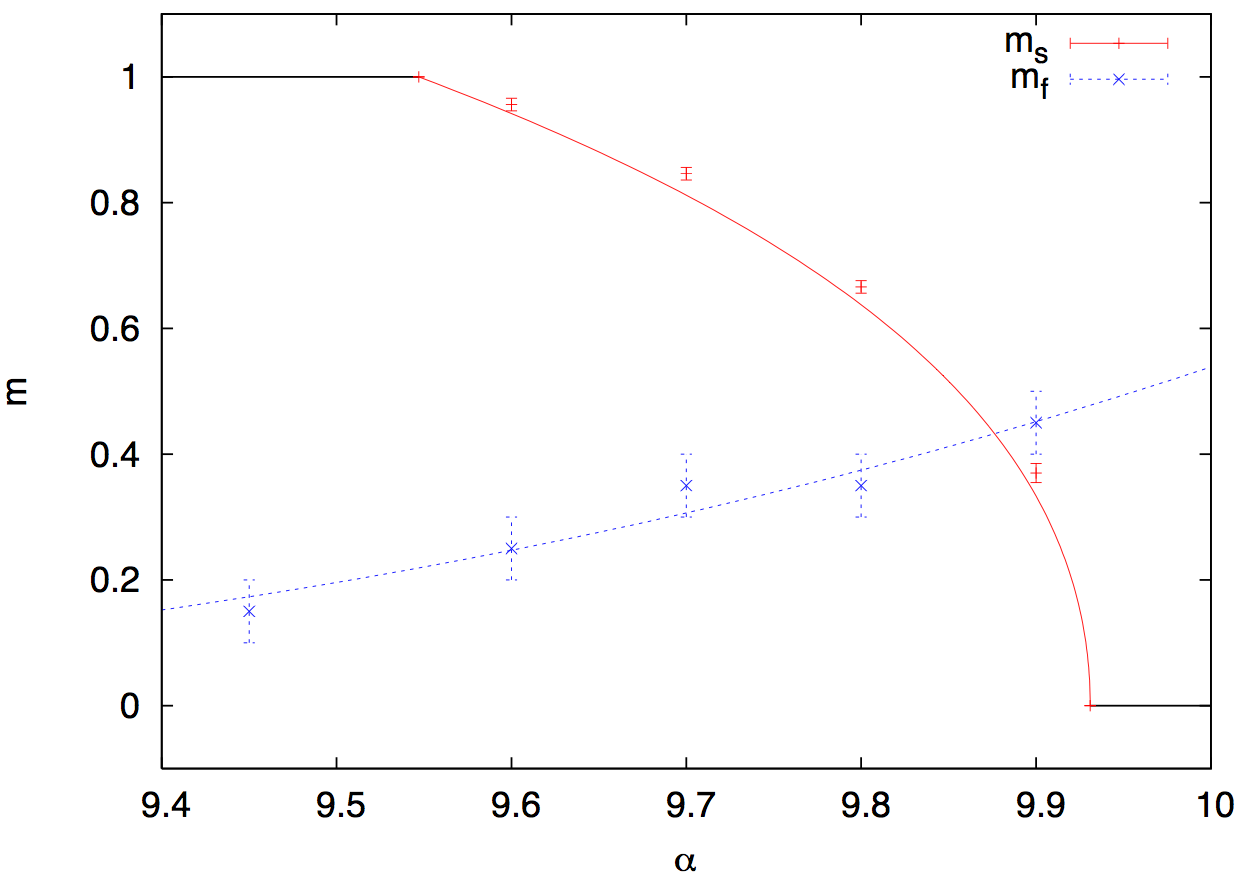}
			\end{center}
			\caption{In black/red the Parisi 1RSB parameter $m_s$ for the thermodynamically relevant clusters, going from the $m_s = 1$ of the clustered phase to the $m_s = 0$ of the UNSAT phase. In blue the parameter $m_f$ describing the biggest clusters containing frozen variables. Figure from \cite{core_clustersAndRSB}.}
			\label{core_clustersAndRSBfig5}
		\end{figure}

		Given a cluster $\psi$, we'll define the \textbf{projection} on the $i^{th}$ spin by

		\begin{equation}
			\pi_\psi(i) = \suchthat{s \in \{\pm 1\}}{\exists \, \sigma \in \psi \text{ s.t. } \sigma_i = s}
		\end{equation}

		and we'll say that the variable $x_i$ is \textbf{frozen} in cluster $\psi$ if $\pi_\psi(i) \neq \{\pm 1\}$, i.e. if its value in the cluster is fixed. 

		\newpage

		Then a theorem from \cite{geometry_achlioptasRicciTersenghi} guarantees that a \textbf{freezing phase transition} exists, i.e. there is\footnote{For K big enough: the authors of  \cite{geometry_achlioptasRicciTersenghi} prove it for $K \geq 9$ and report evidence suggesting that no freezing exists for $K=3$; we'll see that experimental results from \cite{core_clustersAndRSB} suggest that freezing takes place for $K\geq 4$.} a value $\alpha_f$ such that for $\alpha > \alpha_f$ every cluster will w.h.p. contain a majority of frozen variables:
		\begin{equation}
			\alpha_f = \left(\frac{4}{5}+o(1)\right) \, 2^K \log 2
		\end{equation}

		In fact the authors prove that for any $\varepsilon \in (0,1)$ there is a an $\alpha_f(\varepsilon)$ s.t. every cluster will w.h.p.  contain at least $\varepsilon \, N$ frozen variables (and $\alpha_f \equiv \alpha_f(1/2)$).\\

		From the point of view of the Parisi 1RSB parameter, we can ask which value $m_f$ describes the biggest clusters that w.h.p. will have frozen variables: the freezing phase transition is then the point $\alpha_f$ s.t. $m_f(\alpha_f) = m_s(\alpha_f)$, which figure \ref{core_clustersAndRSBfig5} (p.\pageref{core_clustersAndRSBfig5}) shows for 4-sat to be at $\alpha_f \approx 9.88$. 
		The situation presented by figure \ref{core_clustersAndRSBfig5} is not generic though: for $K=4$ we have $\alpha_f > \alpha_c$ while for  $K\geq 6$ it can be shown that $\alpha_d < \alpha_f < \alpha_c$.\\

		But we can do better: given a frozen variable $x_i$ and a satisfying assignment $\sigma$, we can ask how far we have to go in the solution space to find an assignment $\tau$ where $x_i$ takes a different value.

		\newpage

		We start by defining the following notions of rigidity:

		\begin{enumerate}
			
			\item[(a)] $x_i$ is \textbf{$f(N)$-rigid} if for every satisfying assignment $\tau$ we have $\tau_i \neq \sigma_i \imply d(\sigma,\tau)>f(N)$. We'll (slightly) change our notion of frozen variable to: $x_i$ is \textbf{frozen} if it is $\log(N)$-rigid. 
			
			\item[(b)] $x_i$ is \textbf{$f(N)$-loose} if there is a satisfying assignment $\tau$ s.t. $\tau_i \neq \sigma_i \wedge d(\sigma,\tau)\leq f(N)$. We'll say that $x_i$ is \textbf{fluid} if it is $\log(N)$-loose.
		
		\end{enumerate}
		
		Then the authors of \cite{core_algorithmicBarriers} and \cite{survey_cojaOghlan} proved the following result:

		\begin{theorem} (Frozen and fluid variables)\\
			Let $(\mathcal{I},\sigma)$ a random instance-solution pair\footnote{First choose a uniformly random satisfiable instance $\mathcal{I}$, then choose a uniformly random satisfying assignment $\sigma$ for $\mathcal{I}$. The distribution of $(\mathcal{I},\sigma)$ is not uniform over all instance-solution pairs.} s.t. $\alpha$ is in the region of eq'n \ref{achlioptasShatteringRegion} (p.\pageref{achlioptasShatteringRegion}). Then w.h.p. the number of $\Omega(N)$-rigid (and thus frozen) variables in $\sigma$ will be at least $\gamma_K \, N$, for a sequence $\gamma_K \rightarrow 1$. 
			Furthermore the $\Omega(N)$ bound is tight, as w.h.p. $\sigma$ will have $\Omega(N)$ variables that are not bound by any constraint (and thus cannot be rigid).\\
			Finally in the region $\alpha < \alpha_d$ every variable of $\sigma$ will w.h.p. be $o(n)$-loose, and in fact fluid.
		\end{theorem}

	\subsection{SAT/UNSAT phase transition}
		Finally the SAT/UNSAT phase transition marks the point where $\mathcal{I}$ goes from being w.h.p. satisfiable to being w.h.p. unsatisfiable. Figure \ref{core_clustersAndRSBfig3} (p.\pageref{core_clustersAndRSBfig3}) shows this to coincide with the point at which the maximum $\Sigma(m=0)$ of the complexity vanishes.\\
		
		Also we see that $\omega(m=m_s)$ gets to coincide with $\omega(m=0)$: the thermodynamically dominant, condensed clusters (sub-exponential in number) decrease in free-entropy density and size up to those of the thermodynamically irrelevant (but exponential in number) clusters, and then the exponential family of clusters composing the whole solution space suddenly disappears at $\alpha_s$.\\

		\begin{figure}[h!]
			\begin{center}
				\includegraphics[width=16cm]{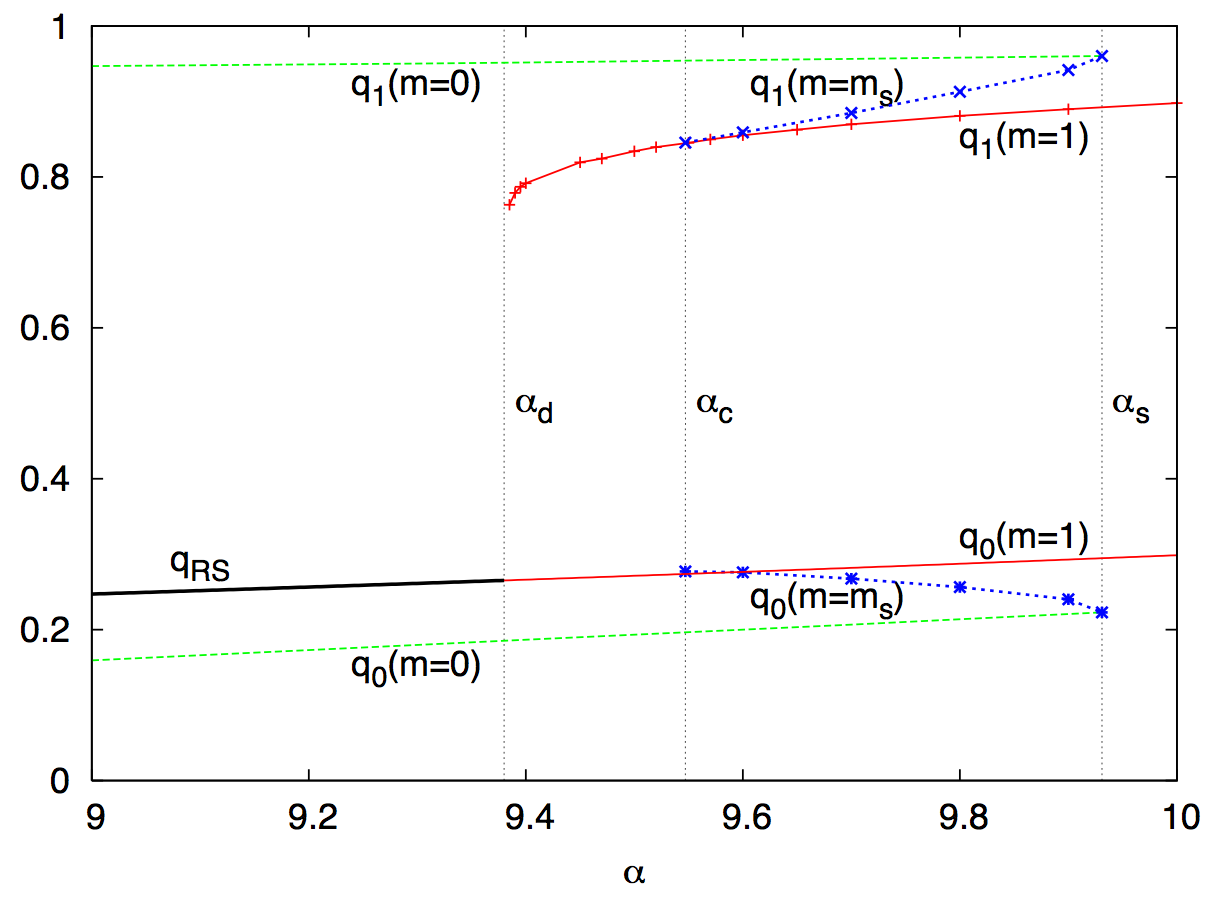}
			\end{center}
			\caption{The RS overlap is shown before the clustering phase transition. The 1RSB overlaps are shown in the clustered phase for $m=1$ and in the condensed phase for $m=m_s$. Figure from \cite{core_clustersAndRSB}.}
			\label{core_clustersAndRSBfig6}
		\end{figure}

		\newpage

		Figure \ref{core_clustersAndRSBfig5} (p.\pageref{core_clustersAndRSBfig5}) shows this to coincide with the point at which $m_s$ vanishes, suggesting that the thermodynamically dominant clusters become described, just before $\alpha_s$, by the same 1RSB solution with $m=0$ that until then described the thermodynamically irrelevant clusters generated at $\alpha_{d,+}$. \\
		
		Figure \ref{core_clustersAndRSBfig6} provides further confirmation of this by showing that the overlaps $q_0(m=m_s), q_1(m=m_s)$ of the 1RSB solution describing the condensed clusters go to coincide with the overlaps $q_0(m=0),q_1(m=0)$ of the 1RSB solution describing the clusters shedded at $\alpha_{d,+}$.\\

		Cavity method calculations from \cite{core_clustersAndRSB} \cite{core_gibbsStates} set
		\begin{equation}
			\alpha_s = 2^K \log 2 -\frac{1+\log 2}{2} + O(2^{-K})
		\end{equation}
		
		The following rigorous result is proven in \cite{achlioptasPeres} via the \textit{second moment method}:
		\begin{equation}
		2^K \log 2 - (K+1)\frac{\log 2}{2}-1-o(1) \leq \alpha_s \leq 2^K \log 2 -\frac{1+\log 2}{2} + o(1)
		\end{equation}
%% END The phases of K-sat

\newpage

%% BEGIN Algorithms for K-sat
\section{Algorithms for K-sat}
	\label{chapter_KsatAlgorithms}
	
	\subsection{The algorithmic barrier}
		
		Unless otherwise stated, this section is based on \cite{core_algorithmicBarriers}\cite{core_gibbsStates}\cite{survey_cojaOghlan}\cite{survey_randomSatisfiability}.\\

		The search for a clustering phase transition was initiated by the following empirical observation: all known efficient (i.e. poly-time w.h.p.) algorithms for K-sat stopped finding solutions at densities $\approx \frac{2^K}{K}$, much lower than the well known SAT/UNSAT threshold; in fact, no efficient algorithm performed, asymptotically in K, better than the naive \textbf{Unit Clause Propagation}\footnote{If there is a unit clause, satisfy it, otherwise assign a random value to a random variable}. 
		In 2010, just 2 years after the proof of existence of the clustering phase transition at $\alpha_d \approx \frac{2^K}{K}\log K$ appeared in \cite{core_algorithmicBarriers}, the algorithm Fix, which succeeds w.h.p. up to $\alpha_d$ was introduced and rigorously analysed in \cite{cojaOghlanFix}. 
		The following table \cite{cojaOghlanFix}\cite{survey_cojaOghlan} gives \textbf{algorithmic barriers}\footnote{I.e. densities above which the algorithms stop succeeding w.h.p. (in polytime, when relevant).} for the best known efficient K-sat solvers.\\

		\begin{figure}[h!]
			\begin{center}
			\includegraphics[width=12cm]{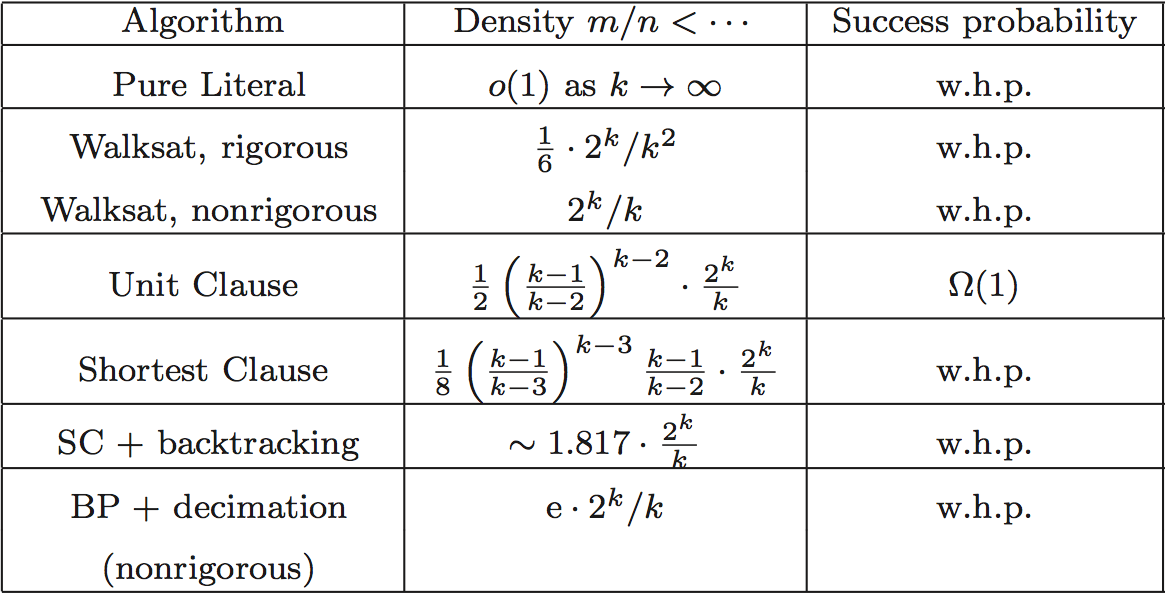}
			\end{center}
			\label{KsatAlgorithmsTable}
		\end{figure}

		So what is the intuition behind these barriers? If a problem has non-zero complexity, local algorithms will easily get stuck in the exponential multitude of local minima.
		This is the common problem of Walksat, Unit Clause, Shortest clause, all of which start failing at $\approx \frac{2^K}{K}$, just before the clustering phase transition. 
		This clustering, with its exponential number of small, far-away dominant clusters, high energy barriers and long-range correlations, is widely believed to be the ultimate barrier for local algorithms. Indeed no efficient algorithm is rigorously proven to succeed past it.\\ 

		We'll now move onto the most important non-local algorithms, Belief Propagation and Survey Propagation: neither of them is rigorously known to succeed on densities higher than $\alpha_d$ (in fact we'll see that Belief Propagation is guaranteed to fail w.h.p. at densities higher than $O(\frac{2^K}{K})$), but the ideas involved in their formulation and analysis are invaluable for a thorough understanding of the connection between the phase transitions of K-sat and the efficiency of algorithms to solve it.

	\subsection{Belief propagation}
		Unless otherwise stated, this section is based on \cite{gibbs_factor}\cite{core_mezardZecchina}\cite{BSP_newLook}, with some insights from \cite{core_clustersAndRSB}\cite{survey_randomSatisfiability}.\\

		The following \textbf{DPLL algorithm}\footnote{And in fact a much wider family of DPLL algorithms obtained by replacing step 1 with better heuristics.} forms the basis of many K-sat solvers, and succeeds w.h.p. in linear time on constraint densities up to $\alpha = O(2^k / k)$:\\

		DPLL($\mathcal{I}$):
		\begin{enumerate}

			\item Apply the \textit{pure literal rule}
				\footnote{Satisfy all pure literals, i.e. literals that appear always with the same polarisation, i.e. literals the complement of which doesn't appear in the formula.} 
			and satisfy any \textit{unit clause}
				\footnote{I.e. a clause with just one literal.} 
			until no pure literal and/or unit clauses remain. Call the result $\mathcal{I}'$.

			\begin{enumerate}
				
				\item Exit returning SATISFIABLE if $\mathcal{I}'$ is empty.
				
				\item Exit if a contradiction is generated.
				
			\end{enumerate}
			
			\item Select a variable $\sigma_i$ appearing in $\mathcal{I'}$ and a random value $s_i \in \{\pm 1\}$
			
			\item DPLL($\left. \mathcal{I}' \right|_{\sigma_i=+s_i}$)
		
			\item DPLL($\left. \mathcal{I}' \right|_{\sigma_i=-s_i}$)
		
		\end{enumerate}

		A way to improve step 2 of DPLL is to compute the marginal distribution 
		\begin{equation}
		\mu_N(\sigma_i) = \sum\limits_{\sigma_j \text{ s.t. } j\neq i} \mu_N(\sigma_1,...,\sigma_N)
		\end{equation}
		and then setting $s_i$ to its most likely value under the marginal $\mu_N(\sigma_i)$.
		%\begin{enumerate}
		%\item[(a)] $s_i = +1$ with probability $\mu_N(\sigma_i = +1)$
		%\item[(b)] $s_i=-1$ with probability $\mu_N(\sigma_i = -1)$
		%\end{enumerate}
		The purpose of the \textbf{Belief Propagation (BP) algorithm} is that of computing that marginal efficiently by exploiting the factor graph. BP returns the exact marginal if the factor graph is a tree, and a (hopefully converging) series of approximations if the factor graph has loops. 		
				
	\subsubsection{An typical example of BP}

\begin{figure}[h!]
		\begin{center}
		\includegraphics[width=12cm]{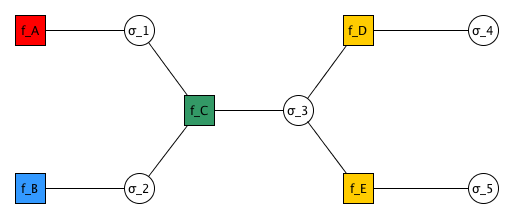}
		\end{center}
		\caption{Factor graph for the example of BP done in this section.}
		\label{BPexampleFactorGraph}
		\end{figure}

		First we see a typical example of computation of marginals, taken from \cite{gibbs_factor}. Suppose that $N=5$ and that the factor graph is given by figure \ref{BPexampleFactorGraph} (p.\pageref{BPexampleFactorGraph}): then the measure factors as
		\begin{equation}
			Z_5 \cdot \mu_5(\sigma_1, ... ,\sigma_5) = f_A(\sigma_1) f_B(\sigma_2) f_C(\sigma_1,\sigma_2,\sigma_3) f_D(\sigma_3,\sigma_4) f_E(\sigma_3,\sigma_5) 
		\end{equation}

		Using distributivity of sum and product (which is what BP is all about) we write the marginal as
		\begin{equation}
			Z_5 \cdot \mu_5(\sigma_1) = f_A(\sigma_1) \left(\sum_{\sigma_2}\sum_{\sigma_3} f_B(\sigma_2) 
			  f_C(\sigma_1,\sigma_2,\sigma_3) \left(  \left(  \sum_{\sigma_4} f_D(\sigma_3,\sigma_4) \right) \left(  \sum_{\sigma_5} f_E(\sigma_3,\sigma_5) \right) \right) \right)
		\end{equation}

		We see that to compute the marginal $\mu_5(\sigma_1)$ we need $f_A(\sigma_1)$ and $f_{BCDE}(\sigma_1,\sigma_2,\sigma_3) \downarrow \sigma_1$, where we defined the \textbf{summary operator} $\downarrow$ by
		\begin{equation}
			f(\left.\sigma\right|_{\partial a}) \downarrow \sigma_i \eqdef \sum_{\substack{\sigma_j \text{ s.t. }\\ j \in \partial a \backslash \{i\}}} f(...)
		\end{equation} 

		We have also defined the shorthand 
		\begin{equation}
			f_{BCDE}(\sigma_1,\sigma_2,\sigma_3) \equiv f_B(\sigma_2) f_C(\sigma_1,\sigma_2,\sigma_3) \left( \left(  \sum_{\sigma_4} f_D(\sigma_3,\sigma_4) \right) \left(  \sum_{\sigma_5} f_E(\sigma_3,\sigma_5) \right) \right)
		\end{equation}

		In turn to compute $f_{BCDE}(\sigma_1,\sigma_2,\sigma_3)$ we need $f_B(\sigma_2)$, $f_C(\sigma_1,\sigma_2,\sigma_3)$ and $f_{DE}(\sigma_3,\sigma_4,\sigma_5) \downarrow \sigma_3$, where we have defined another shorthand
		\begin{equation}
			\label{equation_fDE_decomposition}
			f_{DE}(\sigma_3) \equiv  \left(  \sum_{\sigma_4} f_D(\sigma_3,\sigma_4) \right) \left(  \sum_{\sigma_5} f_E(\sigma_3,\sigma_5) \right)
		\end{equation}

		Finally to compute $f_{DE}(\sigma_3,\sigma_4,\sigma_5)$ we need $f_D(\sigma_3,\sigma_4) \downarrow \sigma_3$ and $f_E(\sigma_3,\sigma_5) \downarrow \sigma_3$. 

		\begin{figure}[h!]
		\begin{center}
		\includegraphics[width=7cm]{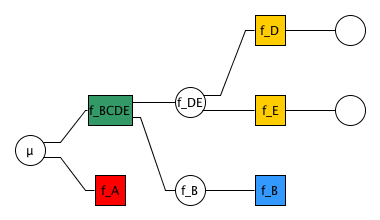}
		\end{center}
		\caption{Dependencies for the computation of the marginal $\mu_5(\sigma_1)$.}
		\label{BPexampleDependencies}
		\end{figure}

		\begin{figure}[h!]
		\begin{center}
		\includegraphics[width=12cm]{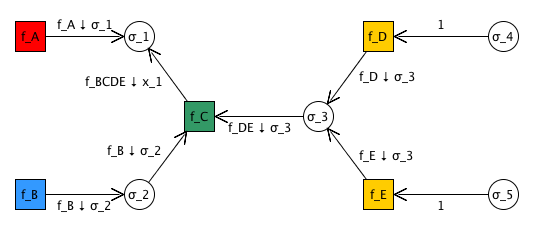}
		\end{center}
		\caption{Messages passed in the execution of BP.}
		\label{BPexampleExecution}
		\end{figure}

		\newpage

		The previous decomposition of the dependences suggest a computation for $\mu_5(\sigma_1)$ based on the idea of passing messages $\eta_{a \rightarrow i}$ and $\eta_{i \rightarrow a}$ on directed edges $f_a \rightarrow \sigma_i$ and $\sigma_i\rightarrow f_a$: this is illustrated in figure \ref{BPexampleExecution} (p.\pageref{BPexampleExecution}) and reported in detail below.

		\begin{enumerate}

		\item[1.] Start with messages from the leaves: $\eta_{4 \rightarrow D}(\sigma_4) = 1$, $\eta_{5 \rightarrow E}(\sigma_5) = 1$

		\begin{eqnarray}
			\eta_{B \rightarrow 2}(\sigma_2) =  f_B(\sigma_2) \downarrow \sigma_2 = \sum\limits_{\substack{\sigma_j \text{ s.t. }\\ j \in \partial B \backslash \{2\}}} f_B(\sigma_2) \cdot 1 = f_B(\sigma_2)\\
			\eta_{A \rightarrow 1}(\sigma_1) =  f_A(\sigma_1) \downarrow \sigma_1 = \sum\limits_{\substack{\sigma_j \text{ s.t. }\\ j \in \partial A \backslash \{1\}}} f_A(\sigma_1) \cdot 1 = f_A(\sigma_1)
		\end{eqnarray}

		Notice that we can write  $1 = \prod\limits_{\emptyset}... \; = \prod\limits_{j \in \partial B \backslash \{2\}} \eta_{j \rightarrow B} = \prod\limits_{j \in \partial A \backslash \{1\}} \eta_{j \rightarrow A} \;$.

		\item[2a.] Then proceed to compute messages

		\begin{eqnarray}
			\eta_{D \rightarrow 3}(\sigma_3) = f_D(\sigma_3,\sigma_4) \downarrow \sigma_3 &= \sum\limits_{\substack{\sigma_j \text{ s.t. }\\ j \in \partial D \backslash \{3\}}}  f_D(\sigma_3,\sigma_4) \cdot \eta_{4 \rightarrow D}(\sigma_4)\\ 
			& =\sum\limits_{\sigma_4 = \pm} f_D(\sigma_3,\sigma_4) \cdot \eta_{4 \rightarrow D}(\sigma_4)\\
			\eta_{E \rightarrow 3}(\sigma_3) = f_E(\sigma_3,\sigma_5) \downarrow \sigma_3 &= \sum\limits_{\substack{\sigma_j \text{ s.t. }\\ j \in \partial E \backslash \{3\}}}  f_E(\sigma_3,\sigma_5) \cdot \eta_{5 \rightarrow E}(\sigma_5)\\ 
			&= \sum\limits_{\sigma_5 = \pm} f_E(\sigma_3,\sigma_5) \cdot \eta_{5 \rightarrow D}(\sigma_5)
		\end{eqnarray}

		Notice that we can write $\eta_{4 \rightarrow D}(\sigma_4) \; = \prod\limits_{j \in \partial D \backslash \{3\}} \eta_{j \rightarrow D}$ and similarly for $\eta_{5 \rightarrow E}$.

		\item[2b.] Also compute message
		\begin{equation}
			\eta_{2\rightarrow C}(\sigma_2) = f_B(\sigma_2) \downarrow \sigma_2 = \eta_{B \rightarrow 2}(\sigma_2)
		\end{equation}

		Notice that we can write $\eta_{B \rightarrow 2}(\sigma_2) = \prod\limits_{b \in \partial 2 \backslash \{C\}} \eta_{b \rightarrow 2}(\sigma_2)$.

		\item[3.] Continue by computing the message (see eq'n \ref{equation_fDE_decomposition})
		\begin{equation}
			\eta_{3 \rightarrow C}(\sigma_3) = f_{DE}(\sigma_3) \downarrow \sigma_3 = \prod \limits_{d \in \partial 3 \backslash \{C\} } \eta_{d \rightarrow 3}(\sigma_3)
		\end{equation}

		\item[4.] Finally compute the message 
		\begin{eqnarray}
			\eta_{C \rightarrow 1}(\sigma_1) &= f_{BCDE}(\sigma_1,\sigma_2,\sigma_3) \downarrow \sigma_1 = \sum\limits_{\substack{\sigma_j \text{ s.t. }\\ j \in \partial C \backslash \{1\}}} f_C (\sigma_1,\sigma_2,\sigma_3) \cdot \prod\limits_{j \in \partial C \backslash \{1\}} \eta_{j \rightarrow C}(\sigma_j)\\
			&= \sum\limits_{\sigma_2 = \pm}\sum\limits_{\sigma_3 = \pm} f_C (\sigma_1,\sigma_2,\sigma_3) \cdot \eta_{2 \rightarrow C}(\sigma_2)\cdot \eta_{3 \rightarrow C}(\sigma_3)
		\end{eqnarray}

		\end{enumerate}
		And now we're done: all we need to do to compute the marginal $\mu_5(\sigma_1)$ is observe that
		\begin{equation}
		\begin{aligned}
			Z_5 \cdot \mu_5(\sigma_1) & = \prod\limits_{a \in \partial 1} \eta_{a\rightarrow 1}(\sigma_1)\\
			Z_5 & = \sum\limits_{\sigma_1 = \pm} Z_5 \cdot \mu_5(\sigma_1)
		\end{aligned}
		\end{equation}

		One thing worth noting is that the messages passed on directed edges are functions, which might seem a little too abstract  from an implementative point of view. What has to be kept in mind, though, is that the messages are functions $\eta_{a \rightarrow i}(\sigma_{i})$ or $\eta_{i \rightarrow a}(\sigma_{i})$ of a binary spin variable $\sigma_i$, and thus what we're passing on are effectively pairs of real numbers.

		\newpage

	\subsubsection{The BP algorithm for tree factor graphs}

		The previous example clearly spells out what the BP algorithm for tree factor graphs should be. We want to compute $Z_N \cdot \mu_N(\sigma_t)$:

		\begin{enumerate}

			\item \textbf{Building the schedule:} the factor graph is explored breadth-first starting from variable node $\sigma_t$: whenever a node $v$ is explored as a child of some node $u$, we set a direction $u \leftarrow v$ to the edge $uv$. The directed graph (a tree) obtained at the end will have exactly one outgoing edge for all nodes, except for node $\sigma_t$ (the root), which has only incoming edges.

			\item \textbf{Computing the messages:} starting from leaves (no incoming edges) of the tree, the outgoing messages of all nodes are computed, the outgoing message of node $u$ being computed only when all incoming messages have been already computed.

			\item \textbf{Computing the marginal:} the exact marginal for $\sigma_t$ is computed as $Z_N \cdot \mu_N(\sigma_t)  = \prod\limits_{a \in \partial t} \eta_{a\rightarrow t}$. 

		\end{enumerate}

		The rules for computing the messages (also known as \textbf{message-passing fixed point eq'ns}) are:
		\begin{equation}
			\label{BPmessageRules}
			\begin{aligned}
				\eta_{a \rightarrow i}(\sigma_i) & =  \sum\limits_{\substack{\sigma_j \text{ s.t. }\\ j \in \partial a \backslash \{i\}}} f_a(\left.\sigma\right|_{\partial a}) \; \cdot \prod\limits_{j \in \partial a \backslash \{i\}} \eta_{j \rightarrow a}(\sigma_j)\\
				&&\\
				\eta_{i \rightarrow a}(\sigma_i) & = \prod\limits_{b \in \partial i \backslash \{a\}} \eta_{b \rightarrow i}(\sigma_i)
			\end{aligned}
		\end{equation}
			
	\subsubsection{The BP algorithm for general factor graphs}	
		The situation becomes more complicated for factor graphs with loops: there isn't a natural way to build the schedule, and the computation will not be exact. On the other hand there are a number of well-studied scheduling algorithms for message passing, and a number of results on convergence of the computation to marginals: we'll only cover the general principles here, and refer the reader to \cite{gibbs_factor} for detailed descriptions and results.

		\begin{enumerate}

			\item \textbf{Building the schedule:} each edge of the factor graph is replaced with two directed edges (instead of one), and each directed edge is initialised with some random message\footnote{A random function of the spin involved in the edge, i.e. random pair of real numbers.}. The exact details depend on the scheduling algorithm.

			\item \textbf{Computing the messages:} the messages are updated with the rules of eq'n \ref{BPmessageRules}: all incoming edges have values on them at any time and are updated according to some scheduling algorithm\footnote{The simplest algorithm updates all messages in subsequent generations, but much more efficient schedules exists.}. 
			This time the node $\sigma_t$ is not treated differently from the others.

			\item \textbf{Computing the marginal:} an approximation to the marginal at $\sigma_t$ can be computed at any sweep of the algorithm as $Z_N \cdot \mu_N(\sigma_t)  \approx \prod\limits_{a \in \partial t} \eta_{a\rightarrow t}(\sigma_t)$: the approximation is expected to be accurate and convergent in the RS phase, but a better algorithm (Survey Propagation) will be required in the 1RSB and RSB phases.

		\end{enumerate}
			
		\newpage

	\subsubsection{The BP algorithm for K-sat}
		\label{BPalgorithmForKsat}

		The functions on the factor nodes of K-sat all have the exponential form
		\begin{equation}
			\left. \mu \right|_{a}(\left. \sigma \right|_{\partial a}) = \exp\left[ - \beta \, W_a(\left. \sigma \right|_{\partial a}) \right]
		\end{equation} 
		so it is convenient to express the messages in exponential form as well: all the multiplicative operations involved in BP message processing become additive operations over the exponents.\\

		Concretely we write the messages as
		\begin{equation}
			\label{ksatBPMessages}
			\begin{aligned}
				\eta_{i \rightarrow a}(\sigma_i) & = \exp\left[ \beta \,(h_{i\rightarrow a}\sigma_i+\omega_{i \rightarrow a}) \right] \\
				\eta_{a \rightarrow i}(\sigma_i) & = \exp\left[ \beta \,(u_{a \rightarrow i}\sigma_i+\omega_{a \rightarrow i}) \right]
			\end{aligned}
		\end{equation} 

		and the BP rules \ref{BPmessageRules} for updating the messages become \cite{core_mezardZecchina}\cite{core_clustersAndRSB}
		\begin{equation}
			\label{ksatBPRules}
			\begin{aligned}
				h_{i \rightarrow a} &= \sum\limits_{b \in \partial i \backslash \{a\}} u_{b \rightarrow i}\\
				\exp\left[\beta \,u_{a \rightarrow i}\sigma_i\right] &= \sum\limits_{\substack{\sigma_j \text{ s.t. }\\ j \in \partial a \backslash \{i\}}} \exp\left[ \beta \left(  -W_a(\left. \sigma \right|_{\partial a}) + \sum\limits_{j \in \partial a \backslash \{i\}} h_{j \rightarrow a} \sigma_{j} \right) \right] 
			\end{aligned}
		\end{equation}

		Under eq'ns \ref{ksatBPRules}, we consider the message passed on edge $i \rightarrow a$ to be the \textbf{cavity-field} $h_{i\rightarrow a}$, and the message passed on edge $a \rightarrow i$ to be the \textbf{cavity-bias} $u_{a \rightarrow i}$.\\

		An interpretation for the BP messages is given by\footnote{Approximations are exact in tree factor graphs, by Markov property of $\mu_N$ (see Hammersley–Clifford theorem).}
		\begin{equation}
			\begin{aligned}
			\label{BPmessageInterpretation}
				\frac{1}{z_{a \rightarrow i}} \eta_{a \rightarrow i}(\sigma_i) & \approx \text{ marginal law of }\sigma_i\text{ when all factor nodes in }\partial i \backslash \{a\} \text{ are removed.}\\
				\frac{1}{z_{i \rightarrow a}} \eta_{i \rightarrow a}(\sigma_i) & \approx \text{ marginal law of }\sigma_i\text{ when factor node } a \text{ is removed.}
			\end{aligned}
		\end{equation}
		where $z_{u \rightarrow v} = \eta_{u \rightarrow v}(+1) + \eta_{u \rightarrow v}(-1)$ is the normalisation constant\footnote{We'll often write $\eta_{u \rightarrow v}$ when we don't need to distinguish the direction of the message.}. 
		Thus only the parameters $h_{i\rightarrow a}$ and $u_{a \rightarrow i}$ are of relevance, as the $\omega_{u\rightarrow v}$ can be eliminated through normalisation (but the message passing equations for normalised messages are more complicated).\\

		The computation of the marginal for $\sigma_i$ can then be written as
		\begin{equation}
			\begin{aligned}
			\label{BPKsatMarginal}
				\mu_N(\sigma_i) \approx \dfrac{\exp\left(\beta h_i \sigma_i\right)}{2 \cosh\left(\beta h_i \right)} 
			\end{aligned}
		\end{equation}
		where we have defined the \textbf{local field} at $\sigma_i$ by 
		\begin{equation}
			\label{BPKsatLocalField}
			 h_i = \sum\limits_{a \in \partial i} u_{a \rightarrow i}
		\end{equation}

		\newpage

	\subsection{BP guided decimation}
		\label{BPbetterLook}
		Unless otherwise stated, this section is based on \cite{montanari2007solving}\cite{cojaOghlanBP}\cite{ricciTersenghiBP}.

	\subsubsection{BP guided decimation algorithm}
		\label{BPGuidedDecimationAlgorithm}
		
		To understand the limits of BP, we will now see the \textbf{BP-guided decimation} algorithm. Starting with an instance $\mathcal{I}_0 \eqdeftemp \mathcal{I}$ of K-sat we proceed by \textit{decimation}, i.e. we assign values to one spin at a time by running BP on a sequence of progressively simplified CNF formulae $\mathcal{I}_0,\mathcal{I}_1,...,\mathcal{I}_t,...,\mathcal{I}_N$.
		We'll denote by $U_t$ to be the set of indices of spins that have been fixed after step $t$, and $\left. \tau \right|_{U_t}$ the family, indexed by $U_t$, of values that we've assigned to those spins. 
		The spins $\left. \sigma \right|_{U_t}$ will be called the \textbf{fixed spins/variables} at time $t$.\\

		Set $U_0 = \emptyset$, $\tau_{|_{U_{0}}} = ( \, )$ and for $t=1,...,N$ do:
		
		\begin{enumerate}
		
			\item choose a random $i_t \in V \backslash U_{t-1}$ and set $U_t := U_{t-1} \cup \{i_t\}$ 
		
			\item run BP on the factor graph of $\mathcal{I}_{t-1}$ to approximate the marginal of $\sigma_{i_t}$ in it:
			\begin{equation}
			\mu_{N-t+1}(\sigma_{i_t}) \equiv \mu_N( \sigma_{i_t}  | \tau_{|_{U_{t-1}}} ) 
			\end{equation}
		
			\item fix the spin value $\tau_{i_t} := \pm1$ with probability $\mu_{N-t+1}(\pm1)$
		
			\item simplify the CNF formula $\mathcal{I}_t := \left. \mathcal{I}_{t-1} \right|_{\sigma_{i_t} = \tau_{i_t}}$
			\end{enumerate}
			The algorithm can stop for only two reasons:
			\begin{enumerate}
		
			\item[(a)] at some point the partial assignment $\left. \tau\right|_{U_{t-1}}$ is not compatible with any solution: in this case BP fails to compute the marginal $\mu_{N-t+1}(\sigma_{i_t})$ because $Z_{N-t+1} \mu_{N-t+1}(\sigma_{i_t}) = 0$ and cannot be normalised;
		
			\item[(b)] a value has been assigned to all spins: this means that BP managed to compute all marginals, and thus the assignment $\sigma = \tau$ is a satisfying assignment for $\mathcal{I}$. 
		
		\end{enumerate}

		In fact the mechanism that makes BP fail is equivalent to the \textbf{Unit Clause Propagation (UCP)} algorithm, which sequentially progressively simplifies all unit clauses in a CNF $\mathcal{I}_t$\footnote{Just for this specific instance we'll go back to the boolean formulation.}:

		\begin{enumerate}
			
			\item pick a unit clause $C_a = \zeta_i \in \{x_i, \neg x_i\}$ 
			
			\item set $x_i = 0,1$ to satisfy clause $C_a$, i.e. to get $z_i = 1$
			
			\item for all other clauses $C_b$
			
			\begin{enumerate}
			
				\item if $z_i$ appears in a clause $C_b$, remove clause $C_b$ from the CNF (as it is now satisfied)
			
				\item if $\neg z_i$ appears in a clause $C_b$, remove $\neg z_i$ from $C_b$. If $\neg z_i$ was the only literal of $C_b$, return instance $\mathcal{I}_t$ to be unsatisfiable (i.e. a contradiction has been discovered)
			
			\end{enumerate}
		
		\end{enumerate}
		
		Running UCP until no unit clauses remain provides a family of spins whose values are \textbf{directly implied} at time $t$ by the fixed spins $\left. \sigma \right|_{U_t}$. We'll say that a spin $\sigma_i$ is \textbf{frozen}\footnote{Slightly different concept of frozen from the one of section \ref{section_frozen}} at time $t$ if either $i \in U_t$ or $\sigma_i$ is directly implied at time $t$, and denote by $W_t$ the set of indices of all frozen spins at time $t$ (thus $W_t \backslash U_t$ is the set of indices of all spins which are directly implied at time $t$).

	\subsubsection{Frozen variables and success probability for BP}
		Unless specified otherwise, this section is based on \cite{ricciTersenghiBP}.\\

		Start by defining the set of \textbf{newly frozen spins/variables} at time $t$ by $\left. \sigma \right|_{Z_t}$, where
		\begin{equation}
			Z_t \eqdef W_t \backslash W_{t-1} 
		\end{equation}
		Then $|Z_{\theta N}|$ is the amount of newly frozen variables when the fraction of fixed spins is $\theta$.\\

		Now  consider the subgraph $G_{\theta N}$ of the factor graph of $\mathcal{I}_{\theta N}$ obtained by only considering variables in $Z_{\theta N}$ and their adjacent factor nodes: if $|Z_{\theta N}|$ remains bounded in the thermodynamic limit then $G_{\theta N}$ will w.h.p. be a tree (see \cite{ricciTersenghiBP}); on the other hand if $|Z_{\theta N}|$ diverges then $G_{\theta N}$ will w.h.p. contain loops (see \cite{ricciTersenghiBP}\cite{montanari2007solving}). 
		The following theorem then connects the asymptotic behaviour of $Z_{\theta N}$ to the success probability of BP:

		\begin{theorem} 
			If $G_t$ is a tree then no contradiction will arise at time $t$, i.e. UCP will find no contradiction. As a consequence if $|Z_{\theta N}|$ remains bounded in the thermodynamic limit then BP will succeed w.h.p. at fraction $\theta$ of fixed variables.
		\end{theorem}

		Finally define the \textbf{fraction of frozen spins/variables}\footnote{The expectation is taken over all sources of randomness, including the set of fixed variables.}

		\begin{equation}
			\phi(\theta) \eqdef \frac{1}{N}\expect{\left| Z_{\theta N} \right|}
		\end{equation}
		to get, under reasonable regularity conditions, that 
		\begin{equation}
			\lim\limits_{N \rightarrow \infty} Z_{\theta N} = \dfrac{d \phi}{d\theta}
		\end{equation} 
		Thus BP succeeds w.h.p. in the region where $\phi(\theta)$ has bounded derivative, i.e. where the amount of newly frozen variables is bounded in the thermodynamic limit.\\

		The \textit{cavity method}\footnote{We'll present it in section \ref{cavityMethod}, but we'll not go over the computation of $\phi(\theta)$.} provides a fixed point equation for $\phi(\theta)$ which can be approximately solved via iterative methods:
		%which can be solved (approximately) using a message-passing procedure akin to BP and called \textit{population dynamics}. 
		for $\alpha$ low enough, any initial condition will converge to the same fixed point $\phi(\theta)$; for $\alpha$ high enough, on the other hand, the fixed point equation has 3 distinct solutions $\phi_{-}(\theta)$, $\phi_{0}(\theta)$ and $\phi_{+}(\theta)$, satisfying $\phi_{+}(\theta) \leq \phi_{0}(\theta) \leq \phi_{-}(\theta)$ at all $\theta$.\\

		We'll denote the value of $\alpha$ after which multiple fixed points appear by $\alpha_{\star}$, and we will be interested in $\phi_{\pm}(\theta)$ only: careful analysis of the initial conditions for population dynamics that lead to $\phi_{\pm}(\theta)$ shows that $\phi_{+}(\theta)$ is the solution for the RS regime, while $\phi_{-}(\theta)$ is the solution for the 1RSB regime. Both $\phi_{\pm}(\theta)$ have vertical slope and are discontinuous\footnote{Or multivalued, depending on the point of view.} at some fraction of fixed variables $\theta_{\pm}'$, with $\theta_{-}' \leq \theta_{+}'$.\\

\begin{figure}[h!]
			\begin{center}
				\includegraphics[width=12cm]{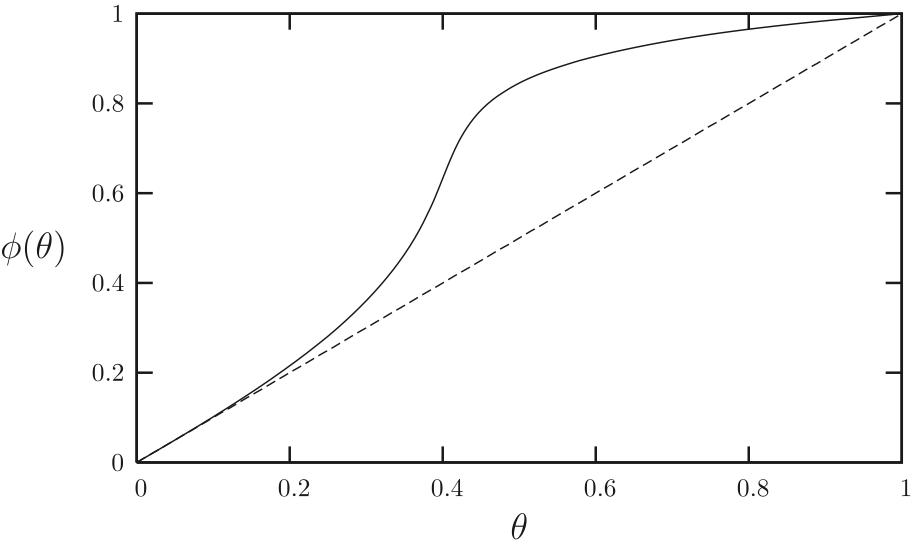}\\
				\includegraphics[width=12cm]{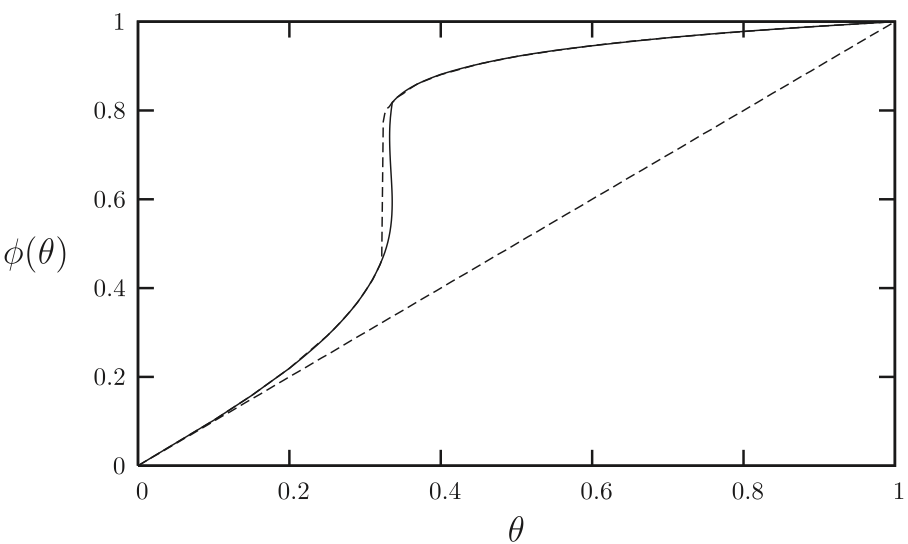}\\
			\end{center}
			\caption{Fraction of frozen variables $\phi(\theta)$ for random 4-sat, as a function of the fraction of fixed variables $\theta$. In the figure above we have $\alpha = 7.0 < \alpha_d^{(res)}$ and there is only one curve for $\phi(\theta)$ (the solid line). In the figure below we have $\alpha = 8.4 > \alpha_d^{(res)}$ and two curves appear for $\phi(\theta)$, superimposed on the range $\theta> \theta_{+}'$: the curve below (solid line) is the RS solution $\phi_{+}(\theta)$ and is discontinuous at $\theta_{+}'$, while the curve above (dashed line) is the 1RSB solution $\phi_{-}(\theta)$ and is discontinuous at $\theta_{-}'$ (where it starts being defined). For the dependence of $\theta_{\pm}'$ on $\alpha$ see figure \ref{ricciTersenghiBPfig79} (p.\pageref{ricciTersenghiBPfig79}). At small $\theta$, $\phi(\theta)$ is close to $\theta$ and its slope is close to 1: almost all frozen variables are just fixed variables, and the only newly frozen variables to be expected are the ones just fixed. At some intermediate value of $\theta$ the slope of $\phi(\theta)$ reaches its maximum (diverging in the bottom figure): enough variables have been fixed to induce sizeable cascades of unit clauses. At high $\theta$ the slope of $\phi(\theta)$ tends to zero: the expected number of newly frozen variables tends to zero, as most of the variables fixed are already directly implied. Figure from \cite{ricciTersenghiBP}.}
			\label{ricciTersenghiBPfig8}
		\end{figure}

		Figure \ref{ricciTersenghiBPfig8} (p.\pageref{ricciTersenghiBPfig8}) shows $\phi(\theta)$ for a value $\alpha < \alpha_{\star}$ and $\phi_{\pm}(\theta)$ for a value $\alpha > \alpha_{\star}$.
		The bottom plot in figure \ref{ricciTersenghiBPfig79} (p.\pageref{ricciTersenghiBPfig79}) shows, amongst other things, the lines $\theta_{\pm}'(\alpha)$ for 4-sat, and allows to estimate the critical point where they both originate to be $(\alpha_{\star}, \theta_{\star}) \approx (8.05,0.35)$. 

	\subsubsection{A condensation phase transition for the residual free-entropy density}
		\label{condensationPhaseTransitionResidulaEntropy}

		Unless specified otherwise, this section is based on \cite{ricciTersenghiBP}. \\

		It would be neat if $\alpha_{\star}$ marked the point where BP stops being effective, but this is not the case: as we'll see in section \ref{BPalgorithmicBarrier}, the algorithm behaves well up to a higher constraint density $\alpha_a$, very near to another phase transition $\alpha_c^{(res)}$ which we'll now explore.\\

		Firstly we need to generalised the quenched average $p_N$ of the log-partition function to the so-called \textbf{decimated regime}, i.e. to ensembles where a fraction $\theta$ of the spins is allowed to be fixed: in this context it is called the \textbf{residual free-entropy density} (or simply \textbf{residual entropy}) and it is denoted by $\omega(\theta)$\footnote{The expectation is taken over all sources of randomness, including the set of fixed variables.}:
		\begin{equation}
			\begin{aligned}
				Z_N[\tau_{|_{U}}] &\eqdef \sum_{\sigma}\prod_{a \in F} \exp\left[-\beta W_a(\left. \sigma \right|_{\partial a}) \right] \cdot 1_{\sigma_{|_{U}} = \tau_{|_{U}}}\\
				\omega(\theta) & \eqdef \lim\limits_{N \rightarrow \infty} \frac{1}{N} \expect{ \log Z_N[\tau_{|_{U}}] }
			\end{aligned}
		\end{equation}
		where we have put $U \equiv U_{\theta N}$. Please note that $\omega(0) = \lim\limits_{N \rightarrow \infty} p_N$ is the usual quenched average.\\

		Computation of the residual entropy proceeds through the cavity method, and it turns out that it can be expressed entirely in terms of the fraction of frozen variables $\phi(\theta)$, i.e. as $\omega(\theta) = \hat{\omega}\left(\phi(\theta)\right)$: 
		
		\begin{enumerate}
			
			\item[(a)]  in the region $\alpha < \alpha_{\star}$  the fixed point $\phi(\theta)$ is unique and there is no ambiguity for $\omega(\theta)$
			
			\item[(b)] in the region $\alpha_{\star}< \alpha < \alpha_{c}^{(res)}$ there are multiple solutions for $\phi(\theta)$, but again there is no ambiguity for $\omega(\theta)$ because the expressions $\hat{\omega}(\phi_{\pm}(\theta))$ coincide
			
			\item[(c)] in the region $\alpha > \alpha_{c}^{(res)} $ the two distinct expressions $\hat{\omega}(\phi_{\pm}(\theta))$ concur in determining $\omega(\theta)$ as
			
			\begin{equation}
				\omega(\theta) = \max\left[ \hat{\omega}(\phi_{-}(\theta)), \hat{\omega}(\phi_{+}(\theta)) \right]
			\end{equation}
		
		\end{enumerate}

\begin{figure}[h!]
			\begin{center}
				\includegraphics[width=12cm]{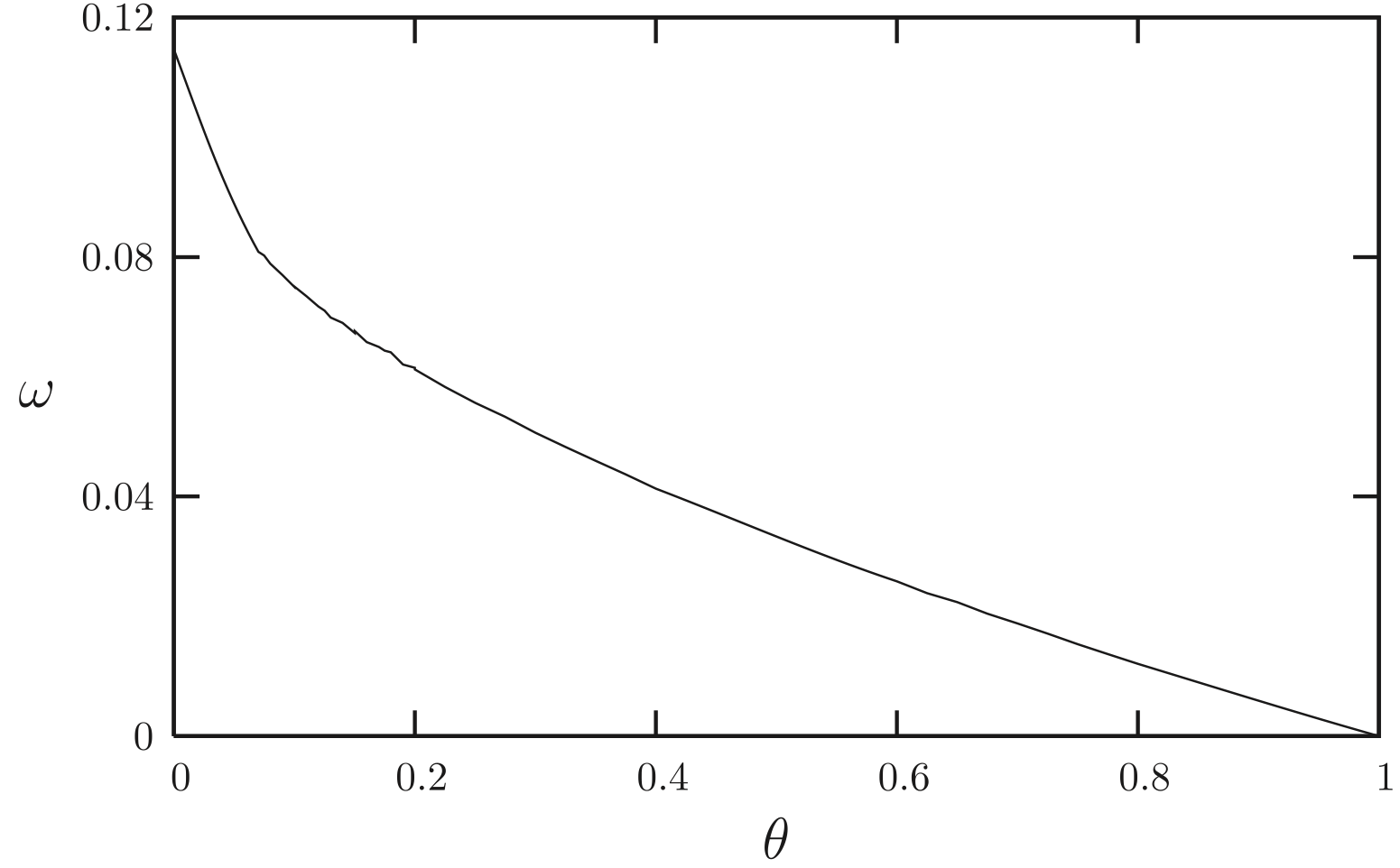}\\
				\includegraphics[width=12cm]{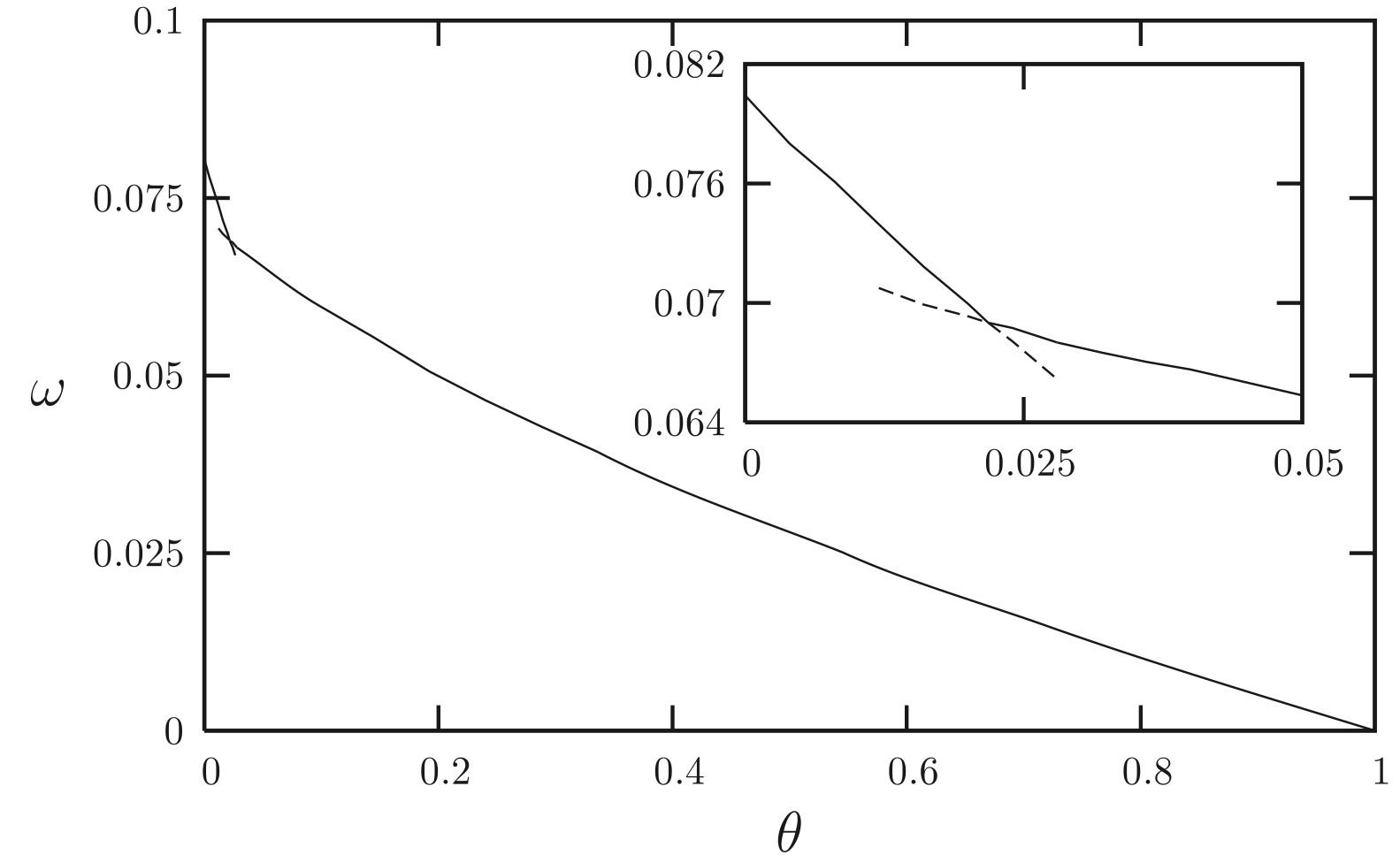}\\
			\end{center}
			\caption{Plots of the residual entropy for random 4-sat. In the figure above we have $\alpha = 8.8 < \alpha_{c}^{(res)}$ and the residual entropy is a smooth function of $\theta$. In the figure below we have $\alpha = 9.3 > \alpha_{c}^{(res)}$ and a singular point appears (in detail in the inset). The dashed lines in the inset correspond to the continuation of the two solutions in their region of coexistence. Figure from \cite{ricciTersenghiBP}.}
			\label{ricciTersenghiBPfig6}
		\end{figure}

		Both $\hat{\omega}(\phi_{\pm}(\theta))$ are decreasing, and for $\alpha > \alpha_{c}^{(res)}$ they intersect at a unique point $\theta_c(\alpha)$, where the derivative of $\omega(\theta)$ develops a discontinuity: the point $\alpha_c^{(res)}$ marking the appearance of this discontinuity is called the \textbf{condensation point} of the residual entropy. 
		Figure \ref{ricciTersenghiBPfig6} (p.\pageref{ricciTersenghiBPfig6}) shows plots of the residual entropy for 4-sat below and above its condensation point $\alpha_{c}^{(res)} \approx 9.05$.\\

		So what happens after $\alpha_c^{(res)}$ to $\omega(\theta)$, and why is it called a \textit{condensation point}? Careful analysis of $\hat{\omega}(\phi_{\pm}(\theta))$ shows that for $\alpha > \alpha_c^{(res)}$ there exists a region $[\theta_{-}(\alpha), \theta_{+}(\alpha)]$ where both solutions coexists, and that 
		\begin{enumerate}
			
			\item[(a)] $\omega(\theta) = \hat{\omega}(\phi_{+}(\theta))$ for $\theta \in [\theta_{-}(\alpha), \theta_c(\alpha)]$ (i.e. the RS solution dominates)
			
			\item[(b)] $\omega(\theta) = \hat{\omega}(\phi_{-}(\theta))$ for $\theta \in [\theta_{c}(\alpha), \theta_{+}(\alpha)]$ (i.e. the 1RSB solution dominates)
			
		\end{enumerate}

		\newpage

\begin{figure}[h!]
			\begin{center}
				\includegraphics[width=12cm]{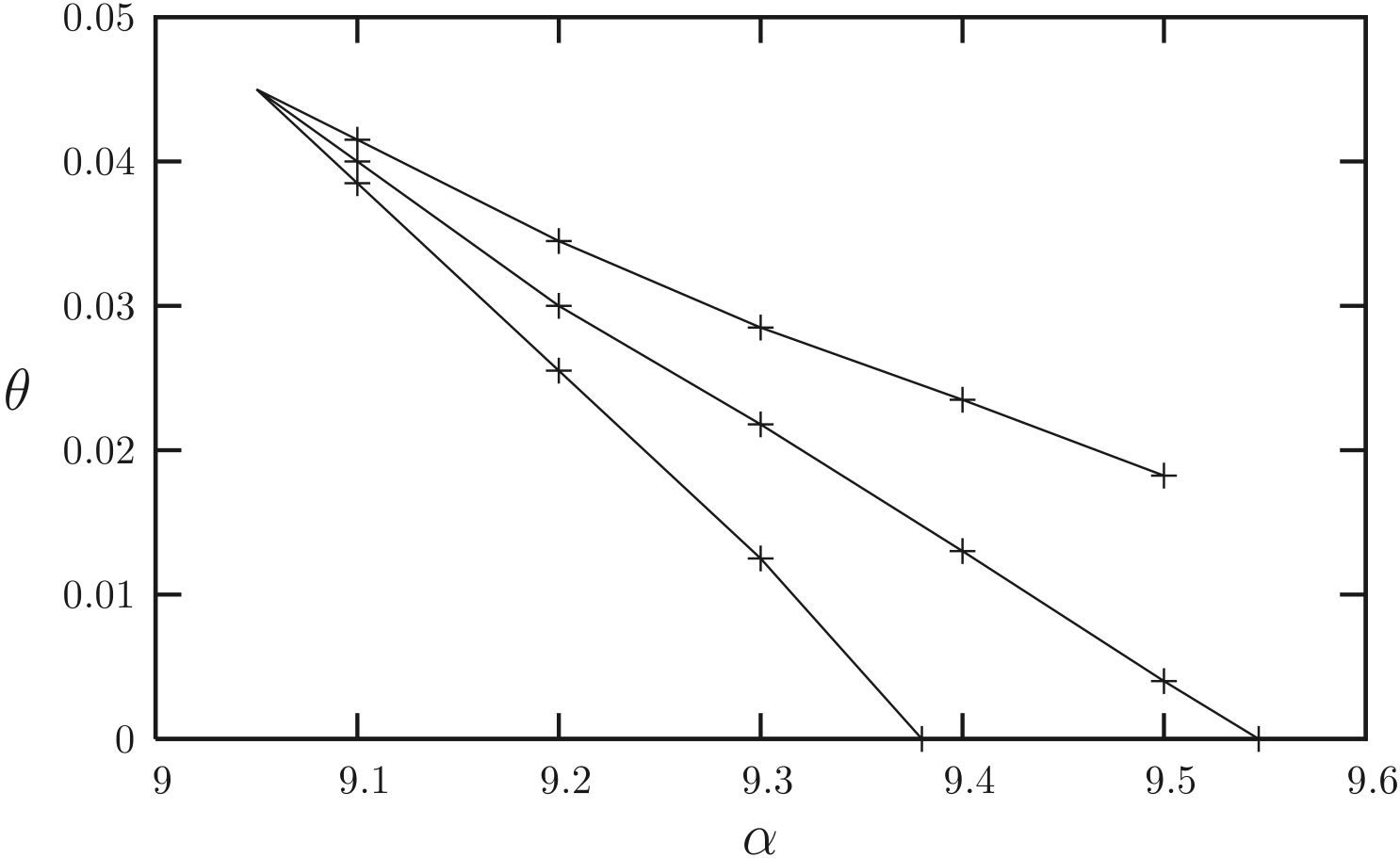}\\
				\includegraphics[width=12cm]{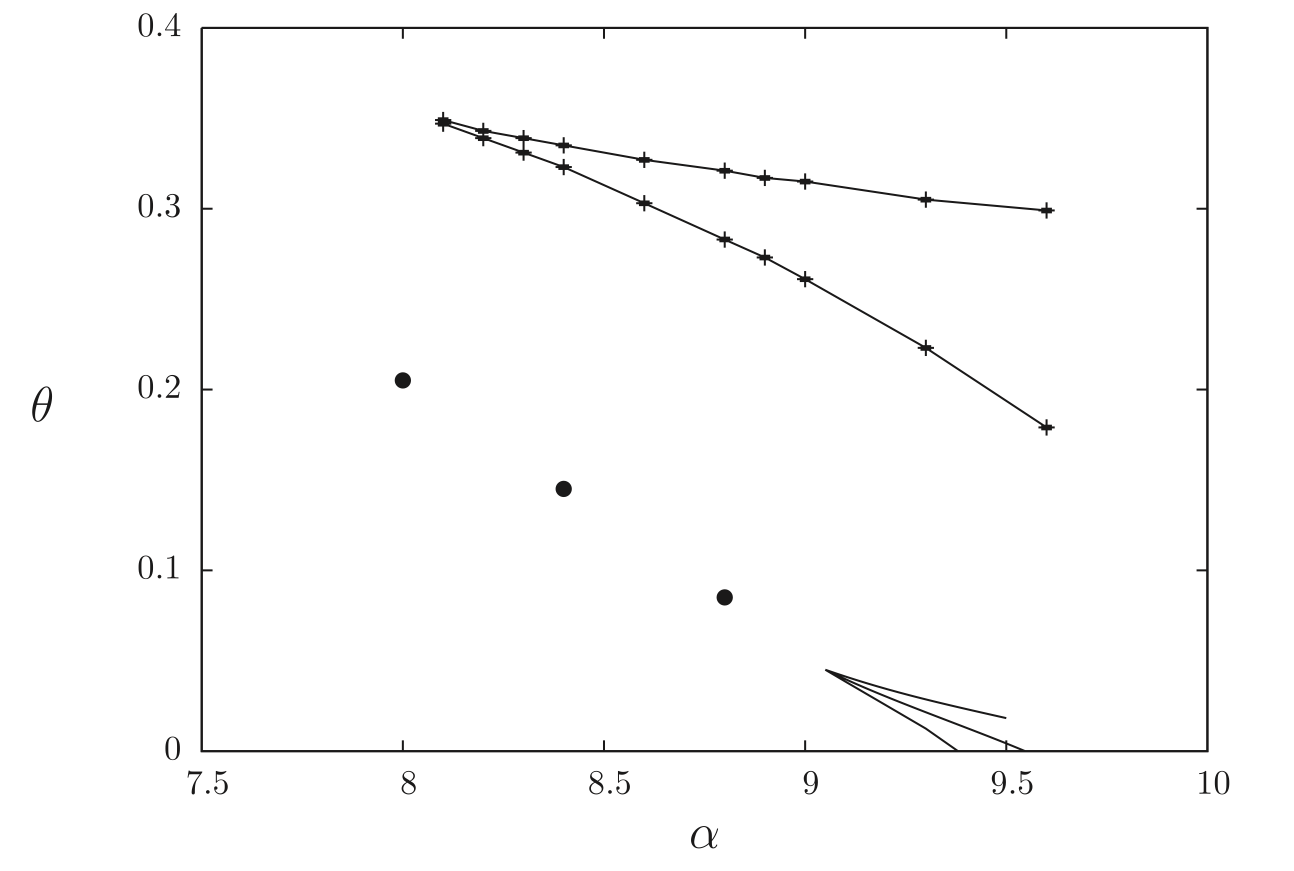}
			\end{center}
			\caption{Some phase transitions in the $(\alpha,\theta)$ plane of random 4-sat. In the figure above we have a plot of the three curves $\theta_{-}(\alpha)$, $\theta_{c}(\alpha)$ and $\theta_{+}(\alpha)$: notice the critical point $(\alpha_{c}^{(res)}, \theta_{c}^{\star}) \approx (9.05,0.045)$ in the top left corner. Also notice, in the bottom right corner, $\theta_{-}(\alpha)$ and $\theta_c(\alpha)$ touching $\theta = 0$ at $\alpha_d \approx 9.38$ and $\alpha_c \approx 9.547$ respectively. In the figure below we have a larger region of the $(\alpha,\theta)$ plane, showing the three $\theta_{\pm,c}(\alpha)$ curves in the bottom right corner and the $\theta_{\pm}'(\alpha)$ curves in the top half: notice the critical point $(\alpha_{\star}, \theta_{\star}) \approx (8.05,0.35)$ in the top left corner. Figure from \cite{ricciTersenghiBP}.}
			\label{ricciTersenghiBPfig79}
		\end{figure}

		Figure \ref{ricciTersenghiBPfig79} (p.\pageref{ricciTersenghiBPfig79}) shows the three curves $\theta_{\pm,c}(\alpha)$ of 4-sat stemming from the critical point $(\alpha_{c}^{(res)}, \theta_{c}^{\star}) \approx (9.05,0.045)$. 
		The \textbf{condensation curve} $\theta_c(\alpha)$ touches $\theta = 0$ at $\alpha_c \approx 9.547$, asserting the status of $\theta_c(\alpha)$ as the condensation phase transition in the $(\alpha,\theta)$ plane
			\footnote{I.e. it generalises the condensation phase transition from the condensation point on the $\alpha$ line to the condensation curve in the $(\alpha,\theta)$ plane.}. 
		The \textbf{clustering curve} $\theta_{-}(\alpha)$ touches $\theta = 0$ at $\alpha_d \approx 9.38$, asserting the status of $\theta_{-}(\alpha)$ as the clustering phase transition in the $(\alpha,\theta)$ plane. These interpretations are confirmed by the following considerations.\\

		In the region $[\theta_{-}(\alpha), \theta_{c}(\alpha)]$ the residual entropy is dominated by the RS solution $\phi_{+}(\theta)$, but the 1RSB solution exists and the complexity (at Parisi 1RSB parameter $m=1$) can be computed as:
		\begin{equation}
			\begin{aligned}
				\hat{\omega}(\phi_{+}(\theta)) - \hat{\omega}(\phi_{-}(\theta)) = \lim\limits_{N \rightarrow \infty} \expect{ \log \frac{Z_{+}}{Z_{-}} } = \lim\limits_{N \rightarrow \infty} \expect{ \log \mathcal{N}(\omega(\theta)) } = \left. \Sigma(\omega(\theta);\alpha) \right|_{m=1}
			\end{aligned}
		\end{equation}

		We thus have a well-defined, positive complexity, and recalling section \ref{section_clustering} we conclude the region between the curves $\theta_{-}(\alpha)$ and $\theta_{c}(\alpha)$ to be the clustering phase of K-sat in the $(\alpha,\theta)$ plane.\\

		In the region $[\theta_{c}(\alpha), \theta_{+}(\alpha)]$ the residual entropy is dominated by the 1RSB solution $\phi_{-}(\theta)$, and the complexity (at Parisi 1RSB parameter $m=1$), still defined, becomes negative:  recalling section \ref{section_condensation} we conclude the region between the curves $\theta_{c}(\alpha)$ and $\theta_{+}(\alpha)$ to be the condensation phase of K-sat in the $(\alpha,\theta)$ plane.\\

		\begin{figure}[h!]
			\begin{center}
				\includegraphics[width=12cm]{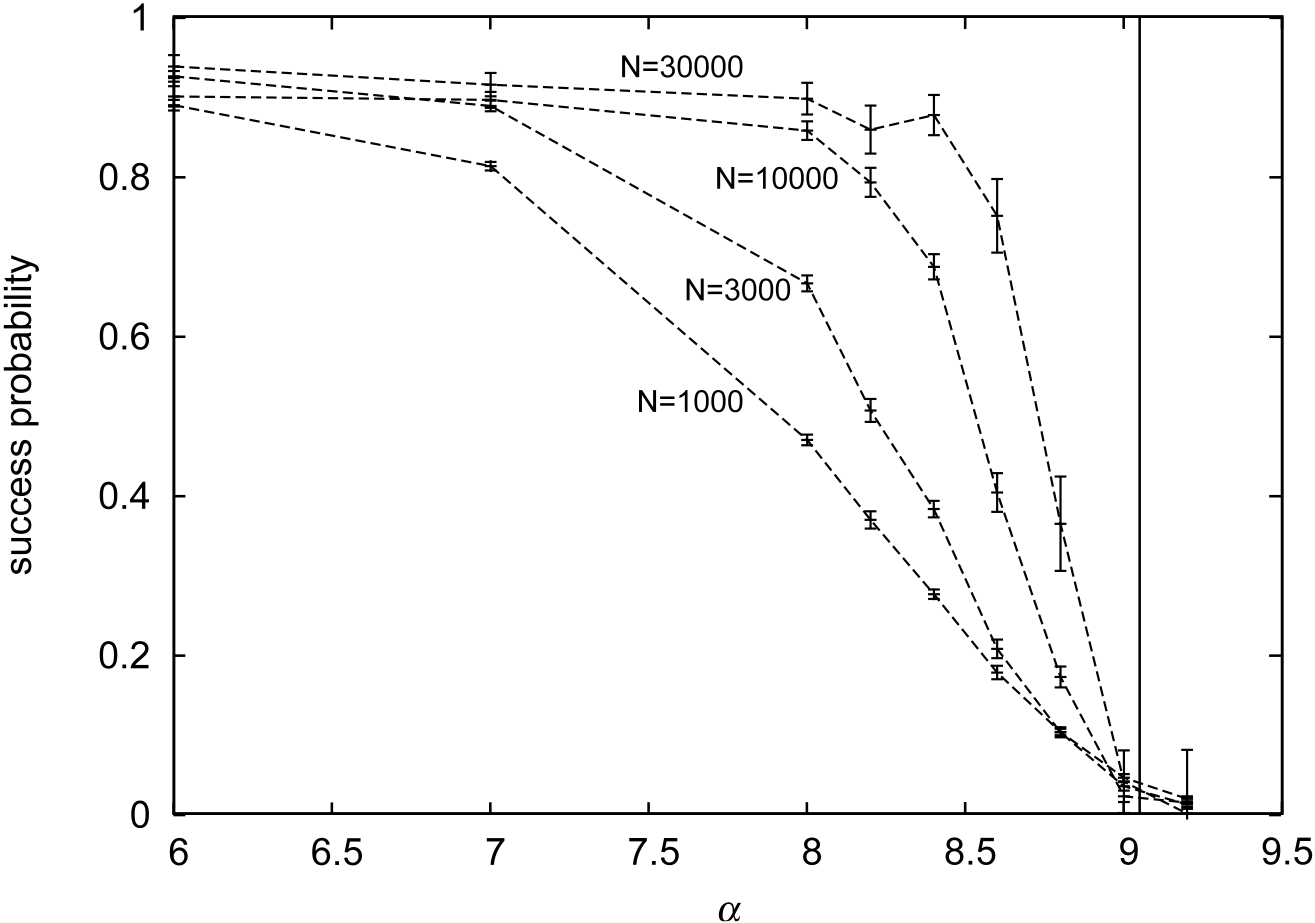}
			\end{center}
			\caption{Success probability for BP-guided decimation as a function of $\alpha$ for random 4-sat. The vertical line marks $\alpha_c^{(res)}$, value beyond which a condensation phase transition appears for the residual entropy (see e.g. figure \ref{ricciTersenghiBPfig79} (p.\pageref{ricciTersenghiBPfig79})). Figure from \cite{ricciTersenghiBP}.}
			\label{ricciTersenghiBPfig10}
		\end{figure}

		\newpage

		\newpage

		\newpage

		\newpage

	\subsubsection{The algorithmic barrier for BP}
		\label{BPalgorithmicBarrier}

		Now that we've seen the clustering and condensation phases in the $(\alpha ,\theta)$ plane\footnote{The $(\alpha,\theta)$ plane is the natural phase space for BP-guided decimation, and all other applications of BP to compute marginals of partially reduced instances of K-sat.}, we're ready to explore the algorithmic barrier $\alpha_a$ for BP, i.e. the set of results (both rigorous and experimental) connecting the success probability of BP to the geometry of its phase space. 
		Firstly it's interesting to consider the main hypothesis \cite{cojaOghlanBP} backing the introduction of BP in the context of decimation algorithms.
		\begin{hyp}
			\label{BPhyp} Define $\mu_{N-t+1}(\sigma_{i_t} ; r)$ to be the marginal computed by BP when the factor graph of $\mathcal{I}_{t-1}$ is restricted to the ball of radius $2r$ around node $\sigma_{i_t}$. Then we expect BP to be effective in computing the marginal $\mu_{N-t+1}(\sigma_{i_t})$ if the following conditions of locality hold:
			\begin{enumerate}
			\item $\forall \, \epsilon > 0 \; \exists \, r_\epsilon$ s.t. $\forall \, t \; \left| \; \mu_{N-t+1}(\sigma_{i_t}) - \mu_{N-t+1}(\sigma_{i_t} ; r_\epsilon) \; \right| \leq \epsilon$
			\item there is a function $r(N)$ s.t. $\mu_{N-t+1}(\sigma_{i_t}) \approx \mu_{N-t+1}(\sigma_{i_t}; r(N))$ w.h.p.
			\end{enumerate}
		\end{hyp}

		The validity of hypothesis \ref{BPhyp} for K-sat is established in \cite{montanariCounting} up to $\alpha_u = \frac{2 \log K}{K} + o(1)$ (called the \textit{Gibbs uniqueness} phase transition for K-sat) by considering correlation decay in tree factor graphs, and using the fact that for $1 < r \ll \log N$ the ball around any variable node is w.h.p. a tree. \\

\begin{figure}[h!]
			\begin{center}
			\includegraphics[width=16cm]{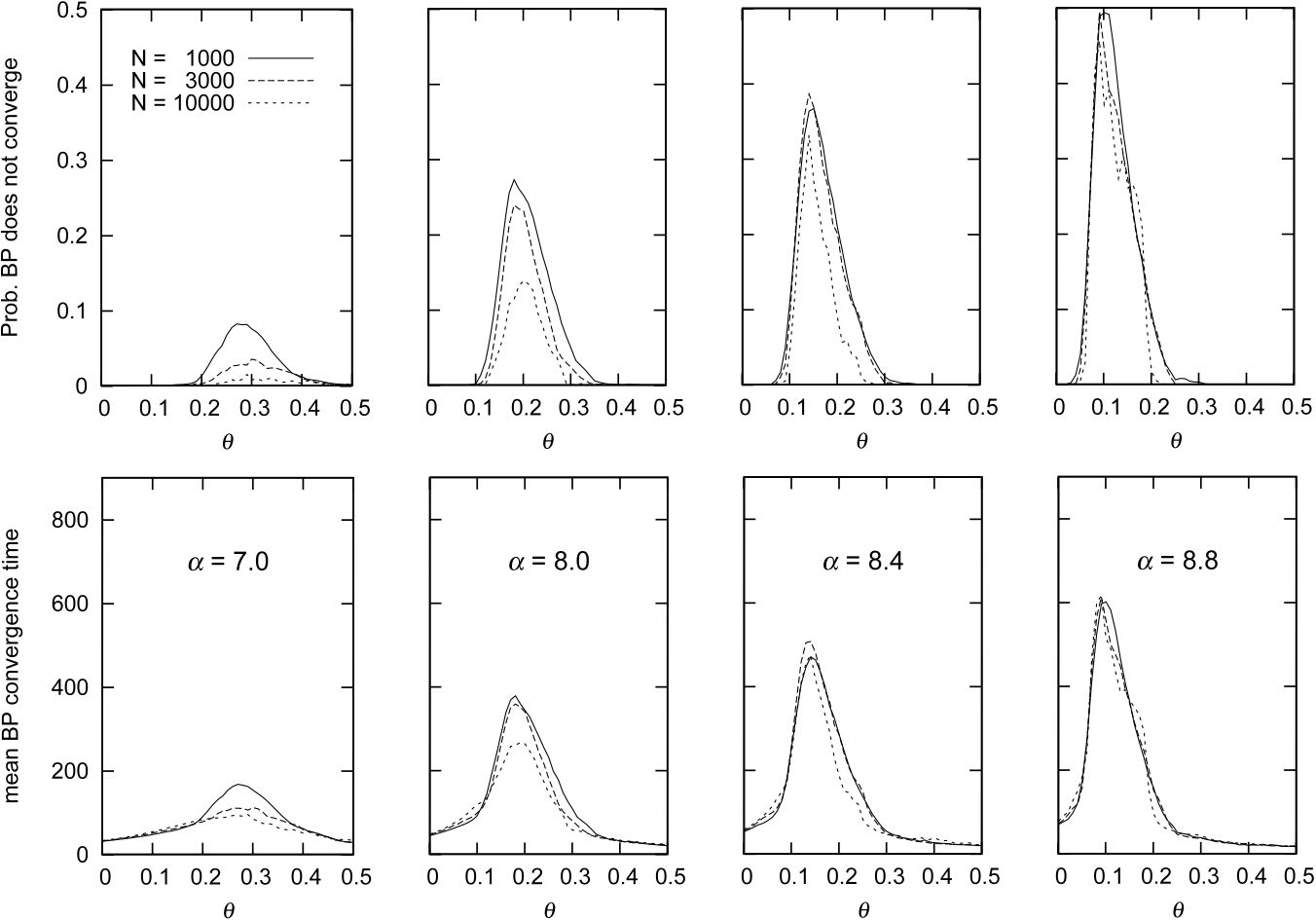}
			\end{center}
			\caption{Probabilities of non-convergence within 1000 iterations (panel above) and mean convergence time (panel below) for BP-guided decimation in random 4-sat, at various values of $\alpha$ approaching the (conjectured) algorithmic phase transition $\alpha_{a} \approx 9.05$. Both the probability of non-convergence and the mean convergence time peak at a value $\theta_{max}$, with sharper and taller peaks as $\alpha$ approaches $\alpha_a$. 
			%Values of $\theta_{max}(\alpha)$ of slowest convergence for $\alpha = 8.0,8.4,8.8$ are shown as solid dots in the phase diagram of figure \ref{ricciTersenghiBPfig79} (p.\pageref{ricciTersenghiBPfig79}). Since the maximum running time is expected to diverge at $\alpha_{a} \approx \alpha_{c}^{(res)}$, in correlation with the appearance of a condensation phase transition for the residual entropy, we expect the $\theta_{max}(\alpha)$ curve of maximum running time to be the precursor, in the $\alpha < \alpha_c^{(res)}$ region, of the $\theta_{c}(\alpha)$ curve: figure \ref{ricciTersenghiBPfig79} indeed supports this. 
			Figure from \cite{ricciTersenghiBP}.}
			\label{ricciTersenghiBPfig12}
		\end{figure}

		Figures \ref{ricciTersenghiBPfig10} (p.\pageref{ricciTersenghiBPfig10}) and \ref{ricciTersenghiBPfig12} (p.\pageref{ricciTersenghiBPfig12}) show a common scenario in numerical experiments for BP-guided decimation over small $K$: there is an $\alpha_a$ near to the condensation point $\alpha_c^{(res)}$ s.t.
		
		\begin{enumerate}
			
			\item the success probability is positively bounded below in the thermodynamic limit for all $\alpha < \alpha_a$
			
			\item the success probability vanishes in the thermodynamic limit for all $\alpha > \alpha_a$
		
		\end{enumerate}
		
		This has prompted the further hypothesis that, in the \textbf{undecimated regime} $\theta = 0$, BP-guided decimation should perform well up to the condensation phase transition $\alpha_c = 2^K \log 2 - \frac{3}{2}\log 2+O(2^{-K})$. 
		Unfortunately the following rigorous result from \cite{cojaOghlanBP} shows that this is just an artefact of the small $K$ analysis
		\footnote{The only ones around: it's really hard to do any numerical simulation even for $K \geq 10$}:

		\begin{theorem} (Success probability of BP)\\
			There is a constant $\rho > 0$ s.t. for all $\rho \frac{2^K}{K} \geq \alpha \geq 2^K \log 2$ we have 
			\begin{equation}
			\text{Success probability for BP-guided decimation over }\mathcal{I} \leq \exp\left[ - \Omega(N) \right] \text{ w.h.p.}
			\end{equation}
			Furthermore this is not an artefact of the specific implementation of the algorithm: hypothesis \ref{BPhyp} fails as well for $\rho \frac{2^K}{K} \geq \alpha \geq 2^K \log 2$, regardless of the choice of function $r(N)$. 
		\end{theorem}

		The proof of the theorem begins with the observation that, upon success, BP-guided decimation doesn't \textit{just} find a solution,  but it finds a uniformly distributed one. The proof then pushes this observation further to an analysis of the BP operator on a random decimated instances of K-sat: these are shown to obey some specific \textit{quasi-randomness} properties that turn out to be sufficient for a probabilistic analysis of the computation.\\

		In conclusion an algorithmic barrier $\alpha_a = O(2^K/K)$ exists for BP on K-sat, exactly the same as for all other solvers (except for Fix). It is related to the condensation phase transition only for small $K$: for large $K$ the barrier not only doesn't reach the condensation point, but in fact appears before the clustering phase transition $\alpha_d$.

	\subsection{From BP to SP}
		The solution to computing marginals in the clustering and condensation phases is the \textit{Survey Propagation} algorithm, which we'll see in section \ref{section_SP}: the clever idea behind it is to overcome the shattering of the solution space by considering the uniform measure over clusters rather than the uniform measure over spin configurations. Where BP computes global marginals, SP computes distributions\footnote{Which makes messages all the more complicated (and interesting), as now we have to pass distributions around.} of marginals restricted to clusters: the tool used to compute those distribution is the \textit{cavity method}, to which we'll now turn.

		%\subsection{The \texttt{Fix} algorithm of \cite{cojaOghlanFix}}
		%Before we dive into the cavity method, it's worth spending a few brief words on the \texttt{Fix} algorithm of \cite{cojaOghlanFix}. The algorithm is inspired by Unit Clause, which given a factor node inspects all the neighbouring variable nodes to determine whether the factor node is a unit clause: in this sense it is, like Shortest clause and Walksat, a depht-1 algorithm. \texttt{Fix} does the same as Unit Clause, but goes to dept-3 in the neighbourhood of the factor node. The analysis of \texttt{Fix} given in \cite{cojaOghlanFix} is much more complicated than the classic ones for depth-1 algorithms, and entails both probabilistic (e.g. martingales) and combinatorial methods.
%% END Algorithms for K-sat

\newpage

%% BEGIN The Cavity Method and Survey Propagation
\section{The Cavity Method and Survey Propagation}
	\label{cavityMethod}

	\subsection{The Cavity Method}
		Unless otherwise stated this section is based on \cite{core_mezardParisiZecchina}\cite{core_mezardZecchina}\cite{core_gibbsStates}\cite{core_clustersAndRSB}. The cavity method was developed in the study of spin glasses and is, at least in principle, equivalent to the replica method. It does, however, posses two key advantages over the latter:
		
		\begin{enumerate}
		
			\item[(a)] it is a standard probabilistic analysis of the factor graph, which makes explicit hypotheses on the correlation of the variable nodes. This makes it easier to turn into a rigorous method (or at least distinctly more rigorous than the replica method).
			
			\item[(b)] the replica method starts by averaging over the disorder, while the cavity method works at fixed disorder (and only takes the average in the end, when statistical properties have to be computed with it). This makes it suitable for actual algorithms.
		
		\end{enumerate}

		Consider a spin $\sigma_N$ connected to some number $k$\footnote{A poisson r.v. with mean $\alpha K$ (in the thermodynamic limit).} of factor nodes $f_{a_1},...,f_{a_k}$, each $f_a$ being connected to $K-1$ other spins $\sigma_{i(a,1)},...,\sigma_{i(a,K-1)}$. The spins $\sigma_{i(a,l)}$ are then very near to each other\footnote{And thus strongly correlated.}, but (and this is the key point) upon removal of $\sigma_N$ (a.k.a. creation of a \textbf{cavity}) their distance typically jumps up to $\log(N)$.\\

		In the RS phase, where only one cluster exists, the correlation-decay property of pure states allows us to claim that large distance implies vanishing correlation, and to conclude that the spins $\sigma_{i(a,l)}$ are, upon removal of $\sigma_N$, uncorrelated in the thermodynamic limit: recalling the interpretation of eq'n \ref{BPmessageInterpretation} (p.\pageref{BPmessageInterpretation}), one can then justify writing the marginal for $\sigma_N$ as the product of the $\eta_{a \rightarrow N}$, themselves defined in terms of the $\eta_{i(a_j,l)\rightarrow a_j}$.
		This is the \textbf{cavity method} derivation
			\footnote{In the RS phase. Notice that this derivation starts failing at $\alpha_{d,+}$, way before the clustering phase transition.} 
		of the BP message-passing eq'ns \ref{BPmessageRules} (p.\pageref{BPmessageRules}): in fact the BP messages embody the cavity method itself, the induction process behind the latter being encoded in the fixed-point equations governing the formers. From now on we'll phrase all our results in terms of messages (cavity-fields and cavity-biases).\\

		In the d1RSB and 1RSB phases, on the other hand, one cannot claim a connection between large distance and vanishing correlation, as shown in section \ref{section_clustering}. 
		The issue is the presence of multiple pure states, and the solution is to apply the cavity method while keeping separate messages for each each pure state: this is equivalent to running BP on each pure state via the following procedure (which is given in more detail in \cite{core_clustersAndRSB}):
		
		\begin{enumerate}
		
			\item restrict the factor graph to a tree-like neighborhood $T(\sigma_N)$ of $\sigma_N$, and let $B$ be its boundary.
			
			\item impose boundary conditions to the messages from boundary nodes in $B$ that isolate a single pure state $\psi$.
			
			\item run BP on $T(\sigma_N)$ (constrained to the boundary conditions) and obtain a family of cavity-fields $(h^{(\psi)}_{i \rightarrow a})_{ia}$ and of cavity-biases $(u^{(\psi)}_{a \rightarrow i})_{ia}$ for the pure state $\psi$.
			
		\end{enumerate}
		
		Each edge $i\rightarrow a$ of the factor graph will then have a family of cavity-fields $(h^{(\psi)}_{i \rightarrow a})_{\psi}$ and cavity-biases $(u^{(\psi)}_{a \rightarrow i})_{\psi}$ indexed by the pure states, and eq'n \ref{BPKsatLocalField} (p.\pageref{BPKsatLocalField}) will compute $\mu_N^{(\psi)}$.

	\subsubsection{From messages to surveys}

		The first step in dealing with the multi-cluster scenarios of d1RSB and 1RSB is to take the BP message passing eq'ns \ref{ksatBPRules} (p.\pageref{ksatBPRules}) and reinterpret them in a probabilistic way, by considering them as the deterministic case of more general message-passing eq'ns involving probability distributions. 
		The idea is that the BP messages are accurate descriptions of the marginals when restricted to a single pure state (which has the right correlation decay properties by definition), and that the case of many pure states can be treated by passing around PDFs of messages over states.\\

		One then changes the messages into:
		\begin{equation}
			\begin{aligned}
				P_{i \rightarrow a}(h) &\eqdef \text{ probability density for a cavity-field }h^{(\psi)}_{i \rightarrow a}\text{ to take value }h \\
				Q_{a \rightarrow i}(u) &\eqdef \text{ probability density for a cavity-bias }u^{(\psi)}_{a \rightarrow i}\text{ to take value }u 
			\end{aligned}
		\end{equation}
		and recovers BP as the 1-state, deterministic case
		\begin{equation}
			\begin{aligned}
				P_{i \rightarrow a}(h) &= \delta(h - h_{i \rightarrow a})\\
				Q_{a \rightarrow i}(u) &= \delta(u - u_{a \rightarrow i})
			\end{aligned}
		\end{equation}
		The $P_{i \rightarrow a}(h)$ messages are called \textbf{$h$-surveys}, while the $Q_{a \rightarrow i}(u)$ messages are called \textbf{$u$-surveys}.\\

		The BP message-passing fixed-point eq'ns \ref{ksatBPRules} (p.\pageref{ksatBPRules}) (or rather their normalised version) then become the 1-state, deterministic limit of the following \textbf{1RSB distributional fixed-point equations} for surveys:
		\begin{equation}
			\label{1RSBDistributionalFixedPointEquation}
			\begin{aligned}
				P_{i \rightarrow a}(h) &= 
				\dfrac{1}{Z_i[(Q_{b \rightarrow i})_b]} \;
				\int \prod_b dQ_{b \rightarrow i}(u_{b\rightarrow i}) \;
				\delta \left( h - \sum_b u_{b\rightarrow i} \right) z_i[(u_{b\rightarrow i})_b]^{m}\\
				& \text{ where we have taken } b \textbf{ to range over } \partial i \backslash \{a\}\\
				Q_{a \rightarrow i}(u) &= 
				\dfrac{1}{Z_a[(P_{j \rightarrow a})_j]} \;
				\int \prod_j dP_{j \rightarrow a}(h_{j\rightarrow a}) \;
				\delta \left( u - f\left[(h_{j\rightarrow a})_j\right] \right) z_{a'}[(h_{j\rightarrow a})_j]^{m}\\
				& \text{ where we have taken } j \textbf{ to range over } \partial a \backslash \{i\}\\
			\end{aligned}
		\end{equation}
		where $\left(u_{a \rightarrow i} - f\left[(h_{j\rightarrow a})_j\right] \right)$ is obtained from eq'n \ref{ksatBPRules} (p.\pageref{ksatBPRules}). The effect of the Parisi 1RSB parameter $m$ is the same as in the replicated free-entropy density of section \ref{Parisi1RSBParameterSection} (p.\pageref{Parisi1RSBParameterSection}): it favours clusters described by a specific 1RSB solution, concentrating the measure over them.\\

		$Z_{i}$ and $Z_{a}$ are normalisation factors and we have defined the following 
		\begin{equation}
		\label{RSFreeEntropyShifts}
		\begin{aligned}
		z_i[(u_{b\rightarrow i})_b] & \eqdef \dfrac{\exp\left[+\sum_{b} u_{b \rightarrow i}\right]+\exp\left[-\sum_{b} u_{b \rightarrow i}\right]}{\prod_b \cosh u_{b \rightarrow i}}\\
		z_{a'}[(h_{j\rightarrow a})_j] & \eqdef 1+\exp\left(-2 f\left[(h_{j\rightarrow a})_j\right] \right) \\
		\end{aligned}
		\end{equation}

		Note that $P_{i \rightarrow a}(h)$ and $Q_{a \rightarrow i}(u)$ are themselves r.v.s (depending on the r.v. $\mathcal{I}$):  we define $\mathcal{P}(P)$ and $\mathcal{Q}(Q)$ to be their respective PDFs, and will abuse notation by writing $\mathcal{P}(h)$ and $\mathcal{Q}(u)$ for the distributions of cavity-fields and cavity-biases after having taken the randomness of $\mathcal{I}$ into account.

	\subsubsection{The cavity method: free-entropy density, overlaps and marginals}
		The quantities $z_i$, $Z_i$ and $Z_a$ are called \textbf{free-entropy shifts}, and contribute to the cavity method formulation of the RS and 1RSB free-entropy densities, which we'll now see.\\
		 
		The cavity method estimate for the \textbf{RS free-entropy density} is
		\begin{equation}
			\begin{aligned}
				\omega & = -\alpha K \expect{\log z_{ia}(h,u)} + \alpha \expect{\log z_{a}(h_1,...,h_K)} + \expect{\log z_{i}(u_1,...,u_k)}
			\end{aligned}
		\end{equation}
		where $k$ is poisson of mean $\alpha K$, and $u,u_1,u_2,...$ are iid distributed according to $\mathcal{Q}(u)$, and $h,h_1,h_2,...$ are iid distributed according to $\mathcal{P}(h)$. 
		The \textbf{RS free-entropy shifts} $z_{ia}$, $z_{a}$ and $z_i$ are defined by eq'n \ref{RSFreeEntropyShifts} and the following
		\begin{equation}
			\begin{aligned}
				z_{ia}(h,u) & \eqdef 1+\tanh h \tanh u\\
				z_{a}(h_1,...,h_r) & \eqdef 1-\prod\limits_{i = 1}^{r} \frac{\exp[-h_i]}{2 \cosh[h_i]}
			\end{aligned}
		\end{equation}

		The cavity method estimate for the \textbf{1RSB (replicated) free-entropy density} is
		\begin{equation}
			\label{1RSBFreeEntropyDensity}
			\begin{aligned}
				\Phi(m) & =  -\alpha K \expect{\log Z_{ia}(P,Q)} + \alpha \expect{\log Z_{a}(P_1,...,P_K)} + \expect{\log Z_{i}(Q_1,...,Q_k)}
			\end{aligned}
		\end{equation}
		where $k$ is poisson of mean $\alpha K$, and $Q,Q_1,Q_2,...$ are iid distributed according to $\mathcal{Q}(Q)$, and $P,P_1,P_2,...$ are iid distributed according to $\mathcal{P}(P)$. 
		The \textbf{1RSB free-entropy shifts} $Z_{ia}$, $Z_{a}$ and $Z_i$ are defined by eq'n \ref{RSFreeEntropyShifts} and the following
		\begin{equation}
			\begin{aligned}
				Z_{ia}(P,Q) & \eqdef \int dP(h) dQ(u) \; z_{ia}(h,u)^m\\
				Z_{a}(P_1,...,P_r) & \eqdef \int \prod\limits_{i=1}^{r}dP_{i}(h_i) \; z_a(h_1,...,h_r)^m\\
				Z_{i}(Q_1,...,Q_l) & \eqdef \int \prod\limits_{a=1}^{l}dQ_{a}(u_a) \; z_i(u_1,...,h_l)^m
			\end{aligned}
		\end{equation}

		Furthermore the surveys allow computation of the overlaps: the intra-state overlap of the dominant RS cluster is given by
		\begin{equation}
			q_{RS} = \expect{\tanh^2 h}
		\end{equation} 
		while the overlaps for the d1RSB and 1RSB phases are given by
		\begin{equation}
			\begin{aligned}
				q_{0}(m) = \expect{ \left(\int \, \tanh h \, dP(h) \right)^2} & & q_{1}(m) = \expect{ \int \, \tanh^2 h \, dP(h)}
			\end{aligned}
		\end{equation} 

		Finally we can obtain the \textbf{surveys of local fields} as
		\begin{equation}
			\label{surveyLocalField}
			P_{i}(h) = 
			\dfrac{1}{Z_i[(Q_{b \rightarrow i})_b]} \;
			\int \prod_b dQ_{b \rightarrow i}(u_{b\rightarrow i}) \;
			\delta \left( h - \sum_b u_{b\rightarrow i} \right) z_i[(u_{b\rightarrow i})_b]^{m}
		\end{equation} 
		where we have taken $b$ to range over $\partial i$ this time.

		\newpage

	\subsection{Survey Propagation}
		\label{section_SP}
		Unless otherwise stated, this section is based on the seminal papers \cite{core_mezardParisiZecchina}\cite{core_mezardZecchina}.

	\subsubsection{Encoding the surveys}

		We want to use the surveys in an algorithm, but how do we practically carry PDFs around? The trick is to work in the zero temperature limit $\beta \rightarrow \infty$, where the second rule of  \ref{ksatBPMessages} (p.\pageref{ksatBPMessages}) simplifies to a minimisation operation.  
		The cavity-biases are thenn only allowed to take a finite number of standard values
			\footnote{If the random initialisation of BP is done within those standard values, that is.} 
		(e.g. in 3-sat at zero temperature we have $u_{a \rightarrow i} \in \{0,\pm 1\}$) and the u-surveys can be encoded as histograms.

	\subsubsection{The SP algorithm}
		The SP algorithm is an evolution of BP, adapted to the multi-cluster scenario of the d1RSB and 1RSB phases: the messages passed around are now the u-surveys $Q_{a \rightarrow i}(u)$, and instead of computing a local field $h_i$ (which gave us a marginal via eq'n \ref{BPKsatLocalField} (p.\pageref{BPKsatLocalField})) we compute a distribution $P_i(h)$ of local fields (which gives us a distribution of marginals). The algorithm works at a fixed value of the Parisi 1RSB parameter $m$, i.e. focusing the measure on a specific family of clusters.

		\begin{enumerate}
		\item[0.] Initialise all the u-surveys at random
		\item Select a random factor node $a$
		\item For each $i \in \partial a$, compute the h-survey $P_{i \rightarrow a}(h)$ by using eq'n \ref{1RSBDistributionalFixedPointEquation} (p.\pageref{1RSBDistributionalFixedPointEquation})
		\item For each $i \in \partial a$, update the u-survey $Q_{a \rightarrow i}(u)$ by using eq'n \ref{1RSBDistributionalFixedPointEquation} (p.\pageref{1RSBDistributionalFixedPointEquation})
		\item Test convergence. If convergence is not reached, go to step 1. Otherwise:
		\begin{enumerate}
		\item compute the surveys of local fields $\left(P_i(h)\right)_i$ by using eq'n \ref{surveyLocalField} (p.\pageref{surveyLocalField}) and the u-surveys which were just computed
		\item compute the free-entropy density $\Phi(m)$ by using eq'n \ref{1RSBFreeEntropyDensity} (p.\pageref{1RSBFreeEntropyDensity}), the u-surveys and the h-surveys which were just computed
		\item return $\left(P_i(h)\right)_i$ and $\Phi(m)$
		\end{enumerate}
		\end{enumerate}

		The SP algorithm can be used in a decimation procedure, called \textbf{SID}, similar to BP-guided decimation: steps 2 and 3 of the procedure from section \ref{BPGuidedDecimationAlgorithm} are changed into
		\begin{enumerate}
		\item[2.] Run SP, obtain the surveys $\left(P_i(h)\right)_i$ and for each survey compute $w_i^{\pm},w_i^{0}$:
		\begin{equation}
		\begin{aligned}
		w_i^{+} = \int \limits_{0^+}^{+\infty} dP_i(h) &,& w_i^{-} = \int \limits_{-\infty}^{0^-} dP_i(h) &,& w_i^{0} = 1-w_i^{+}-w_i^{-}
		\end{aligned}
		\end{equation}
		\item[3a.] Check if the system is in the \textbf{paramagnetic phase} (i.e. for every spin $\sigma_i$ we have $w_i^{0} = 1$). If it is, try running a some fast local search algorithm (like simulated annealing or Walksat), and if the algorithm finds a solution return it. Otherwise proceed to step 3b
		\item[3b.] Select and fix the\footnote{Or a random one if more than one exists with same bias.} most biased spin (i.e. largest $\left| w_i^{+} - w_i^{-} \right|$)
		\end{enumerate}

	\subsubsection{Convergence of SP}
		SP converges to different solutions depending on the value of $m$ used, i.e. depending on which clusters we are concentrating our measure onto. It will usually be necessary to run SP at different values of $m$ to find the best solution: the most efficient way to do this is to start at a high value of $m$ (like $m = 1$ or a bit above $m_s$, if known) and progressively lower the value of SP, using the u-surveys computed at higher-$m$ runs as initial condition for lower-$m$ runs, to speed up convergence. The value of $m$ describing the thermodynamically relevant clusters is then, as usual, the one minimising $\Phi(m)/m$ (where we use the estimate for $\Phi(m)$ returned by SP). The following experimental results about convergence are reported in \cite{core_mezardZecchina}\footnote{There they are reported as a function of the parameter $y = \frac{\partial \Sigma}{\partial \epsilon}$, where $\epsilon$ is the density of violated clauses, and a (fairly straightforward) rephrasing in terms of the parameter $m$ has been done.}.\\

		For $\alpha < \alpha_{d,+}$ SP always converges to the trivial paramagnetic solution ($Q_{a \rightarrow i}(u) = \delta(u)$ and $P_i(h) = \delta(h)$ for all $i$), irrespective of the value of $m$: this is to be expected, as there is only one pure state. Fast local search algorithms can be used effectively in this phase.\\

		For $\alpha_d < \alpha < \alpha_c$ the behaviour depends on $m$:
		\begin{enumerate}
		\item[(a)] for $m$ low enough the algorithm converges to the trivial paramagnetic solution.
		\item[(b)] for $m$ near enough to 1 the algorithm converges to a unique non-trivial solution, describing the surveys amongst the thermodynamically dominating clusters.
		\end{enumerate}

		For $\alpha_c < \alpha < \alpha_s$ the behaviour depends even more strongly on $m$:
		\begin{enumerate}
		\item[(a)] for $m$ low enough the algorithm converges to the trivial paramagnetic solution.
		\item[(b)] for intermediate values\footnote{It's reasonable to assume this means for values of $m \approx m_s$, but the authors don't make this observation explicitly.} of $m$ the algorithm converges to a unique non-trivial solution, describing the surveys amongst the thermodynamically dominating clusters.
		\item[(c)] For larger values of $m$ the algorithm stops converging. The range of values for which the algorithm converges to a non-trivial solution is reported to be sufficient, in the numerical experiments of \cite{core_mezardZecchina}, for the free-entropy density to converge to the value expected for the thermodynamically relevant clusters.
		\end{enumerate}  

	\section{Conclusion}

		Starting from the Sherrington-Kirkpatrick model of spin glass, we have presented the main contributions given, between the years 2002 and 2010, to the application of spin glass theory to understanding the ensamble properties of the solution space of K-sat. From the Hamiltonian formulation to the condensation phase transition, we have reviewed, amalgamated and consolidated a decade of work
			\footnote{Not counting the spin glass works it's based on, which span almost 30 years.,}
		to provide a global understanding of the achievements and potential of this field.\\

		An earlier version of this work was presented as the author's 2013 Part III Essay at DAMTP, University of Cambridge.
%% END The Cavity Method and Survey Propagation

\newpage

%% BEGIN Bibliography

%% END Bibliography


\begin{thebibliography}{99}

	% Survey material
	\bibitem[-]{}\underline{\textbf{Topic survey references:}}

	\bibitem{survey_randomSatisfiability} D Achlioptas, \textit{Random satisfiability}, Handbook of Satisfiability 185 (2009).

	\bibitem{survey_cojaOghlan} A Coja-Oghlan, \textit{A statistical mechanics perspective on hard computational problems}, Slides online for the Warwick Statistical Mechanics Seminar of 11 Feb 2010.

	\bibitem{survey_talagrandNew} M Talagrand, \textit{Mean field models for spin glasses}, Springer (2010).

	\bibitem{survey_talagrandOld} M Talagrand, \textit{Spin glasses: A challenge for mathematicians}, Springer (2003).

	\bibitem{survey_mezardParisiVirasoro} M Mézard, G Parisi, M A Virasoro, \textit{Spin glass theory and beyond}, (1987).

	% Core material
	\bibitem[-]{}\underline{\textbf{Core references:}}

	\bibitem{core_algorithmicBarriers} D Achlioptas, A Coja-Oghlan, \textit{Algorithmic barriers from phase transitions}, Foundations of Computer Science, IEEE 49th Annual IEEE Symposium on (2008).

	\bibitem{core_gibbsStates} F Krzakala, A Montanari, F Ricci-Tersenghi, G Semerjian, L Zdeborova \textit{Gibbs states and the set of solutions of random constraint satisfaction problems} P of the Nat. Acad. of Sc. (2007).

	\bibitem{core_clustersAndRSB} A Montanari, F Ricci-Tersenghi, G Semerjian \textit{Clusters of solutions and replica symmetry breaking in random k-satisfiability}, Journal of Statistical Mechanics: Theory and Experiment (2008)[version used: http://arxiv.org/abs/0802.3627v2]. 

	\bibitem{core_mezardParisiZecchina} M Mezard, G Parisi, R Zecchina, \textit{Analytic and algorithmic solution of random satisfiability problems}, Science Vol 297 (2002).

	\bibitem{core_mezardZecchina} M Mezard, R Zecchina, \textit{The random K-satisfiability problem: from an analytic solution to an efficient algorithm}, Physical Review E (2002) [version used: http://arxiv.org/abs/cond-mat/0207194v3].

	% Gibbs measures
	\bibitem[-]{}\underline{\textbf{References on Statistical Physics and Gibbs measures:}}

	\bibitem{gibbs_grimmett} G R Grimmett, \textit{A theorem about random fields}, B of the London Mathematical Society (1973).

	\bibitem{gibbs_monassonZecchina} R Monasson, R Zecchina. \textit{Statistical mechanics of the random K-satisfiability model} (1997).

	\bibitem{gibbs_tong} D Tong, \textit{Lectures on Statistical Physics}, Notes online from the author's DAMTP teaching page.

	\bibitem{gibbs_georgii} H O Georgii, \textit{Gibbs measures and phase transitions}, Walter de Gruyter (2011).

	\bibitem{gibbs_factor} F R Kschischang, B J Frey, H A Loeliger, \textit{Factor graphs and the sum-product algorithm}, IEEE Transactions on Information Theory (2001) 

	% Solution space geometry
	\bibitem[-]{}\underline{\textbf{Additional references on Solution Space Geometry:}}

	\bibitem{geometry_achlioptasRicciTersenghi} D Achlioptas , F Ricci-Tersenghi \textit{On the solution-space geometry of random constraint satisfaction problems}, P of the 38thannual ACM symposium on Theory of computing (2006).

	\bibitem{geometry_clustering} D Achlioptas, \textit{Solution Clustering in Random Satisfiability}, B of the Europ. Phys. J (2008).

	\bibitem{geometry_frozen} D Achlioptas, F Ricci-Tersenghi, \textit{Random formulas have frozen variables}, SIAM Journal on Computing (2009).

	% Replica symmetry breaking
	\bibitem[-]{}\underline{\textbf{Additional references on Replica Symmetry Breaking:}}

	\bibitem{RSB_stringGlasses} F Denef, \textit{String glasses}, Slides available online

	\bibitem{RSB_denef} F Denef, \textit{TASI lectures on complex structures}, http://arxiv.org/pdf/1104.0254.pdf (2011).

	\bibitem{parisiOrderParamSummary} G Parisi, \textit{Order Parameter for Spin-Glasses}, Physical Review Letters (1983)

	\bibitem{parisiOrderParam} G Parisi, \textit{Infinite Number of Order Parameters for Spin-Glasses}, Physical Review L (1979)

	\bibitem{parisiOrderParamFunction} G Parisi, \textit{The order parameter for spin glasses - A function on the interval 0-1}, (1979)

	% Belief propagation, Cavity Method, Survey propagation
	\bibitem[-]{}\underline{\textbf{Additional references on Belief Propagation, Cavity Method, Survey Propagation:}}

	\bibitem{BSP_newLook} E Maneva, E Mossel, M J Wainwright, \textit{A New Look at Survey Propagation and its Generalizations}, (2005) [version used: http://arxiv.org/abs/cs/0409012v3].

	\bibitem{BSP_localEquilibrium} A Braunstein, R Zecchina, \textit{Survey propagation as local equilibrium equations},  Journal of Statistical Mechanics: Theory and Experiment, IOP Publishing (2004) [version used: http://arxiv.org/abs/cond-mat/0312483].

	\bibitem{mezardMontanariInformationPhysicsComputation} M Mezard, A Montanari, \textit{Information, Physics, and Computation}

	\bibitem{cojaOghlanBP} A Coja-Oghlan, \textit{On Belief Propagation Guided Decimation for Random k-SAT}, http://arxiv.org/abs/1007.1328v1 (2010)

	\bibitem{montanari2007solving} A Montanari, F Ricci-Tersenghi, G Semerjian, \textit{Solving constraint satisfaction problems through belief propagation-guided decimation}, http://arxiv.org/abs/0709.1667v2 (2007)

	\bibitem{ricciTersenghiBP} F Ricci-Tersenghi, G Semerjian, \textit{On the cavity method for decimated random constraint satisfaction problems and the analysis of belief propagation guided decimation algorithms}, J. Stat. Mech. (2009)

	\bibitem{cojaOghlanFix} A Coja-Oghlan, \textit{A Better Algorithm for Random k-SAT}, SIAM J Comput (2010)

	\bibitem{cavity_mezardParisi}M Mézard, G Parisi, \textit{The cavity method at zero temperature}, Journ of Statistical Physics (2003).

	% Other
	\bibitem[-]{}\underline{\textbf{Other references:}}

	\bibitem{microstructureUltrametricity} M. Mezard, M Virasoro, \textit{The microstructure of ultrametricity}, Journal de Physique (1985)

	\bibitem{natureSpinGlassPhase} M Mezard, G Parisi, N Sourlas, G Toulouse, M Virasoro, \textit{Nature of the spin- glass phase}, Phisical Review Letters (1984)

	\bibitem{infiniteRSB} A Crisanti, T Rizzo, \textit{Analysis of the infinity-replica symmetry breaking solution of the Sherrington-Kirkpatrick model}, Physical Review E, APS (2002)

	\bibitem{stringGlasses} D Anninos, T Anous, J Barandes, F Denef, \textit{String Glasses} (to appear in the future)

	\bibitem{SKSpinGlass} D Sherrington, S Kirkpatrick, \textit{Solvable Model of a Spin-Glass}, Physical Review Letters (1975)

	\bibitem{zdebrodovaPhaseTransColouring} L Zdebrodova, F Krzakala, \textit{Phase transitions in the coloring of random graphs}, P Rev. E (2007)

	\bibitem{hidingQuietSolutions} L Zdeborova, F Krzakala, \textit{Hiding Quiet Solutions in Random Constraint Satisfaction Problems}, http://arxiv.org/abs/0901.2130v2

	\bibitem{mezardLandscape} M Mezard, M Palassini, O Rivoire\textit{Landscape of solutions in constraint satisfaction problems}, http://arxiv.org/abs/cond-mat/0507451v2

	\bibitem{papadimitriouOnSelecting} C H Papadimitriou, \textit{On selecting a satisfying truth assignment}, FOCS (1991)

	\bibitem{linearUpperBoundsRandomWalk} M Alekhnovich, E Ben-Sasson, \textit{Linear upper bounds for random walk on small density random 3-CNFs} SIAM J. Comput. (2007)

	\bibitem{understandingBP}J S  Yedidia, W T Freeman, Y Weiss, \textit{Understanding Belief Propagation and its generalizations}, IJCAI
	(2001)

	\bibitem{montanariCounting} A Montanari, D Shah, \textit{Counting good truth assignments of random k-SAT formulae}, Proc. 18th SODA, (2007)

	\bibitem{achlioptasPeres} D Achlioptas, Y Peres, \textit{The threshold for random K-SAT is $2^K \log 2 - O(k)$}, J AMS (2004).

	\bibitem{griffithsPDP} R C Griffiths, \textit{On the distribution of points in a Poisson Dirichlet process}, J Appl Probab (1988).

	\bibitem{griffithsPDP} R C Griffiths, \textit{On the distribution of points in a Poisson Dirichlet process}, J Appl Probab (1988).

	\bibitem{DNAEvolution} Durrett, \textit{Probability models for DNA sequence evolution}, Springer (2008)
\end{thebibliography}
\end{document}